\documentclass[11pt]{article}

\usepackage[utf8]{inputenc}

\usepackage{amsmath}
\DeclareMathOperator*{\argmax}{arg\,max}
\usepackage{amssymb}
\usepackage{color,soul}

\usepackage{amsthm}
\usepackage[authoryear]{natbib}
\usepackage{booktabs}
\usepackage{xcolor}
\usepackage{comment}
\usepackage{float}
\bibpunct{(}{)}{,}{a}{}{;}
\usepackage{fullpage}
\usepackage{graphicx}
\usepackage{subcaption}
\usepackage{tikz}
\usetikzlibrary{positioning}
\usepackage{tabularx}

\usepackage{smartdiagram}
\usepackage{multirow}
\usepackage{titling} % repeat title
\usepackage{multirow}
\usepackage[unicode=true,psdextra,hidelinks]{hyperref}
\usepackage{geometry}
\usepackage{listings}
\usepackage{color}
\usepackage{adjustbox}
\usepackage{listings}
\usepackage{color} %red, green, blue, yellow, cyan, magenta, black, white
\usepackage{multirow}
\usepackage{soul}
\usepackage{tocloft}
\usepackage{titlesec}

\definecolor{mygreen}{RGB}{28,172,0} % color values Red, Green, Blue
\definecolor{mylilas}{RGB}{170,55,241}

\usepackage{booktabs}
\usepackage{rotating}

\usepackage{xcolor}

\definecolor{codegreen}{rgb}{0,0.6,0}
\definecolor{codegray}{rgb}{0.5,0.5,0.5}
\definecolor{codepurple}{rgb}{0.58,0,0.82}
\definecolor{backcolour}{rgb}{0.95,0.95,0.92}

\lstdefinestyle{mystyle}{
    backgroundcolor=\color{backcolour},   
    commentstyle=\color{codegreen},
    keywordstyle=\color{magenta},
    numberstyle=\tiny\color{codegray},
    stringstyle=\color{codepurple},
    basicstyle=\ttfamily\footnotesize,
    breakatwhitespace=false,         
    breaklines=true,      
    captionpos=b,         
    keepspaces=true,      
    numbers=left,         
    numbersep=5pt,       
    showspaces=false,     
    showstringspaces=false,
    showtabs=false,       
    tabsize=2
}

\newtheorem{lemma}{Lemma}
\newtheorem{prop}{Proposition}

\usepackage{setspace}
\vspace{-40pt}
\title{Algorithmic Collusion of Pricing and Advertising on E-commerce Platforms\thanks{We thank Eric Bradlow, Pinar Yildirim, Raju Jagmohan Singh, Raghu Iyengar, Leon Musolff, Christophe Van den Bulte, Soheil Ghili, Sylvia Hristakeva, Omid Rafieian, Tianxin Zou, Kanishka Misra, Sridhar Moorthy, Tat Chan, Yufeng Huang, Ilya Morozov, and participants at the 4th Annual AI in Management Conference, 14th Annual Theory and Practice of Marketing Conference,  2nd FTC Conference on Marketing and Public Policy, 2025 Workshop on Platform Analytics, 35th Annual POMS Conference, 2025 ASA Marketing Section Doctoral Dissertation Research Award Session, and Conference on Frontiers in Machine Learning and Economics for their helpful comments. All errors are our own. }}
\vspace{-20pt}
\author{Hangcheng Zhao\thanks{Rutgers Business School. \href{mailto:hangcheng.zhao@rutgers.edu}{hangcheng.zhao@rutgers.edu}} 
\and Ron Berman\thanks{The Wharton School of the University of Pennsylvania.  \href{mailto:ronber@wharton.upenn.edu}{ronber@wharton.upenn.edu}} }
\date{October 2025\\[2ex]  % Adds vertical space after the date}
}
\begin{document}

\maketitle
\thispagestyle{empty}

\newpage
\setcounter{page}{1}
\begin{center}
\Large
\thetitle
\end{center}

\setcounter{page}{1}
\onehalfspacing

\begin{abstract}

When online sellers use AI learning algorithms to automatically compete on e-commerce platforms, there is concern that they will learn to coordinate on higher than competitive prices. However, this concern was primarily raised in single-dimension price competition. We investigate whether this prediction holds when sellers make pricing and advertising decisions together, i.e., two-dimensional decisions. We analyze competition in multi-agent reinforcement learning, and use a large-scale dataset from Amazon.com to provide empirical evidence. We show that when consumers have high search costs, learning algorithms can coordinate on prices lower than competitive prices, facilitating a win-win-win for consumers, sellers, and platforms. This occurs because algorithms learn to coordinate on lower advertising bids, which lower advertising costs, leading to lower prices and enlarging demand on the platform. We also show that our results generalize to any learning algorithm that uses exploration of price and advertising bids. Consistent with our predictions, an empirical analysis shows that price levels exhibit a negative interaction between estimated consumer search costs and algorithm usage index. We analyze the platform's strategic response and find that reserve price adjustments will not increase platform profits, but commission adjustments will, while maintaining the beneficial outcomes for both sellers and consumers.

Keywords: Artificial Intelligence, Algorithmic Collusion, Platforms, Advertising, Sponsored Product Ads, Reinforcement Learning, Q-learning, Consumer Search. 
\end{abstract}

\doublespacing

\section{Introduction}
\label{sec:intro}

Reinforcement learning (RL) algorithms have been successfully applied to marketing applications including automated advertising bidding \citep{cai2017real,wu2018budget}, pricing \citep{dube2017scalable,misra2019dynamic,smith2023optimal}, personalized recommendations and more \citep{schwartz2017customer,liu2023dynamic, rafieian2023optimizing}. 
These algorithms are very suitable for learning in dynamic environments without much prior information, because they can be model-free and learn the reward structure as they explore the outcomes from their actions.
The RL algorithms do not have to know how ad auctions operate or how demand responds to prices, or any other a-priori details about the economic environment in order to learn. In practice, multiple firms publicize that they use reinforcement learning to optimize pricing and advertising strategies in real time simultaneously.\footnote{For example, Alibaba implemented deep reinforcement learning for large-scale dynamic pricing \citep{liu2019dynamic}. Feedvisor promotes their ``continuous real-time optimizations for both pricing and advertising allowing to drive profitable growth''(\url{https://feedvisor.com/feedvisors-integrated-solution/}, accessed Oct 20, 2025), while Profasee's ``Price and Ad Optimizers work together to revolutionize the profit potential of your Amazon store.''(\url{https://profasee.com/all-in-optimizer/}, accessed Oct 20, 2025).}

In recent research that used simulations to analyze competitive pricing using reinforcement-learning, \cite{calvano2020artificial,  hansen2021frontiers} and \cite{johnson2023platform} find that these algorithms can learn to tacitly collude on setting prices that are higher than competitive prices (even without direct communication), which is harmful to consumers. The issue became a major policy worry, and the FTC even stated that ``Price fixing by algorithm is still price fixing.''\footnote{Recently, the Federal Trade Commission and Department of Justice are taking different actions to fight algorithmic collusion. See detail at \url{https://www.ftc.gov/business-guidance/blog/2024/03/price-fixing-algorithm-still-price-fixing}.}

In this paper, we ask whether algorithmic decision making by competing firms is necessarily detrimental to consumers, or whether there are cases when consumers, as well as firms and an intermediating platform, might benefit from them. 
One reason for asking this question is that most of the analyses of algorithmic decision making often focused on one-dimensional learning such as learning to price, or to advertise, but in reality, firms often need to make multi-dimensional decisions. Another reason we focus on this question is that for online platforms (but also for other settings such as grocery chains and media markets), their revenue streams combine sales commissions or another profit margin on sales, as well as ad revenue, and as a result the platform has some flexibility in which stream to emphasize more, which in turn affects what algorithms learn. Of course, there are also many other cases where algorithmic decision making can help consumers.\footnote{Well-known examples are recommendation systems and other algorithms that help match consumers and products (or consumers with other consumers) \citep{mullainathan2017machine,miklos2019collusion}.} However, we are interested in exactly those cases that were previously identified as harmful to consumers.

Our analysis combines two approaches---we use an analytical model to analyze a benchmark case of pure competition (with complete information where algorithmic pricing and bidding are not needed) to show the effect of ad bids and search costs. We then use an extensive empirical simulation to analyze algorithmic pricing and bidding using RL to show that prices and bids could be lower when algorithms learn to coordinate in a setting with high consumer search cost.
We also estimate search costs and algorithmic pricing adoption using a large-scale dataset from Amazon.com and provide evidence consistent with  prices being lower in markets with higher consumer search cost and higher algorithm usage.
To make our results comparable to previous research, we  analyze a specific, yet common, RL algorithm called Q-learning (described in Section \ref{sec:MARL}). We later discuss the generalizability of our findings to other types of learning algorithms.

In Section \ref{sec:model}, we motivate our model through providing details of an online platform (such as Amazon.com) where consumers come to search for products to buy, and sellers come to sell their products. The platform displays ordered links to consumers, with the prominent sponsored links sold through an ad auction, while the other (organic) links are ranked based on product features (price, rating and other features). This setting is quite common in e-commerce and includes large companies such as eBay and Expedia. The sellers on the platform set prices for their products and also decide how much to bid for displaying their product in the sponsored position. When consumers search for a keyword, they see an ordered list of links, and traverse it until they find a product to buy or stop searching. We assume that the purchase decisions follow a standard choice model, with consumers having heterogeneous consideration set sizes, which are a result of heterogeneous search costs. % Other research has shown that this model is also consistent with optimal sequential search and other forms of search \citep{weitzman1979optimal,ursu2018power,lam2021platform}. An important assumption we make is that consumers have heterogeneous search costs (or impatience). This assumption is quite realistic in our opinion and has not been analyzed previously in the context of algorithmic pricing. 
We later show evidence that search costs are indeed high and heterogeneous on Amazon.com, lending credibility to this assumption.

In Section \ref{sec:MARL}, we compare two sellers who use reinforcement-learning algorithms to price and bid to the case of pure competition without learning. Under pure competition (without RL), search costs increase prices in equilibrium, because the sponsored positions become more valuable. This in turn increases equilibrium bids on ads, leading to increased marginal costs and higher prices \citep{armstrong2011paying}.
Our first novel finding in Section \ref{sec:MARL}, where we analyze multi-agent reinforcement learning, %. In this setting, competing sellers use Q-learning to set prices and bids. They do not have prior information about the economic environment (what the demand function is, what the demand parameters or utility functions are, or how ad auctions behave) and in fact do not even have a model---they learn a table mapping their actions (prices and bids) to profits. The agents explore different prices and bids, and over time learn how their actions affect profits. When these algorithms run long enough, they converge to an equilibrium where they mostly use the same price/bid combination, and new information they observe does not affect their learning anymore.  However, we find the opposite---when search costs increase, 
is that the algorithms converge on prices lower than competitive prices. In other words, Q-learning can result in coordinated pricing, but the algorithms coordinate on lower prices that help consumers. This is in stark contrast to findings from previous research \citep{calvano2020artificial,johnson2023platform}, where it was shown that pricing algorithms learn to collude on higher prices.

Our insight comes from the role of advertising. If the algorithms could exchange information and coordinate on the best outcome for them, they would have decided to bid zero on the ads and evenly split the market. Because the ads are a cost for the sellers, it is best for them to agree to lower these costs. This is in contrast to prices, where it is beneficial for both sellers to increase prices. When search costs are high, the benefit from coordinating on lower bids outweighs the benefit from coordinating on higher prices, and the equilibrium outcome is better for consumers. Unlike the case of competing over prices only, when competing on prices and ad bids together, lower equilibrium prices result in higher seller profit, because total demand on the platform increases, as well as advertising costs decrease. 

Tacit coordination by algorithms can also benefit the platform (Section \ref{sec:strategic_platform}) when search costs are high, because the increase in sales generates more commissions that dominate the decrease in revenue from advertising. We find that adjusting the ad reserve prices might not be as effective compared to changing commission rates. This is because increasing the reserve price shifts the algorithms even further towards coordinating on lower bids, further reducing the advertising revenue. However, by increasing their commission rates, the platforms can recoup some of the lost advertising revenue due to lower bids.

% One concern with these results, however, is that they depend on assuming that many consumers have high search costs. 
In  Section \ref{sec:empirical_application}, we analyze data from more than 2,000 product search keywords on Amazon.com, yielding more than 1 million observations per day for 2 million products. Our goal is to see if there is evidence that indeed prices decrease for markets with higher search costs and higher adoption of algorithmic pricing.
We estimate the consumer search costs using the variation in observed product rankings and their sales over time.\footnote{We use moment conditions \citep{lam2021platform, yu2024welfare} to address the simultaneity issue between the ranking of a product on a page and its sales. If a product is ranked higher it will likely have more sales, but in turn, having more sales might rank it higher.} 
% We find that search costs can indeed be high for many products in the data we analyze. 
We estimate an algorithm usage index for different product categories using the correlation patterns in prices set by sellers over time \citep{chen2016empirical}. We find a negative interaction on price levels between estimated search costs and algorithmic pricing adoption, consistent with our model's prediction that prices are lower in markets with higher adoption of algorithmic pricing and higher consumer search costs.

%Finally, we also examine the possible response of the platform to prices being set by such algorithms. We analyze the platform's optimal commission rate and advertising reserve price. 

To summarize, unlike much of the past research, we find that consumers as well as the platform might benefit from tacit coordination by algorithms. These findings can be useful for sellers who are concerned about adopting algorithmic pricing with multi-dimensional decision-making. They can also help platforms decide if they should encourage sellers to use algorithmic pricing and bidding, and even whether they should offer these algorithms themselves. %The platforms have information about consumer search costs, and are best situated to benefit from our findings. When the platform needs to respond strategically, our findings provide guidance about the possible methods to consider (e.g., adjusting commission rates, adjusting reserve prices in the ad auction, or disclosing information about the wining bids). Our findings can also help the platform address regulatory concerns from policymakers, given that consumers on the platform can also benefit.
Finally, our results provide good news for consumers because lower prices by algorithms increase consumer welfare. From a regulatory perspective, this area of inquiry is currently of high relevance for modern digital platforms, making our findings extremely relevant.

\section{Contribution and Related Work}
\label{sec:literature}
Our research contributes to a growing stream of literature that examines how algorithms affect competition through pricing. In this stream, a central concern is that reinforcement learning or related heuristics may lead to tacit collusion and supra-competitive prices. For example, \cite{calvano2020artificial} and \cite{hansen2021frontiers} demonstrate for oligopolies that algorithms such as Q-learning and multi-armed bandits can lead to equilibrium prices above competitive levels. \cite{johnson2023platform} investigate platforms’ ability to steer demand towards lower-priced sellers and find that algorithms result in supra-competitive prices in the absence of non-neutral platform intervention. The phenomenon of algorithmic collusion is also documented in scenarios where algorithms make sequential decisions \citep{klein2021autonomous} and where Q-learning algorithms compete with simple heuristic algorithms \citep{wang2023algorithms}. By contrast, despite concerns expressed by policymakers, \cite{miklos2019collusion} find that better forecasting algorithms can lead to lower prices and higher consumer surplus. To uncover the mechanism behind such algorithmic collusion, \cite{asker2022artificial} study the effect of algorithm design on collusion. Similarly, \cite{banchio2022artificial} find that synchronous algorithms are less likely to converge on collusive outcomes.  \cite{banchioartificial} reveal that spontaneous coupling can sustain collusion in prices and market shares. Our research adds to these prior work by considering multi-dimensional decisions, and by analyzing the cost aspect of the decision making. We show that collusion does not always lead to predictable outcomes in one direction for consumers, but can actually benefit (or harm) consumers depending on which force (cost or revenue) is stronger given consumer search heterogeneity.

On the empirical side, \cite{musolff2022algorithmic} documents that repricing algorithms follow Edgeworth cycles to effectively coax competitors into raising their prices, which decreases competition. \cite{brown2021competition} find that algorithms allow for more frequent price changes, generate price dispersion, and increase price levels. \cite{assad2024algorithmic} find that algorithmic-pricing software significantly increased the margin in Germany’s retail gasoline market. \cite{calder2023coordinated} find that algorithms set more responsive prices, leading to higher rents and lower occupancy in the U.S. multifamily rental housing market. Our research adds further evidence by considering the effects of consumer heterogeneity in search costs which affects how valuable sponsored ads are. We find empirical evidence that prices might be lower when algorithms are used to price and bid and consumer search costs are high, compared to a case of full-information competition. This finding provides a more nuanced view of the potential impact of algorithmic pricing when other forces are at play.

Substantively, our paper builds upon research in search advertising and sponsored ads \citep{edelman2007internet, varian2007position,sayedi2018exclusive,choi2019monetizing,long2022designing}. These works typically assume that the value of winning an ad auction is exogenously given. \cite{athey2011position} introduced consumer search, which endogenizes advertisers' valuation. \cite{chen2011paid, kang2021sponsored} incorporated the pricing decision into the sponsored ads auction. \cite{armstrong2011paying} demonstrated how higher search costs raise bids, ad expenditures, and ultimately prices --- an effect closely related to the mechanisms we examine. Our research builds on these (mostly theoretical) insights by considering the impact of learning algorithms in these settings, and by providing empirical evidence for these predictions. In these settings, platform algorithms that rank products by price, rating, or advertising bids influence both seller profits and consumer surplus \citep{lam2021platform,lee2021entry,farronato2023self}. In these works, classic models of consumer search \citep{weitzman1979optimal,ursu2018power,honka2017simultaneous,morozov2021estimation} are often used and emphasize heterogeneity in search costs, which we find have a substantial impact on the findings in our work.

\section{Institutional Background and Model Setup}
\label{sec:model}

\subsection{Institutional Setting}
\label{subsectoin:empirical_background}

%Amazon.com's Marketplace, the world's largest retail digital platform, is a primary example of the setting we consider.\footnote{In 2022, Amazon reported nearly \$514 billion in net sales revenue worldwide, and held approximately 40\% of the e-commerce market share. Source: Statista, \url{https://www.statista.com/topics/846/amazon/\#topicOverview}} 
We analyze platforms similar to Amazon.com that list and sell products from third-party sellers. In our main analysis, if the platform  directly sells a product to consumers, we assume that consumers will treat it as another seller. We discuss self-preferencing (where Amazon might promote its own  products more) in the Conclusion.

We consider consumers who start by typing a keyword into the search box. The platform then lists relevant products with both organic and sponsored listings. We focus on the default ``Featured'' ranking, where the platform ranks organic products based on multiple factors like sales performance, customer ratings, and conversion histories.\footnote{Consumers can also choose other sorting options, including ``Price: Low to High'', ``Price: High to Low'', ``Avg. Customer Review'', ``Newest Arrivals'', and ``Best Sellers''.} We assume consumers consider products sequentially in the order displayed to them. They have heterogeneous search costs and might stop their search without considering later products on the search results page. Consumers who have high search costs (e.g., impatient), generally visit only the top links (i.e., mostly sponsored) and only consider buying the products that appear there.\footnote{For example, \cite{farronato2025vertical} find that a median consumer had a consideration
set of 15 products and only 23.3\% of consumers considered a product located in the middle of the
results page.} By contrast, consumers with low search costs or frictions (e.g., patient), will consider also organic products which appear lower on the list.

\begin{figure}[!ht]
    \caption{Example of First Two Rows of an Amazon Search Result Page with 60 Products}
    \label{fig:Amazon_sponsored_product_example}
\begin{center}
    \includegraphics[width=0.7\textwidth]{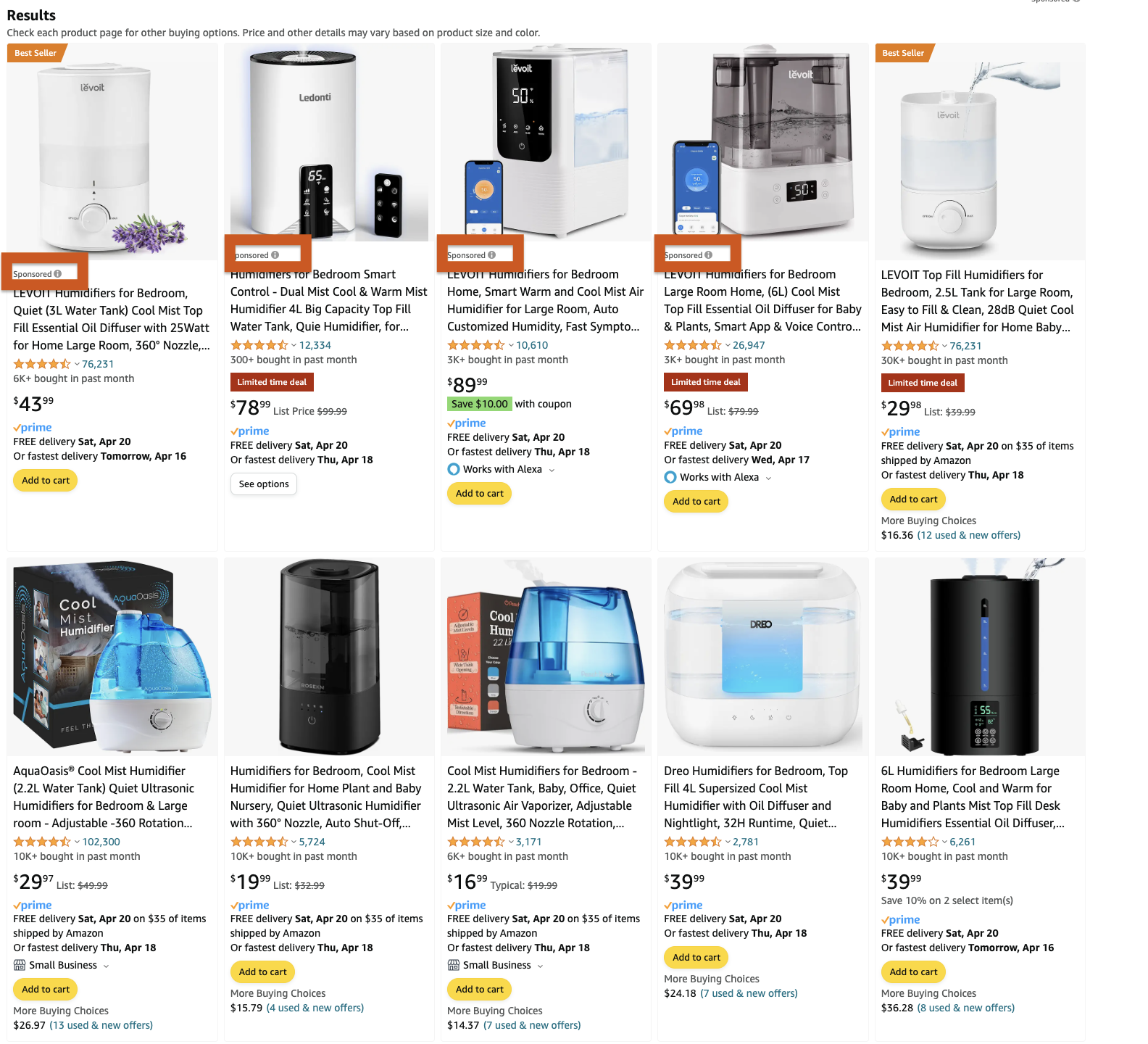}
\end{center}    
First two rows of the search results for ``humidifiers for bedroom.'' Orange rectangles emphasize sponsored listings that link to products by sellers who won the ad position in an auction.
\end{figure}

Sponsored products are distinguished from organic by the label ``Sponsored,'' and a product can appear in both sponsored and organic positions. %Within a listing page, there are 60 products laid out in 5 columns and 12 rows for most product categories, and 22 products in the `Electronics' category displayed in a vertical layout of 1 column. 
Figure \ref{fig:Amazon_sponsored_product_example}
 and Web Appendix Figure \ref{fig:Amazon_sponsored_product_example_22} present examples.

We focus on %There are other advertisingWithin the search results page, there are several advertising formats, which include ``Sponsored Product Ads'' in the ``search result'' section, as well as ``Sponsored Display'' and ``Sponsored Brands,'' which usually appear in a carousel in different locations on the page. We focus on 
Sponsored Product Ads format because they account for about 78\% of total seller ad spend on Amazon.com.\footnote{Statistic from \textit{Jungle Scout}. See \url{https://www.junglescout.com/blog/amazon-sponsored-product-ads/}, Accessed June 8, 2024.} and appear prominently on the search result page.

Amazon uses real-time ad auctions to determine which ads to display and their order. The ranking is based on a combination of the advertisers' bids and the ads' relevance.\footnote{Amazon Ads, How does bidding work for Sponsored Products? \url{https://advertising.amazon.com/library/videos/campaign-bidding-sponsored-products}.}  
Sellers are motivated to bid for sponsored positions to attract more consumer attention earlier in their search. Sellers bid for each keyword based on their willingness to pay per click (CPC) and pay only when a consumer clicks their ad.\footnote{Amazon uses generalized second-price auctions, where for each consumer click on an ad in the $r$-th sponsored position, the seller pays Amazon an amount equal to the ``realized'' bid of the $(r+1)$-th highest bidder, which is a relevance adjusted bid.}

Sellers also pay the platform a share of the revenue from units sold as commission.\footnote{Amazon discloses that commission rates differ for various product categories, typically being 15\%. See  \url{https://sell.amazon.com/pricing\#referral-fees}, accessed June 8, 2024.} Other fees that Amazon can collect from sellers (e.g., shipping and return administration), are included in the manufacturing (marginal), as they do not alter search result rankings nor consumer search.

Sellers need to pick product prices and if, and how much, to bid for sponsored ads to maximize profits. They face two trade-offs regarding prices. First, they want to avoid strong price competition with other sellers and prefer higher equilibrium prices. Second, they also want the prices not to be too high, because otherwise consumers might choose to buy from another website (an outside option good). Regarding ad bids, the more sellers bid, the higher their chances are of winning the sponsored link and reaching consumers who only consider the top positions. However, they will want to avoid competing too aggressively on bids as they incur costs that lower profits.

In our main analysis we do not consider the platform's incentives to change commission fees or ad auction reserve prices, and discuss these in Section \ref{sec:strategic_platform}.

Because of these two-dimensional tradeoffs for bids and prices, when sellers use profit-maximizing algorithms to guide pricing and bidding decisions, it is unclear whether equilibrium prices and bids will be high or low. On the one hand, sellers want higher price to increase profits, which increase the valuation of winning the sponsored positions leading to higher bids. On the other hand, they would like to bid lower to reduce their marginal costs, leading to lower prices. These two opposing forces are the main factor that impacts the results in our analysis using the model we introduce. A-priori it isn't obvious which force will prevail.
%To explore the impact of competitive algorithmic decision making on prices and bids, we model an e-commerce environment in the next section and then analyze the impact of learning algorithms. We use this model to make predictions about equilibrium prices and bid, and also estimate a version of this model on large-scale data about product searches and sales from Amazon.com.

% 1. What are the tradeoffs (bid high or low, price high or low), and objectives of sellers (want to pay less for ads, want to not compete too much on prices).

% 3. Describe the platform's objectives, and also the tools it has (information, incentives) to affect the objective.

% 4. Create a puzzle, say why it isn't obvious what will happen with algorithmic pricing.

%Here, we argue that advertising alters the order of products consumers see and search, while other costs, such as shipping or restocking fees, are considered part of traditional manufacturing costs. Therefore, we specifically focus on the role of sponsored ads and exclude other costs from our model on Amazon.
%, allowing us to concentrate on the primary driving forces in this context.

\subsection{Model Setup}
Our model has three player types: $n\ge 2$ \textbf{sellers}, who sell one differentiated product on a retail \textbf{platform}, and a unit mass of \textbf{consumers} who choose between purchasing from either one of these sellers on the platform, or an outside good. We analyze this model analytically and also use it for empirical simulation and estimation on data.

\paragraph{Platform}
The platform determines listing rankings on the search results page. Sponsored positions are  determined by ad auctions; organic positions are determined using a recommendation algorithm.  We first consider a simplified case where two sellers compete and there is one sponsored position followed by two organic listings. The sponsored listing is the first product consumers see and is awarded to the winning ad-auction bidder. Figure \ref{fig:Graphical_Representation_Theoretical_Model} in the Web Appendix provides an illustration of the results page for the simplified model.

We assume each search keyword is its own independent market, hence we can focus on one representative keyword only. Each seller has a constant marginal production cost $c_i>0$, and pays the platform a commission rate $\tau \in (0,1)$ of revenues.  Sellers interact repeatedly over time and in each period \( t = 0, 1, \ldots \), they simultaneously decide on the bid \( b_i^t \geq 0 \) to submit in the auction and the price \( p_i^t \geq 0 \) to set for its product.

The platform shows the seller with the highest \textit{realized bid} $\tilde{b}_i^t$ in the sponsored position, when the bids submitted by sellers are $b_i^t$. We approximate the impact of the ads' relevance score and the uncertainty sellers face when deciding their bids, by assuming that the realized bid \( \tilde{b_i^t} \) follows a log-normal distribution: \( \log(\tilde{b_i^t}) \sim \mathcal{N} (\log(b_i^t), \sigma_i) \). 
We use a first-price auction for tractability and to avoid scenarios with multiple equilibria. While this differs from the generalized second price auction that Amazon.com uses, we do not expect this difference to be crucial for the main insights from our analysis.\footnote{Generalized second-price auctions and first-price auctions have similar theoretical strategic incentives and algorithm performance. Both are non-truthful, and algorithms behave similarly and learn to collude on lower bids \citep{rovigatti2023artificial,banchio2022artificial}, unlike in non-generalized second-price auctions. Our assumptions are similar to those imposed in \cite{yu2024welfare}.} We provide more details in Section \ref{sec:theory}.

\paragraph{Consumers} In period $t$ a unit mass of consumers who wish to buy at most one product perform a search. They spend one period in the market and then exit and are replaced by a new cohort. A representative consumer who buys product $i$ in period $t$ obtains utility $u_i^t=a_i-p_i^t + \epsilon_i$,
where $a_i$ capture vertical differentiation and $p_i^t$ is the product's price. The outside good is indexed by 0, with utility $u_0^t=\epsilon_0$. We assume that $\epsilon_0$ and each $\epsilon_i$ are i.i.d. type-I extreme value with common scale parameter $\mu>0$.\footnote{The parameter $\mu$ is also an index of horizontal differentiation, and the case of substitutes is obtained in the limit when $\mu \rightarrow 0$.} 
If product $i$ is in the consideration set \( \mathcal{N}^t \) in time $t$, its demand is:
\begin{equation}
s_i\left(\boldsymbol{p^t}\right)=\frac{\exp \left(\frac{a_i-p_i^t}{\mu}\right)}{\sum_{j \in \mathcal{N}^t} \exp \left(\frac{a_j-p_j^t}{\mu}\right)+1}
\label{equ:logit_demand}
\end{equation}
where $\boldsymbol{p^t}$ is the vector of prices of products in the consumer's consideration set.

Consumers search for products in the order displayed to them, and have heterogeneous search costs. High search costs can lead them to stop their search early and consider only a subset of all products on the page. For the simplified model we follow \cite{armstrong2011paying} and assume that there are two segments of consumers. A fraction \( \theta \) of consumers has high search costs and considers only the product in the sponsored position.  The remaining \(1-\theta\) fraction of consumers considers the products in all positions. For example, if products $i$ and $j$ are presented in order $\{j, i\}$ on the search results page, then the high search cost segment considers only $j$, and the other segment has a consideration set of $\{j, i\}$. %For the case with multiple sponsored and organic positions, to characterize heterogeneous consumer search costs, we assume that the mass of consumers who stop at a particular position follows an exponential distribution. 
If a product appears in both the sponsored and an organic position, we assume that consumers consider it as one product from the organic link.\footnote{Web Appendix \ref{websec:sensitivity_check} describes a sensitivity analysis where a fraction of consumers who consider all products also click the sponsored link.} The search process represented by the two segments can be viewed as the outcome of consumers who perform sequential optimal search and decide to stop when the expected value from continuing is below their search cost, with a heterogeneous distribution of search costs \citep{weitzman1979optimal}. Appendix \ref{subsec:micro-foundation} provides details about the relationship between the simplified model and an optimal search process. Endogenizing $\theta$ would not change our main insights. We also consider a specification where the ranking affects consumer utility directly in Web Appendix \ref{subsubsection:alternative_characterization_of_search_costs} and find similar results.

We denote by $s_i\left(p_i^t\right)$ the market share of product $i$ among consumers who only consider the top position when product $i$ appears in the top position, and denote the market share of product $i$ among consumers considering both sponsored and organic positions by $s_i\left(p_i^t, p_j^t\right)$ . Taking the outside option into account, the market shares of product $i$ are:
\begin{align*}
s_i\left(p_i^t\right) &= \frac{\exp \left(\frac{a_i-p_i^t}{\mu}\right)}{\exp \left(\frac{a_i-p_i^t}{\mu}\right) + 1} & 
s_i\left(p_i^t, p_j^t\right) &= \frac{\exp \left(\frac{a_i-p_i^t}{\mu}\right)}{\exp \left(\frac{a_i-p_i^t}{\mu}\right)+\exp \left(\frac{a_j-p_j^t}{\mu}\right) + 1}
\end{align*}

\paragraph{Sellers} 
In the simplified model, we assume that the two sellers are ex-ante symmetric in their quality and production costs (\( a_i^t = a_j^t \) and  \( c_i^t = c_j^t \)). Hence, the platform will treat the sellers identically for organic rankings and randomize the order of display in the two organic positions. Asymmetric sellers are analyzed in Appendix \ref{subsubsection:asymmetric_sellers}. Our empirical analysis allows for multiple sellers who can be differentiated and asymmetric. The sellers have a common discount factor $\delta \in(0,1)$, and maximize the cumulative discounted profit:
\begin{equation}
    \sum_{t=0}^{\infty} \delta^t \pi_i^t(\boldsymbol{p^t}, \boldsymbol{b^t})
\end{equation}

The one-period profit for seller \( i \) in period \( t \), denoted as \( \pi_i^t(\boldsymbol{p^t}, \boldsymbol{b^t}) \), is given by:
\begin{align}\nonumber
\pi_i^t(\boldsymbol{p^t}, \boldsymbol{b^t}) = & \theta \cdot \operatorname{Pr}(\tilde{b_i^t} > \tilde{b_j^t}) \cdot s_i\left(p_i^t\right) \cdot \left(\left((1-\tau) \cdot p_i^t - c_i\right) - \gamma_i \cdot \mathbb{E}\left[\tilde{b_i^t} \mid \tilde{b_i^t} > \tilde{b_j^t}\right]\right) \\
& + (1-\theta) \cdot s_i\left(p_i^t, p_j^t\right) \cdot \left((1-\tau) \cdot p_i^t - c_i\right)
\label{equ:seller_profit}
\end{align}
where $\boldsymbol{b^t}$ is the vector of bids submitted by all sellers in period $t$. The first term represents the profit from the sponsored position and is affected by the share of ``impatient'' consumers $\theta$, the chance of winning the ad auction  $\operatorname{Pr}(\tilde{b_i^t} > \tilde{b_j^t})$, the market share of the product (vs. the outside good) $s_i\left(p_i^t\right)$ and the profit per sale, which depends on the profit margin and cost of ads. 
The inverse conversion rate, \( \gamma_i \), indicates the number of clicks needed for a sale and is assumed common knowledge. 
The commission rate charged by the platform is denoted as $\tau$. 
Appendix \ref{subsec:appendix_model_detail} derives the expressions for the chance of winning the ad auction $\operatorname{Pr}(\tilde{b_i^t} > \tilde{b_j^t})$ and the expected ad cost conditional on winning the auction and the competitor's bid $\mathbb{E}\left[\tilde{b_i^t} \mid \tilde{b_i^t} > \tilde{b_j^t}\right]$. The second term is the profit from consumers who consider more than just the sponsored product.

\subsection{Full Competition Benchmark}
\label{sec:theory}

%To understand how learning algorithms interact, and the role of search costs in affecting equilibrium prices, bids, and sellers' profit, 
We first analyze a full competition version of the model where the sellers have complete information and do not need to learn about their environment. This setting forms the benchmark which we use to compare to the algorithmic case in repeated games.

We consider two scenarios:
\begin{enumerate}
\item \textit{Full competition with pricing and bidding}: Both sellers have complete information about model parameters and consumer behavior. They compete by setting prices and bids simultaneously. %As there is no closed-form solution because of the demand structure, we find the Nash-Bertrand equilibrium numerically. 
The equilibrium price and bid are denoted $\mathbf{p}^N$ and $\mathbf{b}^N$, respectively. 

    \item \textit{Pricing without bidding}: In this scenario sellers compete only on prices without advertising. Consumer preferences remain the same, with $\theta$ representing the fraction of consumers who focus only on the first position. Each seller's probability of being displayed in the first position is $\frac{1}{2}$ without an auction. We denote the equilibrium price as $\mathbf{p}^{oN}$,\footnote{Here, $N$ stands for the Nash-Bertrand equilibrium, and $o$ indicates the scenario with pricing competition only, but without advertising.}
    and each seller solves:
    \begin{equation}
      \max _{p_i} \pi_i(\boldsymbol{p})=\left(\theta \cdot \frac{1}{2} \cdot s_i\left(p_i\right)+(1-\theta) \cdot s_i\left(p_i, p_j\right)\right) \cdot \left((1-\tau) \cdot p_i-c_i\right)
\end{equation}

\end{enumerate}

\begin{figure}[ht]
\caption{Equilibrium Prices and Bids in One-Shot Game}
\label{fig:theoretical-results} 
\begin{subfigure}{0.51\textwidth}
\caption{Prices}
\centering
\includegraphics[width=\textwidth]{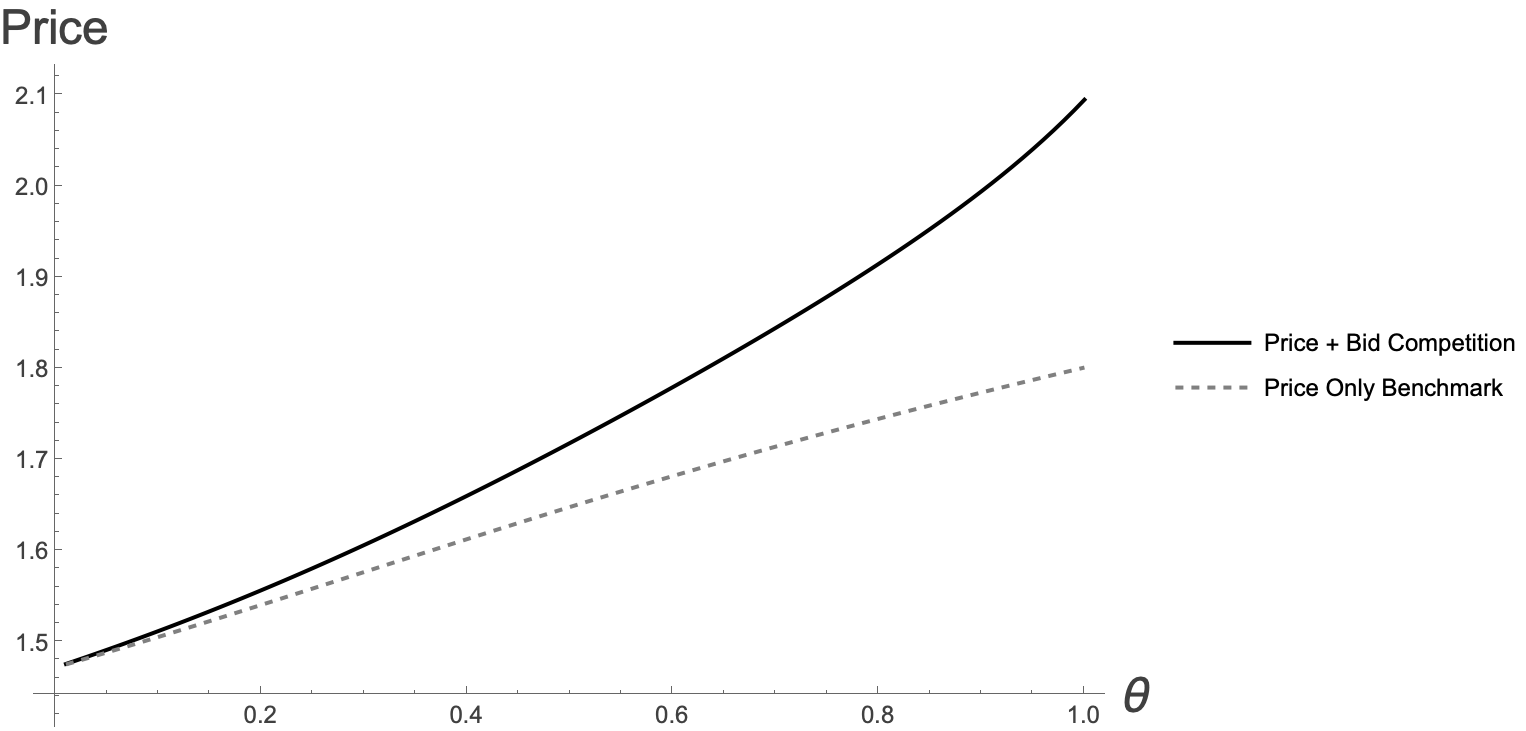}
\label{subfig:theoretical-price} 
\end{subfigure}~
\begin{subfigure}{0.48\textwidth}
\caption{Bids}
\centering
\includegraphics[width=\textwidth]{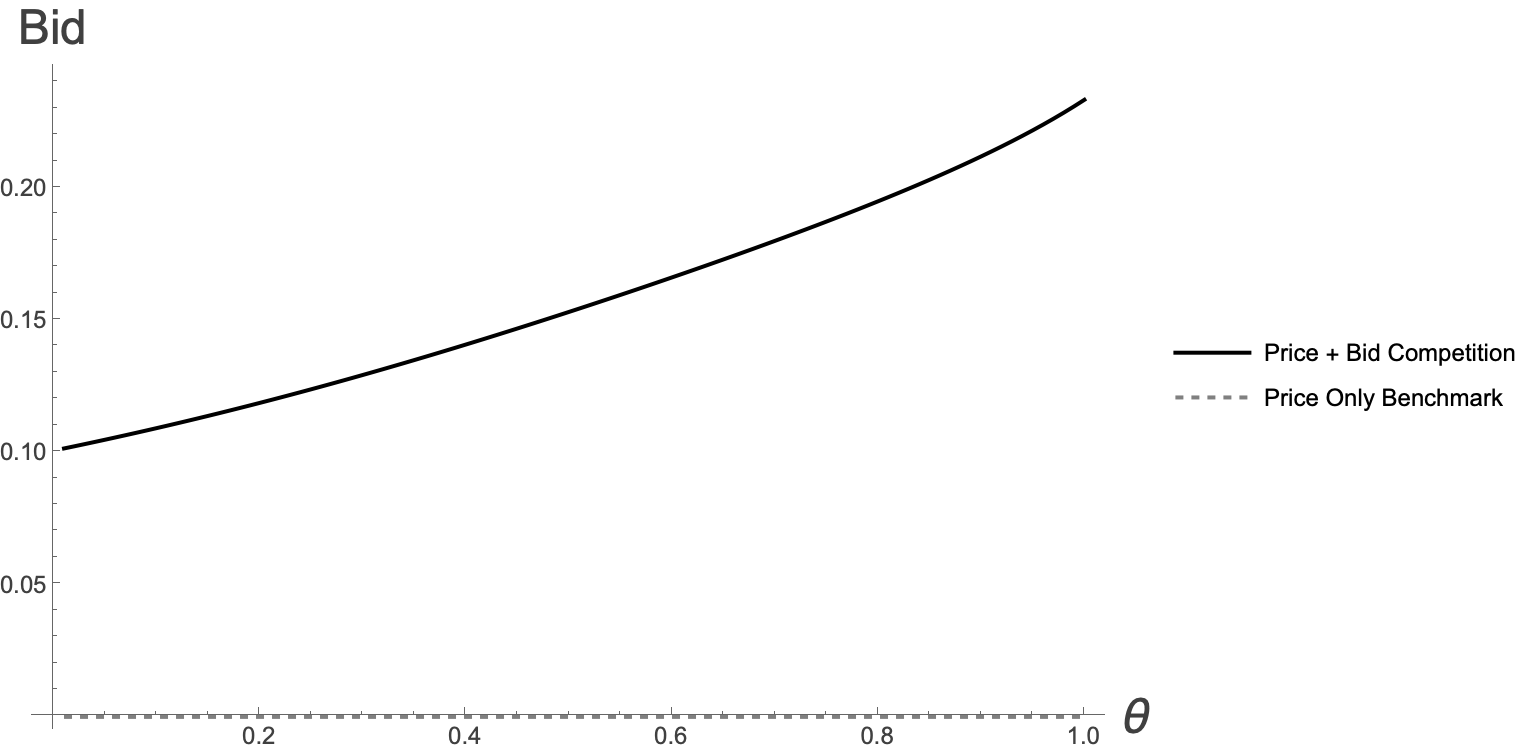}
\label{subfig:theoretical-bid} 
\end{subfigure}

    Panels (a) and (b) show the equilibrium prices and bids as a function of $\theta$ under different scenarios, respectively. The dashed gray line, and black solid line represent the benchmark equilibrium price $\mathbf{p}^{oN}$ (or bid $\mathbf{b}^{oN}$), and the Nash-Bertrand equilibrium price $\mathbf{p}^N$ (or bid $\mathbf{b}^N$), respectively. For this example, we set the parameters as $a_i = a_j = 2$, $c_i = c_j = 1$, and $\mu = \frac{1}{4}$ for comparison with the results in \cite{calvano2020artificial}. For ads-related parameters, we set $\sigma_i = \sigma_j = 0.5$ and $\gamma_i = \gamma_j = 2$.
\end{figure}

We present the equilibrium prices in Figure \ref{subfig:theoretical-price} and the bids in Figure \ref{subfig:theoretical-bid}. The case without advertising has zero bids.

The following lemma, proven in Appendix \ref{sec:appendix_competitive_price}, shows that when sellers bid on advertising, equilibrium prices are higher:
\begin{lemma}
\label{obs:competitive_price}
The equilibrium price in competition with advertising bids is always greater than or equal to the equilibrium price without advertising: $\mathbf{p}^N\ge \mathbf{p}^{oN}$. The equilibrium price without advertising is also increasing in $\theta$.
\end{lemma}

The intuition is that ads increase the marginal cost and drive up prices. When $\theta=0$ and there are no costly ads, the prices in both models coincide. Equilibrium prices increase with the fraction of consumers who only consider the first position $\theta$. 
A higher $\theta$ implies the seller competes more often only with the outside option (and not with another seller), which allows them to charge a higher price. \cite{armstrong2011paying} also similarly show that increased bids to win sponsored positions increase the sellers' advertising expenses and, consequently, their total costs.\footnote{Our proof in Appendix \ref{sec:appendix_competitive_price} also explains why the result only requires that bids are non-zero in equilibrium, and does not depend on the specific auction format.}

Figure \ref{fig:theoretical-results}(b) shows that equilibrium bids $\mathbf{b}^N$ increase with $\theta$. A larger $\theta$ means higher potential profits from winning the sponsored position, thereby incentivizing sellers to bid more aggressively, which leads to pricing with advertising to be higher than pricing without.

% \paragraph{Comparative Statics} We note some of the comparative statics properties here, which can be interesting.
% \begin{enumerate}
% \item The equilibrium bid decreases with respect to the uncertainty in bid realization, i.e., $\frac{\partial b_i}{\partial \sigma_i}<0$. As the uncertainty level goes to infinity ($\sigma_i \rightarrow \infty$), the equilibrium bid goes to zero ($b_i \rightarrow 0$).
% \item The competition price $\mathbf{p}^{N}$ and benchmark price $\mathbf{p}^{oN}$ are increasing with the commission rate $\tau$, i.e., $\frac{\partial \mathbf{p}^{N}}{\partial \tau}>0$ and $\frac{\partial \mathbf{p}^{oN}}{\partial \tau}>0$.
% \item The competition price $\mathbf{p}^{N}$ and benchmark price $\mathbf{p}^{oN}$ can be either increasing or decreasing in $\mu$. When the scale parameter of the idiosyncratic shock, $\mu$, approaches zero ($\mu \rightarrow 0$), both ${p}_i^{N}$ and ${p}_i^{oN}$ approach $a_i$.
% \end{enumerate}

\section{Algorithmic Pricing and Bidding: A Multi-Agent Reinforcement Learning Approach}
\label{sec:MARL}
Turning to our main analysis, we analyze sellers who use
reinforcement learning algorithms (specifically, tabular Q-learning) to make pricing and bidding decisions repeatedly. We use numerical simulations to analyze algorithms that interact in a Multi-Agent setting. First, we introduce the notation for Q-learning in a stationary single-player environment. Then, we extend from single to multiple agents.

% Reinforcement learning, a class of computer science techniques, enables algorithms to learn from their own past experiences in the environment. This learning process becomes more complex with multiple agents, a scenario known as Multi-Agent Reinforcement Learning (MARL). We employ (tabular) Q-learning, a technique motivated by dynamic programming in Markov Decision Environments \citep{watkins1989learning}. Q-learning is a widely-adopted algorithm in computer science, known for its practical applications and superior performance.%Additionally, Q-learning shares its foundational architecture with advanced programs that have demonstrated exceptional, even superhuman, performances. %Second, Q-learning operates with minimal knowledge of the environment. It functions efficiently using a limited number of parameters, each with a clear economic interpretation. The algorithm can recognize the system's current state and available actions, yet it does not require prior knowledge of the payoff functions or state transition probabilities. Lastly, Q-learning is not inherently designed to encourage collusion. While acknowledging that a designer might want to intentionally create an algorithm to collude, we take a neutral perspective that the designer simply wishes to earn high profits in a complex environment.

\paragraph{Single-agent Q-Learning}
An algorithm facing an unknown stationary Markov Decision Environment with a finite set of states  $s_t \in \mathcal{S}$, a finite set of actions $a_t \in \mathcal{A}$, where $t=1, 2, \ldots$ denotes the time periods. Choosing $a_t$ in state $s_t$ yields reward $\pi_t$, and the environment moves to state  $s_{t+1}$ with probability $F\left(\pi_t, s_{t+1} \mid s_t, a_t\right)$. %Q-learning deals with the version of this model where $S$ and $A$ are finite, and $A$ is not state-dependent.
%The objective of the algorithm is to find the policy maximize the expected present value of the stream of rewards $\pi(s_t, a_t)$. Let \( a^*(s) \) denote an optimal policy. Thus, 
%for a state at time \( t \), denoted as \( s_t \) with the initial state \( s_0 \), 
%the policy \( a^*(s) \) maximizes the sum of future expected discounted profits, expressed as $\mathbb{E} \left[\sum_{t=0}^{\infty} \delta^t \pi\left(s_t, a^*\left(s_t\right)\right)\right]$. 
The objective of the algorithm is maximize the expected present value of the stream of rewards  $\mathbb{E} \left[\sum_{t=0}^{\infty} \delta^t \pi\left(s_t, a\left(s_t\right)\right)\right]$.
Q-learning tries to maximize this objective by iteratively estimating an ``action-value function" $Q^*(s, a)$ where $Q^*(s, a)$ gives the expected discounted payoffs of taking action $a$ at state $s$ today and then using the optimal policy function $a^*(s)$ in future periods. Thus,
$$
Q^*(s, a)={\mathbb{E} \left[\pi(s, a)\right]}+{\delta \mathbb{E} \left[V\left(s^{\prime} \mid s, a\right)\right]}
$$
where $V\left(s^{\prime} \mid s, a\right)$ is the future optimal value. The optimal policy $a^*(s)$ is 
$
a^*(s)=\arg \max _{a \in \mathcal{A}} Q^*(s, a) .
$

% Because the state and action spaces are finite, $Q^*(s, a)$ is an $|\mathcal{S}| \times|\mathcal{A}|$ matrix and so $a^*(s)$ is determined by looking at the row corresponding to state $s$ and then choosing the column with the largest element in that row, which corresponds to the optimal action $a^*(s)$.

\paragraph{Learning and Experimentation} Since \( Q^*(s, a) \) is unknown, the Q-Learning algorithm estimates is as follows: Starting with an initial matrix \( Q_0 \), at time \( t \) and in state \( s \), the algorithm determines the action to take. With a probability \( 1-\epsilon_t \), the algorithm operates in \textit{exploitation mode}, choosing the optimal action  according to the current period Q-matrix and state $s$. With probability \( \epsilon_t \), it enters \textit{exploration mode}, and uniformly randomizes over all available actions.  %Such experimentation ensures that Q-learning sufficiently explores all states and actions.
After choosing the action $a_{t}$ in $s_{t}$, the realized payoff $\pi_{t}$ and new state $s^{\prime}$ are observed. The element of the Q-matrix corresponding to $(s, a)$ is updated to 
\begin{equation}
    Q_{t+1}(s, a) = (1-\alpha) {Q_{t}(s, a)}+ \alpha {\left[\pi_{t}(s, a) + \delta \max_{\tilde{a} \in \mathcal{A}} Q_{t}\left(s^{\prime}, \tilde{a}\right)\right]}
    \label{equ:Q_learning}
\end{equation}

where $Q_t(s, a)$ is the previous ``un-updated" element of the Q-matrix, and $\alpha \in(0,1)$ is a learning rate parameter, which captures the extent to which old information is replaced by new information. The probability of experimentation $\epsilon_t$ is 
$
\epsilon_t=e^{-\beta t}
$, where $\beta>0$ is the experimentation parameter.\footnote{This means that initially the algorithms choose purely random actions, as when $t=0$, $\epsilon_t=e^{0}=1$. As time passes, the algorithms take greedy actions more frequently. } %The greater $\beta$, the faster the exploration diminishes. An algorithm's learning is thus characterized by the couple $(\alpha, \beta)$.

Q-learning has theoretical convergence guarantees only in stationary single-player settings. In our multi-agent setting, convergence is not guaranteed because each agent continually changes its strategy by updating its Q-matrix, rendering the environment non-stationary from the other agents' perspectives. Despite this, in our empirical simulations we always observe convergence.
% In stationary single-player environments like the one previously discussed, Q-learning is guaranteed to converge to the actual action-value function \( Q^*(s, a) \) under certain conditions, thereby identifying the true optimal policy \( a^*(s) \) \citep{watkins1992q}. However, we investigate the interactions among multiple players. In Multi-Agent settings, convergence lacks theoretical guarantees. This is primarily because each agent continually changes its strategy by updating its Q-matrix, rendering the environment non-stationary from other agents' perspectives. Despite this, similar to findings in other studies \citep{calvano2020artificial}, we nearly always observe convergence.

\paragraph{Extending to Multiple-seller Pricing and Bidding in E-Commerce}
We extend prior work such as \cite{calvano2020artificial} and \cite{johnson2023platform} by incorporating sponsored ads and consumer search to an implementation of  Multi-Agent Reinforcement Learning (MARL) as follows:

We first consider a symmetric duopoly ($n=2$) with discount factor $\delta=0.95$. Each agent has a marginal cost $c=1$. True demand (unknown a-priori to the algorithms) is given by Equation \eqref{equ:logit_demand} with each firm having the same quality component $a_i =a_j = 2$ and scale parameter $\mu=\frac{1}{4}$. We normalize the quality component of the outside option to be $a_0=0$. The one-period profit is defined as in \eqref{equ:seller_profit}, under the assumption that the demand function remains stationary, and the agent maximizes the sum of future expected discounted profits. In Section~\ref{subsec:robustness} we also investigate the cases of heterogeneous product quality and more than two sellers.

\textbf{Action Space} In each period \( t \), the action space for each agent is a product of the set of possible prices and bids, that is \( a_{it} = (p_{it}, b_{it}) \). The set of prices is discretized into 15 equally spaced values within the range \([ \mathbf{p}^{\min}- \xi(\mathbf{p}^{\max} - \mathbf{p}^{\min}), \mathbf{p}^{\max} + \xi(\mathbf{p}^{\max} - \mathbf{p}^{\min})]\).\footnote{Here, \(\mathbf{p}^{\min}\) and \(\mathbf{p}^{\max}\) represent the minimum and maximum prices across all scenarios, respectively, for all values of \(\theta\). The parameter \( \xi > 0 \), set to \(0.1\), allows prices to extend slightly beyond the maximum and minimum values to include all possible price ranges in every scenario. In our baseline model, the minimum and maximum values are 1.47 and 2.1, respectively. 
}
The bid set contains 10 equally spaced elements in the range \([0, (1+\xi)\cdot\mathbf{b}^{\max}]\), with the upper bound slightly above the maximum bid.\footnote{ The maximum bid $\mathbf{b}^{\max}$ is the Nash-Bertrand bid $\mathbf{b}^{N}$, as the monopoly bid $\mathbf{b}^{M}$ is zero. In our default parameter specifications, the maximum possible bid equals 0.24.} 

% \textbf{State Space} In our baseline scenario, we assume that sellers see all prices of other sellers and an agent only knows their own bid. This is consistent with the real world practice that prices are usually observed by all sellers, especially with the help of different algorithms who  scrape the prices to aim price decision. However, the competitors' bid information is not observed. Thus, the state space \( s_{it} = (p_{it-1}, p_{jt-1}, b_{it-1}) \) consists of the previous period's prices set by all agents and the bids set by the agent itself, resulting in \( 15^n \times 10 \) elements. In section \ref{subsec:robustness}, we also alternative state space where sellers have information about competitors' bid. 

\textbf{State Space} 
In our primary analysis we assume that sellers observe competitor prices but only know their own bid. 
This is consistent with real-world practice: prices on e-commerce platforms are public, and sellers can use automated repricing tools that scrape competitor prices to adjust their own. 
However, bid information in sponsored-ad auctions is usually private and not observable by other sellers.\footnote{For example, Amazon’s Marketplace allows sellers to publicly observe competing prices through various automated repricing tools, yet ad auction bids remain private. The FTC has investigated whether Amazon’s ad-auction system adequately discloses its reserve pricing and auction structure, highlighting concerns about the opacity of auctions on major e-commerce platforms (see \url{https://www.reuters.com/business/retail-consumer/us-ftc-probes-google-amazon-over-search-advertising-practices-source-says-2025-09-12/}, accessed Oct 12, 2025).} We assume that each agent has a one-period memory, which means that only the price and bid from the previous period enter the state space.
The state space $s_{it} = (p_{it-1}, p_{jt-1}, b_{it-1})$,
consists of the previous period's prices set by all agents and the bids set by the agent itself, resulting in \( 15^n \times 10 \) elements. 
In Section~\ref{subsec:robustness}, we also examine an alternative state space $s_{it} = (p_{it-1}, p_{jt-1}, b_{it-1}, b_{jt-1})$ in which sellers observe competitor bids.

Each seller's algorithm independently maintains and  updates its own Q-matrix over time, and makes its pricing and bidding decisions at time \(t\) conditional on the state space. To initialize $Q_0$, and following the fact that at first choices are purely random, 
each cell is set to the discounted payoff that would accrue to player $i$ if opponents randomized their prices and bids uniformly:
$$
Q_{i, 0}\left(s, a_i\right)=\frac{\sum_{a_{-i} \in \mathcal{A}^{n-1}} \pi_i\left(a_i, a_{-i}\right)}{(1-\delta)|\mathcal{A}|^{n-1}} .
$$
The default parameters are set to \(\alpha = 0.15\) and \(\beta = 10^{-5}\) unless specified otherwise.

\textbf{Convergence} We count the simulation round as converged if the induced strategy of each agent remains unchanged for 100,000 periods. For every period's Q-matrix $Q_{i,t}$ and possible state \( s \), we identify the action that corresponds to the highest Q-matrix payoff, inducing a policy function in every period, \( a_{i, t}(s) = \operatorname{argmax}_a\left[Q_{i, t}(s, a)\right] \). We stop the simulation if the policy function remains constant for each agent for 100,000 periods, or after one billion periods elapsed. Subsequently, we calculate payoffs and other relevant metrics by averaging over the last 100,000-period horizon. We repeat the procedure 1000 times for every set of parameters, and report the average across these 1000 iterations for all statistics.

% \subsection{Simulation Results}

% Table \ref{tab:ai_pricing_benchmark} presents the results of replicating the findings in \cite{calvano2020artificial} as a benchmark case where there is no consumer search friction, and every consumer considers all positions (i.e., $\theta=0$). The Nash-Bertrand equilibrium price is 1.47, while the monopoly price is 1.92. Table \ref{tab:ai_pricing_benchmark} also reports the market share-weighted Q-learning prices that the algorithms end up charging, averaging 1.7 and falling between the Nash-Bertrand price and the monopoly price.
% \begin{table}[ht]
%     \caption{Benchmark of Q-learning Outcomes  When There is No Consumer Search Friction}
% \begin{center}
%     \begin{tabular}{ccc}\hline
%       Nash-Bertrand Price   & Monopoly Price & Q-learning Price \\\hline
%       1.47   & 1.92 & 1.7\\\hline
%     \end{tabular}
% \end{center}    
% \label{tab:ai_pricing_benchmark}
    
% This table reports the Q-learning price in comparison to the Nash-Bertrand and monopoly prices when \(\theta=0\).
% \end{table}

\subsection{Reinforcement Learning Simulation Results}
\begin{figure}[ht]
\begin{center}
    \caption{Q-learning Prices vs Competition Prices as a Function of Search Costs}
        \includegraphics[width=0.8\textwidth]{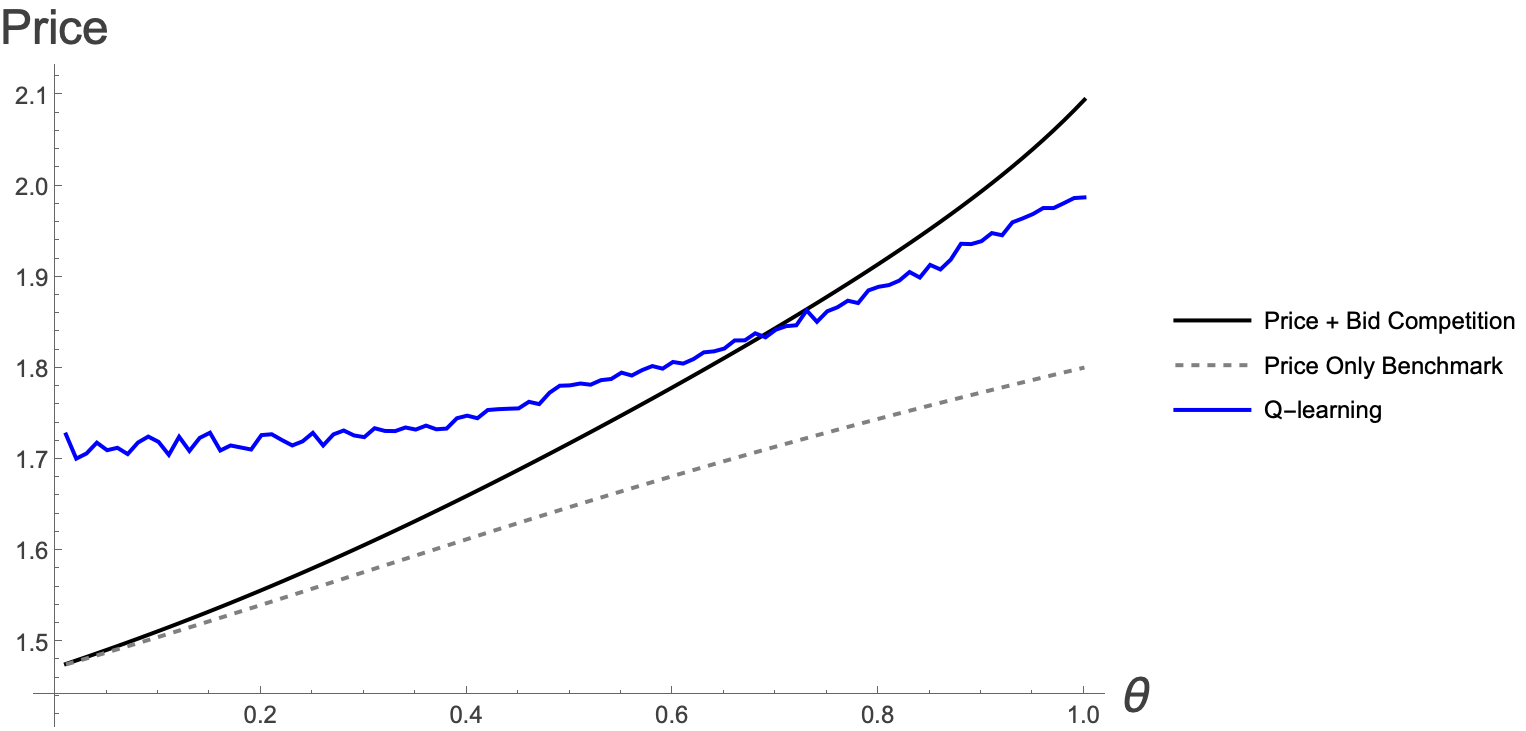}
    \label{fig:qlearning-results}
\end{center}    
    The figure shows the Q-learning simulation prices (blue solid) and full competition prices (with advertising - black solid, without - dashed) as a function of search costs $\theta$. 
    %The dashed gray line, black solid line, and  red solid line represent the theoretical benchmark equilibrium price \(\mathbf{p}^{oN}\) with price only competition, Nash-Bertrand equilibrium price \(\mathbf{p}^N\), and monopoly price \(\mathbf{p}^M\), respectively. 
\end{figure}

Our analysis uses a grid of values of $\theta$ from zero to one in increments of 0.01.  Figure \ref{fig:qlearning-results} displays the Q-learning equilibrium prices as a function of $\theta$, and compares them to the benchmark full competition results.

The case of $\theta=0$ when every consumer considers the products in all positions corresponds to the case considered in the past literature \citep{calvano2020artificial}. The market-share weighted Q-learning equilibrium price is 1.7, which is higher than the Nash-Bertrand price of approximately 1.47, consistent with past results that algorithms learn to set prices higher than competitive when there is no consumer search cost.

When $\theta$ increases, for lower values of $\theta$, Q-learning prices remain above the competitive prices. However, for higher values of $\theta$, Q-learning prices can fall below competitive levels. Whether algorithmic prices fall above or below fully competitive prices depends on the level of search cost heterogeneity in the market. The higher search costs are, the higher the chances that the algorithms will converge to lower than competitive prices.

This finding is one of the major contributions of our paper and is important because it counters the intuition developed by past research. We show that sponsored ads and consumer search costs may cause algorithms to converge to prices that are lower than competitive prices. Unlike other cases of algorithmic collusion, in our case lower prices are beneficial and not harmful to consumers. We will further illustrate the implications for consumer surplus later in section \ref{subsec:impact_sellers_consumers}.

\begin{figure}[ht]
\begin{center}
    \caption{Q-learning Results: Bids}
\includegraphics[width=0.75\textwidth]{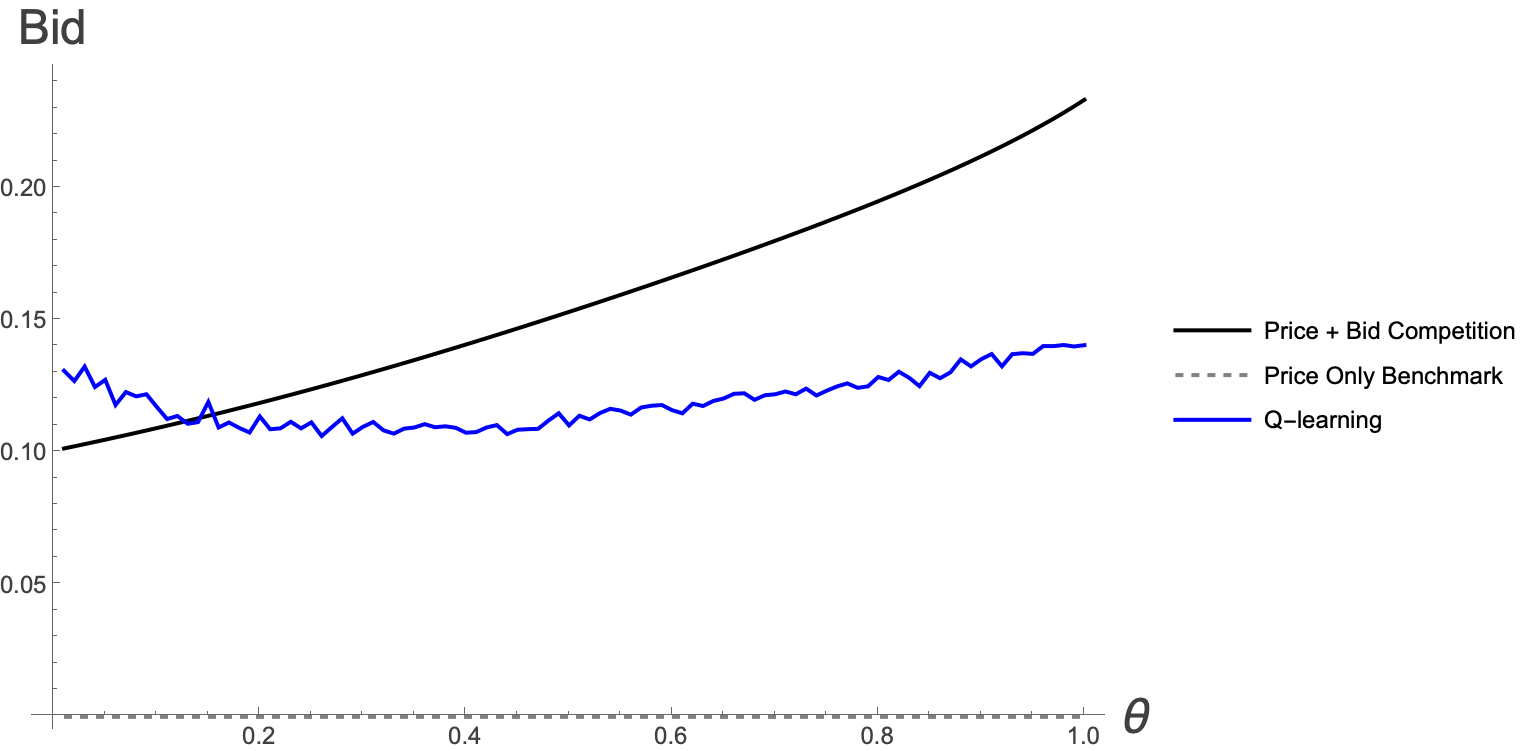}
    \label{fig:qlearnin_bids}
\end{center}    
    The figure shows the Q-learning bids (solid blue) and full competition bids (solid black) as a function of $\theta$. Without advertising the bid is always zero (dashed). 
\end{figure}

\paragraph{Intuition: Bid Coordination} Figure \ref{fig:qlearnin_bids} presents the average equilibrium bids. The bids affect the cost of the sellers. At higher values of $\theta$, average Q-learning bids are significantly lower than those observed with full competition, suggesting that the algorithms learn to coordinate on lower bids. This bid coordination reduces the sellers' advertising expenses, and consequently their overall costs. Such a reduction in costs leads to lower market prices and higher sellers' profit.

% \footnote{As an aside, algorithmic bids are U-shaped because for low $\theta$, the sponsored ad costs will be small, while it is worthwhile to price higher than competitive prices, leading to high bids in the auction. For moderate $\theta$, advertising costs increase with $\theta$, so bids decrease. For high $\theta$, sponsored links are very valuable, leading to higher bids.
% }
% \footnote{For low values of $\theta$, AI bids can be slightly higher than competitive bids. In such scenarios, the algorithms tend to set prices significantly above the competitive level. This high pricing increase the valuation of winning sponsored positions, thereby motivating sellers to bid more aggressively and providing them with the flexibility to do so.}

% Indeed, we observe a rotational pattern among competing agents in terms of bid submission and winning the sponsored positions. This rotational bidding aligns with the ``bidding ring'' phenomenon documented in literature \citep{decarolis2021mad,decarolis2020marketing,choi2023agency}. 

Past research also identified that Q-learning may result in coordination on lower bids in ad auctions \citep{banchioartificial, banchio2022artificial, rovigatti2023artificial}. However, we observe that this force is in opposite to the force that encourages algorithms to increase prices. Because these bids may interact with prices, and the level of this interaction depends on the consumer's search costs and frictions, whether bids decrease or increase in equilibrium depends also on pricing and consumer search costs. 

%Previous research also found that Q-learning algorithms lead to bid coordination, via colluding on lower bids in either a first-price auction \citep{banchio2022artificial} or a generalized-second-price auction \citep{rovigatti2023artificial}, or by segmenting the market and bidding on different keywords \citep{banchioartificial}. These analyses focus on scenarios involving only bidding competition, with the valuation of winning the auction being exogenously fixed. 
%By contrast, in our setting, the valuation of winning the sponsored position is determined by price competition endogenously. Here, the pricing and bidding form an interesting ``interaction'': sellers' bids are affected by the valuation of winning the sponsored position, valuations are determined by prices, and the prices are influenced by the costs (bids). Algorithms have the tendency to increase prices, as this would increase profits. Higher prices would lead to higher bids because the valuation of sponsored position increases. At the same time, algorithms also have the tendency to lower bids, as this decreases costs. Lower costs would lead to lower prices, because the marginal cost decreases. Thus, these are two opposing forces, and the combined effect can go either way, and depends on consumer search costs. Our finding is that when the competition for the sponsored position is strong, the tendency of algorithms to lower advertising costs dominates the tendency to increase prices, ultimately resulting in lower prices than the competitive level.

\subsubsection{Generalizability: Would prices always be lower with multi-agent Q-Learning?}
\label{subsec:theoretical_collusion}

The results we uncovered depended on specific simulation parameter values. We generalize them and prove that we expect lower than competitive prices for any values of the learning algorithm parameters. Our analysis also allows us to explain why we expect these results.

In our second primary contribution, we derive a theoretical bound for the fully collusive values that RL algorithms can potentially reach. We consider two sellers who actively collude on both prices and bids with complete information. In this scenario, the two sellers maximize their joint profits by jointly setting prices and bids, acting as if they are a single monopolist. The reason to perform this analysis (although the scenario is both unrealistic and likely illegal from an anti-trust perspective) is that it provides an upper bound on the profits of the sellers, and hence we can predict that the learning agent equilibria will be between the fully competitive and fully collusive ones. Therefore, if we can prove that for a set of parameter values the collusive prices are lower than the competitive prices, then we would expect this outcome to also generalize to most cases of competing RL algorithms.

The equilibrium collusive price $\mathbf{p}^M$ and bid $\mathbf{b}^M$ are defined as:\footnote{$M$ stands for monopoly.}
\begin{equation*}
    (\mathbf{p}^M, \mathbf{b}^M) = \argmax _{(p_i, b_i)} \pi_i (\boldsymbol{p}, \boldsymbol{b}) +\pi_j (\boldsymbol{p}, \boldsymbol{b}), \qquad s.t. \quad p_j=p_i, b_j=b_i
\end{equation*}

In equilibrium the two sellers would agree to set their bids to the minimum, \( b_i = b_j \rightarrow 0 \),  because this strategy minimizes advertising costs while maintains the same probability of capturing the demand from consumers who consider only the first position. Then, the two sellers would decide on the collusive prices to maximize their aggregate profits given they would both bid as low as possible. %This is also equivalent to the benchmark case where the sellers only collude on prices and there are no sponsored ads, because they set their bids to zero.

The algorithmic prices are expected to be between the theoretical collusive and fully competitive prices. We will show that these prices have a cross-over point for an intermediate value of $\theta$ allowing us to conclude that algorithmic prices will be higher than competitive for low values of $\theta$, but lower than competitive prices for high values of $\theta$.

\begin{figure}[!ht]
\begin{center}
    \caption{Q-learning and Theoretical Collusion}
        
        \label{fig:theoretical-collusion} 
\begin{subfigure}{0.51\textwidth}
\caption{Prices}
\centering
\includegraphics[width=\textwidth]{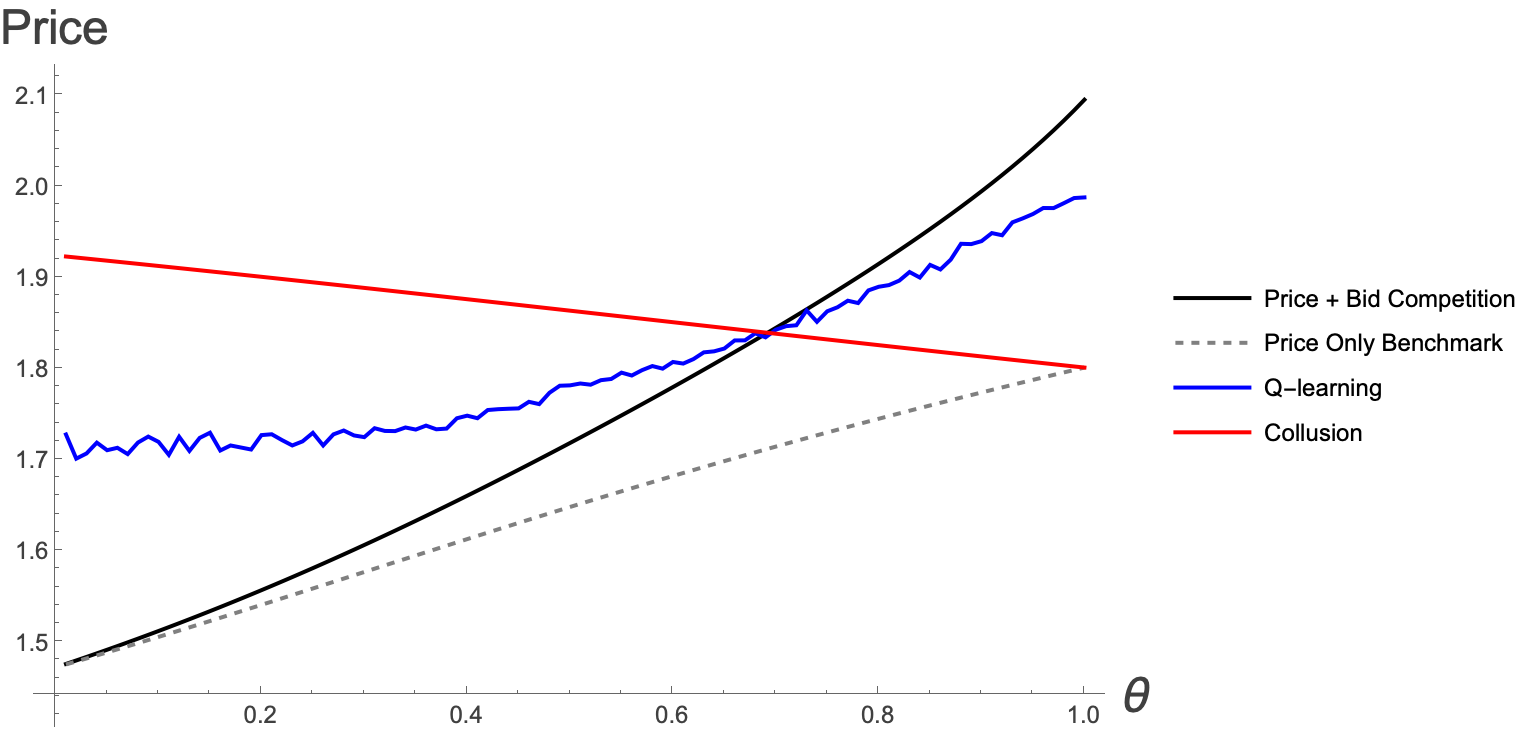}
\end{subfigure}~
\begin{subfigure}{0.49\textwidth}
\caption{Bids}
\centering
\includegraphics[width=\textwidth]{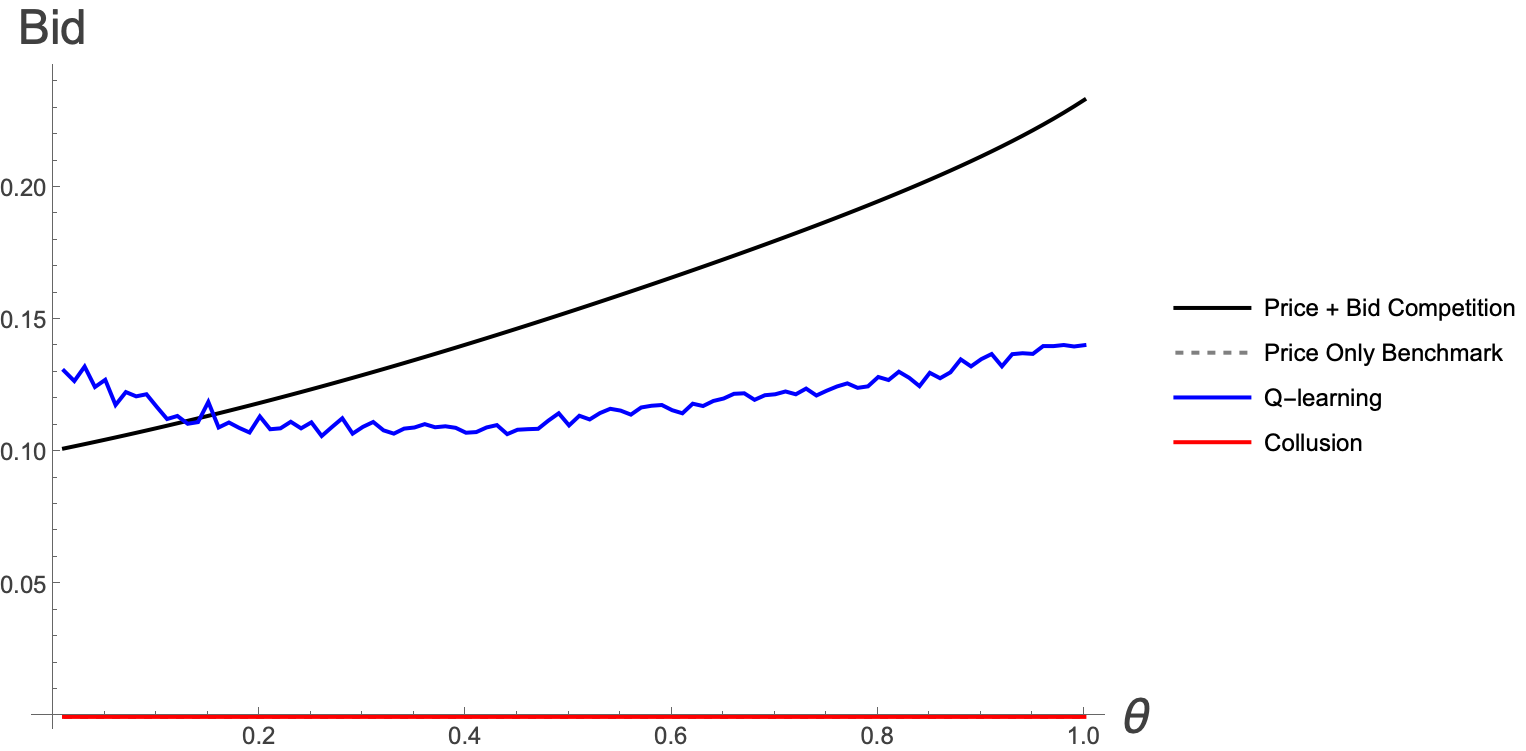}
\end{subfigure}  
\label{fig:qlearning_with_collusion}
\end{center}    
    Panels (a) and (b) show the Q-learning (solid blue) vs competitive (solid black) and collusive (red) prices and bids as a function of $\theta$. 
\end{figure}

Figure \ref{fig:qlearning_with_collusion} illustrates these results where we add the collusive outcome to the previous results. We observe a crossing of the competitive and collusive prices. Comparing the full competition price $\mathbf{p}^N$ with the fully collusive price $\mathbf{p}^M$, we find that for low values of $\theta$, $\mathbf{p}^M > \mathbf{p}^N$, while for high values of $\theta$, $\mathbf{p}^M < \mathbf{p}^N$. There is a crossover value of $\tilde{\theta}$, such that $\mathbf{p}^M(\tilde{\theta}) = \mathbf{p}^N(\tilde{\theta})$.

These results are summarized in the following lemmas:

\begin{lemma}
\label{obs:monopoly_price}
    The monopoly prices $\mathbf{p}^{M}$ decrease with $\theta$.
\end{lemma}

Appendix \ref{sec:appendix_theoretical_collusion} provides the detailed proof. Unlike the case of full competition, collusive prices decrease with $\theta$. When $\theta=0$, all consumers consider both products, and because of the consumer's idiosyncratic preferences, the monopolist has two chances to compete with the outside option. However, when $\theta=1$, the monopolist only has one product to sell (because every consumer only considers one product from the first link). In this sense, the outside option becomes stronger when $\theta=1$, forcing the monopolist to decrease the price and obtain a lower profit margin.

In the limit when $\theta=1$, we prove the following:
\begin{lemma}
\label{obs:limiting_case}
    \textbf{Limiting case}  When $\theta=1$, the monopoly price $\mathbf{p}^M$ is equal to the competitive price without advertising $\mathbf{p}^{oN}$, i.e., $\mathbf{p}^M=\mathbf{p}^{oN}$. 
\end{lemma}
This holds because when \(\theta=1\), consumers in both cases consider only one product, giving both the competitive seller and monopolist the same objective with zero ad costs.

We combine Lemmas \ref{obs:competitive_price}, \ref{obs:monopoly_price} and \ref{obs:limiting_case} to arrive at our main result:

\begin{prop}
\label{prop_crossing}
There exists $\tilde{\theta} \in (0,1)$ such that for $\theta<\tilde{\theta}$,  $\mathbf{p}^M>\mathbf{p}^N$, while for $\theta>\tilde{\theta}$, $\mathbf{p}^M<\mathbf{p}^N$.
\end{prop}

We present detailed proof in Section \ref{appsec:prop_proof}.
% Effectively, we see that search costs and ad auctions create an interaction where collusive prices do not behave as we would intuitively expect.
The proposition guarantees that there is a level of search cost where the fully collusive prices become lower than the fully competitive one, because the savings in ad costs dominate the incentive to raise prices. Since algorithmic pricing and bidding are expected to always be between these two equilibria, we expect our results to generalize to any dynamic pricing algorithm that also buys advertising. 

\subsection{Impact on Sellers and Consumers}
\label{subsec:impact_sellers_consumers}

\begin{figure}[ht]
\caption{Q-learning Results: Sellers' Profit and Consumer Surplus}
\label{fig:qlearnin_profits_cs} 
\begin{subfigure}{0.5\textwidth}
\caption{Sellers' Profit}
\centering
\includegraphics[width=\textwidth]{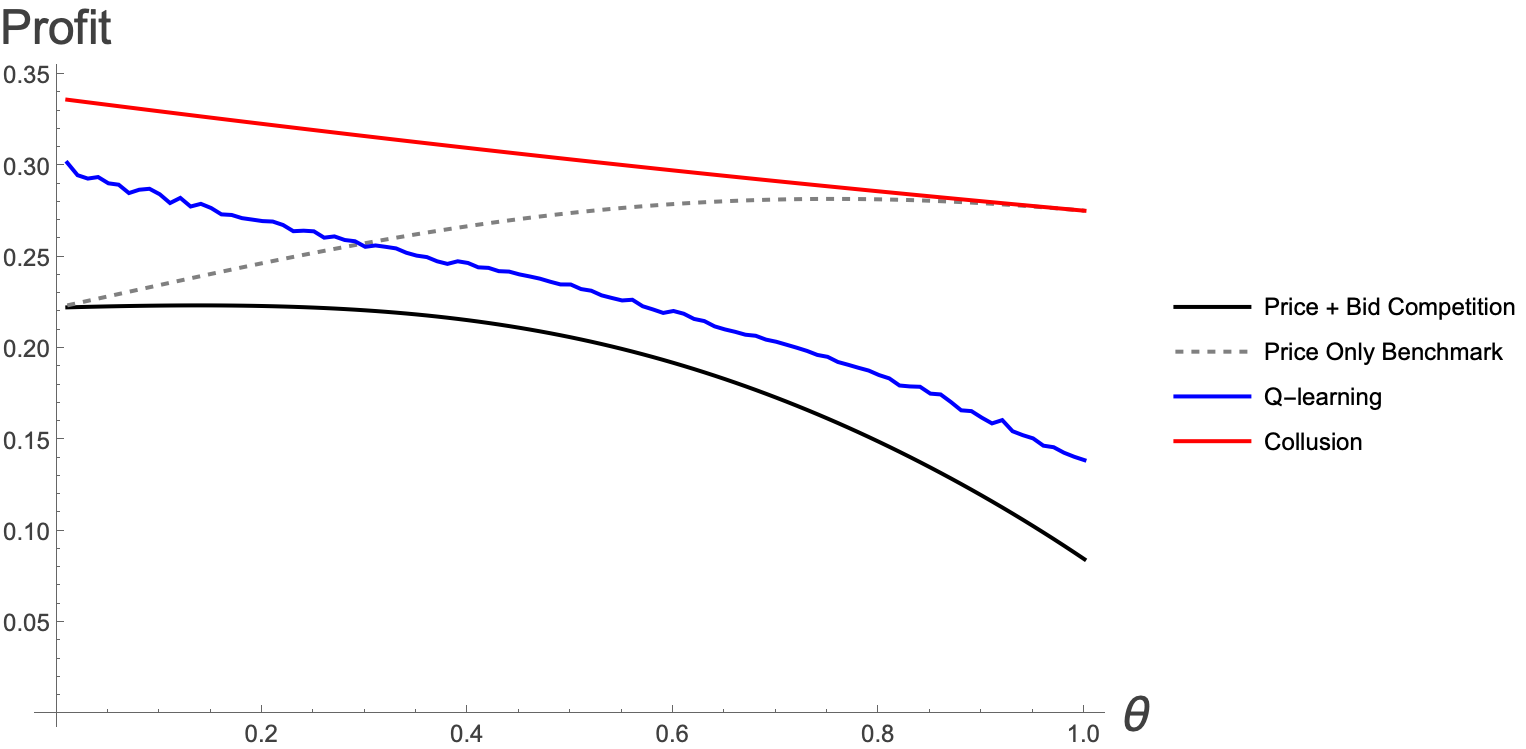}
\end{subfigure}~
\begin{subfigure}{0.5\textwidth}
\caption{Consumer Surplus}
\centering
\includegraphics[width=\textwidth]{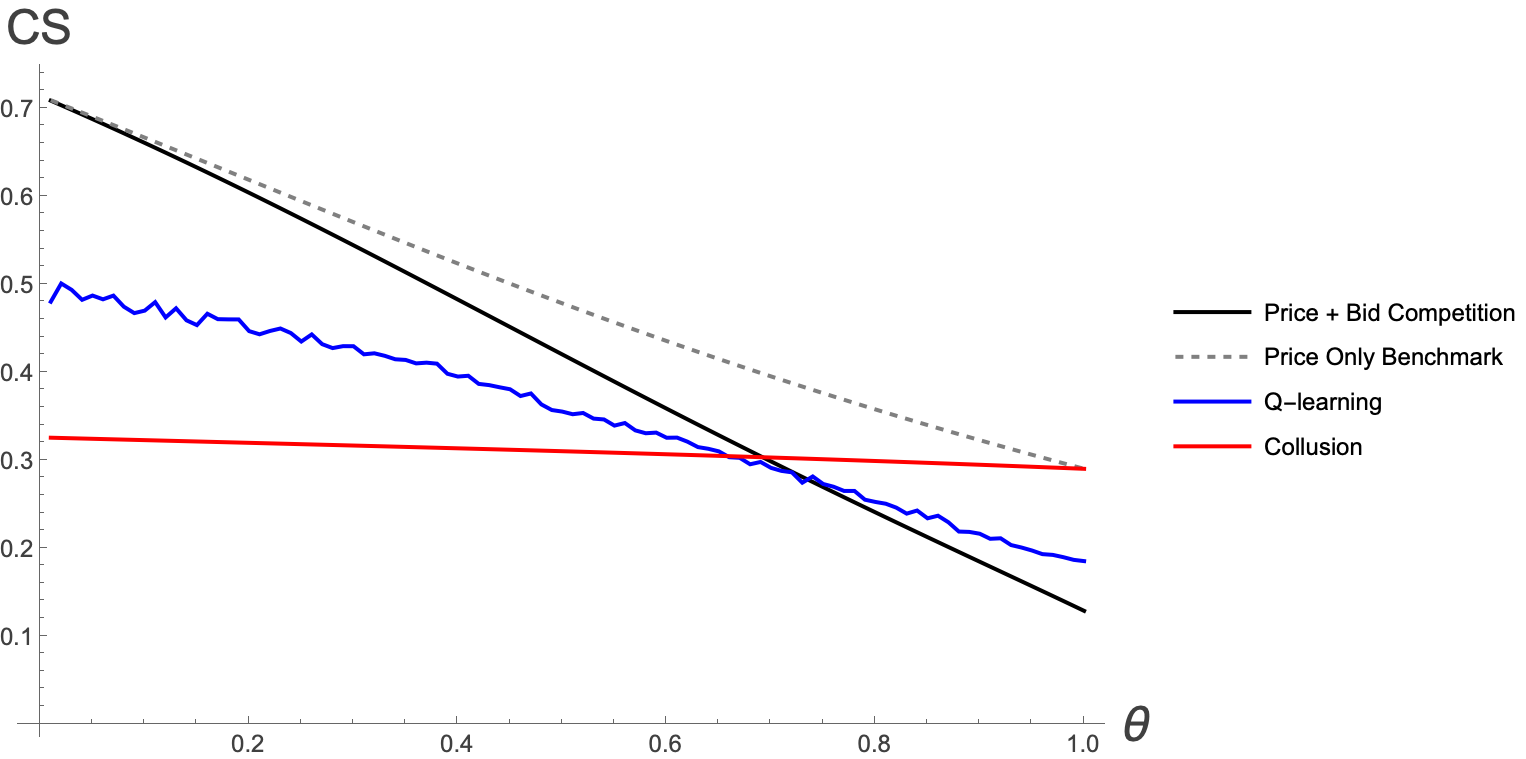}
\end{subfigure}
    The left and right figure show the sellers' profit and consumer surplus, respectively. Solid blue denotes Q-learning simulation, solid black denotes full competition and solid red full collusion. Dashed indicates competition without advertising.
\end{figure}

Figure \ref{fig:qlearnin_profits_cs} presents the Q-learning simulation results for sellers' profit and consumer surplus.\footnote{
Consumer surplus in period $t$ is
\begin{align}\nonumber
CS\left(\boldsymbol{p^t}\right)&=\theta \cdot \sum_j \mathbf{1}\{j \in \mathcal{J}_1(\boldsymbol{\Gamma^t})\} \cdot \mu \log \left[ \exp \left(\frac{a-p_j^t}{\mu}\right)+1\right]+ (1- \theta) \cdot \mu \log \left[\sum_{j} \exp \left(\frac{a-p_j^t}{\mu}\right)+1\right]\\
&=- \theta \cdot \sum_j \mathbf{1}\{j \in \mathcal{J}_1(\boldsymbol{\Gamma^t})\} \cdot \mu \log \left( 1-s_j\left(p_j^t\right) \right)- (1- \theta) \cdot \mu \log \left( 1- \sum_{j} s_j\left(\boldsymbol{p^t}\right)\right)
\label{equ:consumer_surplus}
\end{align}
} Figure \ref{fig:qlearnin_profits_cs}(a) shows that although algorithmic pricing yields lower profit than full collusion, sellers still benefit compared to full competition.
%from Q-learning fall below the theoretical collusion scenario, yet are higher than the levels in full competition. This indicates that the use of learning algorithms can benefit firms compared to fully competitive strategies, for all values of $\theta$. Algorithms, whether used for single-dimensional pricing learning or multi-dimensional learning including both pricing and bidding, facilitate some degree of collusion. While this does not amount to full collusion, collusion typically advantages sellers, yielding higher profits. We also notice that algorithmic profits can be lower than the benchmark case of no ads when $\theta$ is high, because the ad costs outweigh the
%benefits derived from collusion on higher prices.
%The comparison of Q-learning outcomes with the benchmark scenario, where no ads are present, varies depending on the specific values of $\theta$. When consumer search cost is high, sponsored ads increase the sellers' costs, adversely affecting their profits. In such instances, the impact of increased costs outweighs the benefits derived from algorithmic collusion. Thus, sellers' profits are lower than in scenarios with price competition only.

Consumer surplus decreases with prices. Indeed, Figure \ref{fig:qlearnin_profits_cs}(b) shows that for high $\theta$ consumer surplus can be higher when using learning algorithms compared to full competition, because the Q-learning prices are lower than the full competition prices.
%we observe an inverse relationship with prices. Specifically, at higher values of \(\theta\), Q-learning pricing is lower than competitive pricing, leading to increased consumer surplus in the Q-learning scenario compared to the competitive case. 

%Since both consumer surplus and sellers' profit are higher than the competitive case with learning algorithms, we conclude that Q-learning algorithms can yield outcomes beneficial for both sellers and consumers. %This finding challenges the common view that collusion is always harmful to consumers. 
% In our scenario, which incorporates a key component of today's business environment --- sponsored ads, collusion facilitated by Q-learning algorithms does not inevitably lead to negative consequences for consumers. On the contrary, it can lead to outcomes that are more beneficial for consumers compared to those of full competition. 

%Finally, consumer surplus in the benchmark case without sponsored ads exceeds that of any scenario involving ads. This indicates that sponsored ads increase sellers' costs, which are ultimately passed on to the consumers, negatively impacting their welfare. 

The Folk theorem \citep{fundenberg1986folk}, says that in an infinitely repeated game, any outcome between cooperative (our full collusion) and competitive (our full competition) in the one-shot game can be sustained as a subgame-perfect Nash equilibrium (SPE) when the discount factor $\delta$ is high enough. %We believe that this would hold true in our setting, as the sellers change prices and bids at a very high frequency in the e-commerce platform setting.
Consequently, we cannot predict whether the algorithmic prices and profits will be closer to fully collusive or fully competitive, and we again use simulation for the analysis. 
%Thus, the collusive and competitive outcomes in the one-shot game establish the upper and lower bounds of the average payoff in the repeated game. Hence, one would expect the average prices under algorithm in pricing to fall within that range. However, Q-learning has not yet been fully characterized theoretically, to our current knowledge; we cannot predict the exact levels of algorithmic prices without running simulations. Therefore, we need to run simulations, allowing the algorithms to interact with each other to see the final outcomes. More importantly, from the simulation results, we can know the exact level of algorithmic outcomes and their relative relationship to competition and monopoly outcomes --- whether they are closer to a monopoly or to full competition.
% Recall that one of the important goals of this section is to investigate the relationship between algorithmic outcomes and theoretical outcomes, specifically whether they are closer to a monopoly or to competition. 
We plot the ratios of the difference between the algorithmic and competitive prices and profits, divided by the difference between the collusive and competitive prices and profits, respectively. That is, we compute $\frac{p^{Q}-p^N}{p^M-p^N}$ and $\frac{\pi^{Q}-\pi^N}{\pi^M-\pi^N}$. A ratio smaller than 0.5 indicates that the algorithmic outcomes are closer to competition; otherwise, they are closer to collusion. 

\begin{figure}[ht]
\caption{Q-learning Results: Price and Profit Ratios}
\label{fig:qlearnin_ratio} 
\begin{subfigure}{0.45\textwidth}
\caption{Price Ratio}
\centering
\includegraphics[width=\textwidth]{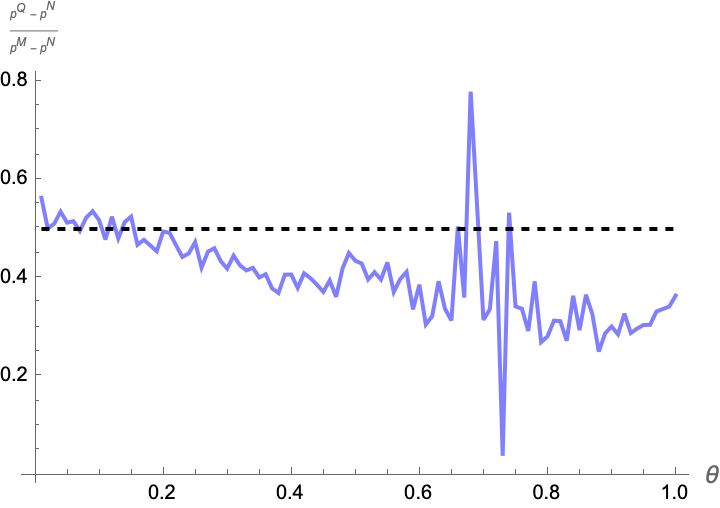}
\end{subfigure}~
\begin{subfigure}{0.45\textwidth}
\caption{Sellers' Profit Ratio}
\centering
\includegraphics[width=\textwidth]{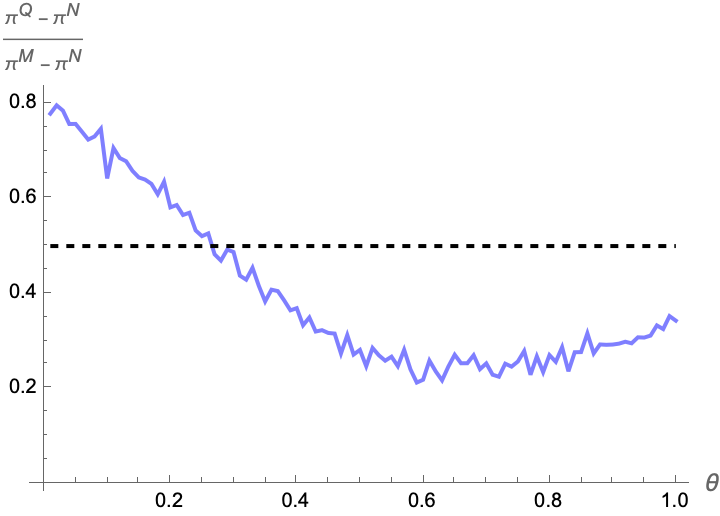}
\end{subfigure}

    The left and right figure show the ratio statistics for  prices and sellers' profit, respectively. The parameter specifications are the same as in Figure \ref{fig:theoretical-results}.  
\end{figure}

Figure  \ref{fig:qlearnin_ratio} shows that when 
$\theta=0$, the algorithms can achieve roughly 60\% of the price increase from the competitive price to the collusive price, and 80\% of the profit increase from the competitive profit to the collusive profit. For low values of $\theta$, we see that the ratios for both price and profit are higher than 0.5, implying that the algorithms generate outcomes closer to monopoly (collusion), which is consistent with previous findings about algorithmic collusion \citep{calvano2020artificial}. However, for high values of $\theta$, in the range where the algorithms benefit both consumers and sellers, the algorithms can only achieve profit increases that are roughly between 30\% and 40\% of the increase from the competitive profit to the collusive profit. This suggests that when algorithms compete in more than just one dimension of pricing, the algorithmic outcomes are closer to competition. This is the opposite of cases of $\theta=0$, where there are no consumer search costs.

%Furthermore, if sellers make sequential decisions in pricing and bidding, we might still observe reduced AI prices due to the cost savings effect.

% \paragraph{How collusion is supported?}

% To understand our results, it is  natural to ask how are lower prices and lower bids than the competitive levels  maintained in the long-term play.   

\subsection{Additional bidding information, differentiated products, multiple sellers, and separate algorithms for pricing and bidding}
\label{subsec:robustness}
We performed robustness analyses for four scenarios.

In Appendix \ref{subsubsection:alternative_bid}, each seller is allowed to also observe the bids of their competitors. 
The motivation is that the platform might experiment with the auction design and disclose winning bids,\footnote{For example, Google has been experimenting with providing additional information to ad auction participants and has recently introduced the ``minimum bid to win'' field for bidders, which provides insight into the bid price needed to win an individual auction (\url{https://support.google.com/authorizedbuyers/answer/2696468?hl=en&utm_source=chatgpt.com}
, accessed Oct 20, 2025)} or that sellers could subscribe to a third-party service that provides bidding information. The additional information will change the sellers' bids, and we are interested in how this will affect the platform's revenue and the sellers' profits. Our results show that the additional information does not impact the platform's revenue and the sellers' profits much. The algorithms converge to almost the same equilibrium but at a slightly higher level of collusion. Hence information-based methods might not be an effective tool for the platform, and it might want to consider more direct incentive-based tools to strategically respond to sellers who use algorithmic pricing (Section \ref{sec:strategic_platform}).

%Thus, we aim to provide guidance to platform managers on how disclosing more or less information about ad auctions can serve as an effective information-based tool to increase profits. Additionally, we explore whether sellers should subscribe to those data services.
 
%We consider the following \textit{full stateful} scenario, where both agents' bids are observed; that is, the state space is \( s_{it} = (p_{it-1}, p_{jt-1}, b_{it-1}, b_{jt-1} ) \). 

Appendices \ref{subsubsection:asymmetric_sellers} and \ref{subsubsection:multiple_sellers} show robustness for multiple sellers and differentiated sellers. In these scenarios with multiple heterogeneous sellers, the market’s average prices can still be lower when algorithm usage is high.
Appendix \ref{subsubsection:separate_algos} also shows that our results remain robust to allowing the sellers to run two separate RL algorithms (one for pricing and one for bidding).

\section{Strategic Response by the Platform}
\label{sec:strategic_platform}
An important aspect not considered in previous work is that the platform can respond strategically to equilibrium prices and bids to maximize its own profits. Since the platform's revenue come from sale commissions and ad auctions, we consider two incentive-based mechanisms: adjusting the commission rate and adjusting the ad-auction reserve bids.

%There are also information-based strategies to affect seller behavior, for example, the platform can disclose more or less information about the winning bids in the ad auctions which will influence the sellers' bids. In terms of objective, the platform might only care about its own profits, but it might also consider consumer surplus and sellers' profits on the platform to encourage more entry and improve long-term business growth.

Unlike most of the research on ad auctions where adjusting auction reserve prices has a clear direction of impact, 
%In our previous discussion, when sellers used algorithms to determine the prices and bids, both algorithmic prices and bids fell below competitive levels. This might adversely impact the platform's revenue from both commission fees and ad revenue. In the online auction setting, search engines and platforms implement various auction designs to increase revenue. Strategies include raising auction reserve prices, limiting the number of sponsored spots, or adjusting auction formats \citep{decarolis2021mad,kobayashi2023dynamic}. However, 
in our setting, due to interactions with pricing competition and demand, %impact of sellers using algorithms on platforms' profits and 
the best direction for the platform's strategic response are unclear. It depends on market parameters and the platform's objectives. 

Appendices \ref{subsec:commission_rate} and \ref{subsec:reserve_price} consider the scenario where the platform's objective is to maximize its own single-period profit which is the sum of sales commissions and ad auction revenue.
%,\footnote{The platform's profit consists of commission fees and advertising revenue, and the platform's profit in a single period is
%\begin{align}\nonumber
%    \pi_p(\boldsymbol{p}, \boldsymbol{b})  = \pi_p^{\text{Ad}}(\boldsymbol{p}, \boldsymbol{b}) + \pi_p^{\text{Com}}(\boldsymbol{p}, \boldsymbol{b}) & = \theta \cdot \sum_j \operatorname{Pr}(\tilde{b}_j > \tilde{b}_{-j}) \cdot s_j(p_j) \cdot \left( \gamma_j \cdot \mathbb{E}[\tilde{b}_j \mid \tilde{b}_j > \tilde{b}_{-j}]\right)\\
%     & + \tau \cdot \theta \cdot \sum_j \operatorname{Pr}(\tilde{b}_j > \tilde{b}_{-j}) \cdot p_j \cdot s_j(p_j)  + \tau \cdot (1-\theta) \cdot \sum_j \cdot p_j \cdot s_j(p_j, p_{-j})
     %\label{equ:platform_profit}
%\end{align}} and 
%We investigate two common incentive-based methods typically used by platforms: adjusting the commission rate (which directly impacts the commission revenues) or the reserve price (which directly impacts the ad auctions revenues). 
We also consider a scenario where the platform's objective is long-term growth by maximizing a weighted average of its own profit, sellers' profit, and consumer surplus in Web Appendix \ref{subsec:total_surplus}. 

We find that increasing the reserve price may not be an effective response for the platform, whereas adjusting the commission fee is. When the platform does make adjustments, sellers and consumers can still continue to benefit (compared to the case of full competition), and this is also the case when the platform’s objective does not only maximizing its own short-term profit.

% The platform's commission   revenue is 
% \begin{align}
%     \pi_p^{\text{Com}}(\boldsymbol{p}, \boldsymbol{b}) = \tau \cdot \theta \cdot \sum_j \operatorname{Pr}(\tilde{b}_j > \tilde{b}_{-j}) \cdot p_j \cdot s_j(p_j)  + \tau \cdot (1-\theta) \cdot \sum_j \cdot p_j \cdot s_j(p_j, p_{-j}) 
% \end{align}

%  The advertising revenue is
%  \begin{equation}
%      \pi_p^{\text{Ad}}(\boldsymbol{p}, \boldsymbol{b})= \theta \cdot \sum_j \operatorname{Pr}(\tilde{b}_j > \tilde{b}_{-j}) \cdot s_j(p_j) \cdot \left( \gamma_j \cdot \mathbb{E}[\tilde{b}_j \mid \tilde{b}_j > \tilde{b}_{-j}]\right)
%  \end{equation}

\section{Empirical Evidence from  Amazon.com}
\label{sec:empirical_application}
Our analysis predicts that when search costs are high on the platform, algorithmic pricing and bidding can benefit consumers and sellers through lowered prices. We collected and analyzed a large-scale dataset from Amazon.com to look for evidence consistent with this prediction. Our analysis is correlational and cross sectional, and hence we cannot fully interpret it as causal evidence.

\subsection{Data}
We collected data from Amazon.com from April 2024 to April 2025. The data includes 2,382 highly searched keywords across all product categories. Our software submitted a search keyword and then navigated to the first search result page mimicking normal user behavior. We recorded the ranking of each product (both sponsored and organic) within the first page. We also recorded other product information, including its price, product ratings, the number of reviews, whether it is an Amazon Prime product, and delivery information. To obtain a representative distribution of the sponsored product listings which vary over time and depending on who wins the ad auctions, we repeated the search for each keyword once every three hours, generating more than 19,000 requests per day. 

We obtained additional data from \textit{Jungle Scout}\footnote{One of the largest e-commerce data intelligence companies.} that included information on the monthly search volume, suggested pay-per-click bid, and related keywords for each keyword. 
We also used \textit{Keepa.com} which provided product information, ratings, reviews, and time listed. %, and daily sales derived from daily sales ranks.\footnote{The dataset contains the daily best sales rank within each product category reported by Amazon, which is generally considered a reliable proxy for sales quantity in the literature \citep{chevalier2003measuring}. We then converted the best seller rank to daily sales using the conversion equation provided by \textit{Jungle Scout}. } 

Table \ref{tab:summary_statistics_keyword} in the Web Appendix presents summary statistics. Figures \ref{fig:by_position_sponsored_combined} and \ref{fig:by_position_sponsored_heatmap_combined} shows that most sponsored listings appear near the top search results, consistent with our assumptions.
%Consistent with our theoretical prediction that advertising raises prices, sponsored products have higher prices than organic products, as shown in Appendix \ref{appsubsec:sponsored_organic_comparison}.

\subsection{Estimation of Search Costs}
To test for evidence of decreasing prices when algorithms are used and search costs increase, we need to estimate consumer search costs in each keyword market. For that we use the variations in organic and sponsored product positions and daily sales. We first introduce the notations and assumptions, then describe the identification strategy, and present our estimation results.

% \paragraph{Model Setup}
\paragraph{Market Definition} We assume that each keyword $k$, is an independent market, and the market size is its search volume. Using the daily bestseller rank information from Amazon\footnote{The daily best sales rank is generally considered a reliable proxy for sales quantity in the literature \citep{chevalier2003measuring}.} and the Rank to Sales estimation tool from \textit{Jungle Scout}, we calculated the imputed daily sales, $q_{jt}$, for a product $j$. The market share of product $j$ in market $k$ at time $t$ (day) is $s_{jkt} = \frac{q_{jt}}{V_{k}/30}$, where $V_{k}$ is the monthly search volume of keyword $k$.

%In each market, products are indexed by $j \in \mathcal{J}_k$. 
\paragraph{Estimating Consumer Consideration Size Distribution} When our scraper searches a keyword $k$, it see $N$ products in the first page of the search results.\footnote{For most of the product categories $N=60$  and $N=22$ for the category `Electronics.'} For each keyword, we conduct a search once every three hours, resulting in a total of $S = 8$ searches each day, with each search indexed by $s$. Let $\mathcal{R}_k^{st}$ be the ordered ranking of products in search $s$ of keyword $k$ in day $t$ and $\mathcal{J}_{n} (\mathcal{R}_k^{st})$ is the set of products in the first $n$ positions.\footnote{Among the $N$ different positions, there are both sponsored and organic positions. The number of sponsored positions is 12 when $N=60$, and 6 when $N=22$.} Consumers face search costs and might stop at a certain position without continuing to the next position on the search results page. We assume that there is a unit mass of consumers and that the mass of consumers who stops at the $n$-th position follows an exponential distribution, i.e., $\lambda \cdot e^{-\lambda \cdot n}$, where $\lambda$ is the rate parameter.\footnote{Hence, the mass of consumers who stop at position $n$ or before is given by the cumulative distribution function of the exponential distribution, $F(\lambda, n) = 1 - e^{-\lambda \cdot n}$. And the mass of consumers with consideration set of size $n$ is $F(\lambda, n) - F(\lambda, n-1) = e^{-\lambda \cdot (n-1)} - e^{-\lambda \cdot n}$.} 

\paragraph{Consumer Choice Model}
Consumer $i$'s utility of purchasing product $j$ at time $t$ is:
\begin{equation}
u_{ijt} = \underbrace{X_{jt}' \cdot \beta - \alpha \cdot p_{jt}  + \gamma Sponsored_{jt} + \xi_{jt}}_{\delta_{jt}} + \epsilon_{ijt}
\end{equation}
where $X_{jt}$ are product characteristics,  $p_{jt}$ is the price of product $j$ at time $t$, and $\xi_{jt}$ is an unobserved demand shock. The idiosyncratic preference shock $\epsilon_{ijt}$ follows a Type I extreme value distribution. The mean utility of the outside option, or not buying any product on the first page of the search results, is normalized to zero. We denote the deterministic part of the utility as $\delta_{jt}$.

As a product may appear in different positions within each search result, each potentially leading to different sales outcomes, we compute the average choice probability across searches. The market share of product $j$ in keyword $k$ in day $t$ is
\begin{equation}
    share_{jkt}=\frac{1}{|S|}\sum_{s=1}^{S}\sum_{n=1}^N \mathbf{1}\{j\in J_{n(\mathcal{R}^s_{kt})}\} \cdot(F(\lambda, n) -F(\lambda, n-1)) \cdot \frac{e^{\delta_{jt}}}{1+\sum_{j' \in J_{n}(\mathcal{R}^s_{kt})}  e^{\delta_{j't}}}
    \label{equ:market_share}
\end{equation}
Here, $F(\lambda, n) -F(\lambda, n-1)$ is the mass of consumers with consideration size $n$. The last term is the standard logit choice probability of product $j$, conditional on it appearing among the top $n$ products in a given search result $\mathcal{R}^s_{kt}$.

% Let $\mathcal{K}_j$ be the set of keywords where product $j$ appears. Thus, the total sales of a product at time $t$ are
% \begin{align}
%       \label{equ:total_sales}
%   q_{jt}&=\sum_{k' \in  \mathcal{K}_j} V_{k'} \cdot s_{jk't}\\\nonumber
% &=\underbrace{\sum_{k' \in  \mathcal{K}_j \cap \bar{\mathcal{K}}  } V_{k'} \cdot s_{jk't}}_{\text{scraped keywords}} + \underbrace{\sum_{k'' \in  \mathcal{K}_j\backslash \bar{\mathcal{K}} } V_{k''} \cdot s_{jk''t}}_{\text{non-scraped keywords}}
% \end{align}

%For now, we argue that the volume of non-scraped keywords is low compared with the scraped keywords, thus its impact on the total daily sales is negligible. %Later, we are going to validate this using additional data from \textit{Jungle Scout} or by performing the estimation iteratively.

\paragraph{Estimation and Identification}
We use the method of moments similar to 
\cite{lam2021platform} and \cite{yu2024welfare} with an instrument to account for the endogeneity between the rank of a product and it sales. The previous period’s organic rank is used as an instrument for the current period's rank.
We assume that the unobserved quality of a product follows an AR(1) process, i.e., ${\xi_{jt}} = \eta_{jt} + \rho \cdot {\xi_{jt-1}}$, where $\xi_{jt-1}$ is the lagged unobserved quality and $\eta_{jt}$ is an uncorrelated contemporaneous shock.
Additionally, we assume that the previous period’s organic rank, $r_{jt-1}$, is uncorrelated with the contemporaneous shock $\eta_{jt}$. The rationale is that the platform cannot predict the next period's shock in the unobserved quality, and even if it could, the platform does not have any incentive to incorporate that into the current period’s organic ranking. Both of these yield the moment condition:
\begin{equation}
\mathbb{E}\left(\begin{array}{l}
\eta_{j t} \cdot \xi_{j t-1} \\
\eta_{j t} \cdot r_{j t-1}
\end{array}\right)=0
\label{equ:moment_condition}
\end{equation}
Estimation proceeds as follows:
\begin{enumerate}
    \item Start with a guess for $\lambda$. Invert \eqref{equ:market_share}, and find solutions   $\tilde{\delta}_{jt}$.

    \item Regress $\tilde{\delta}_{jt}$ on $X_{jt}$, $p_{jt}$ and $Sponsored_{jt}$. Obtain the residual  $\tilde{\xi}_{jt}$.
    \item Given a guess of $\rho$, construct $\tilde{\eta}_{jt}=\tilde{\xi}_{jt}-\rho \cdot \tilde{\xi}_{jt-1}$.
    \item Construct the moment condition according to \eqref{equ:moment_condition}, where $r_{j t-1}$ is the rank of product $j$ at time $t-1$. Search for $\lambda$ and $\rho$ that minimize this condition.
\end{enumerate}

\begin{figure}[!ht]
\begin{center}
\caption{Estimation of Consumer Search Costs}
\label{fig:by_category_search} 
\includegraphics[width=0.8\textwidth]{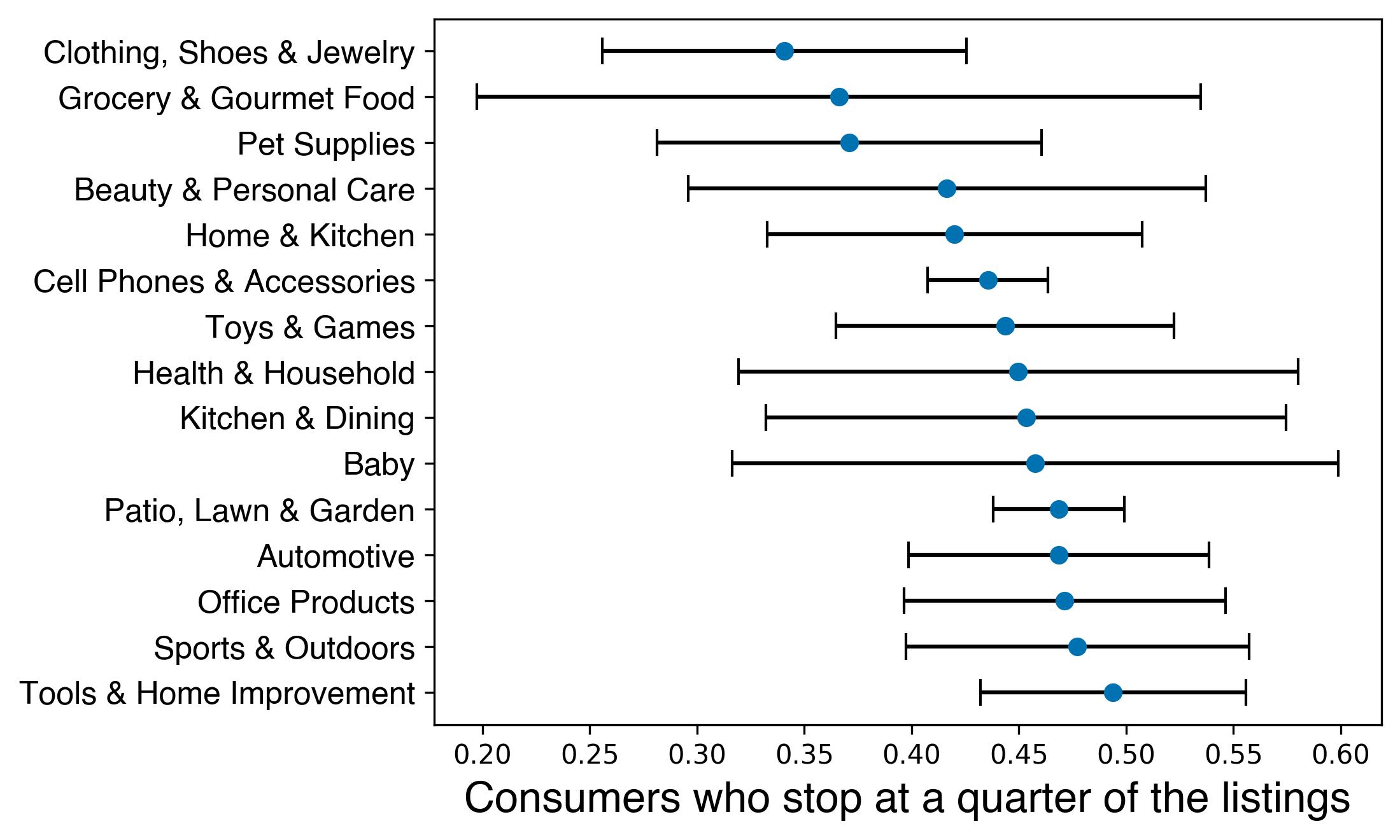}
\end{center}
Fraction of consumers who stop at or before a quarter of the listings in the search results page. The estimate is by keyword/market, and we then obtain the mean and standard deviation for all keywords within each product category in our sample.
\end{figure}

\paragraph{Estimation Results}
Figure \ref{fig:by_category_search} plots the distribution of estimated search costs within each category. There is substantial variation in consumer search costs across categories. For the categories `Clothing, Shoes \& Jewelry,' `Pet Supplies,' and `Beauty \& Personal Care,' consumers tend to stop their product search later in the search results pages. By contrast, for `Office Products,' `Sports \& Outdoors,' and `Tools \& Home Improvement,' consumers search fewer products before making a purchase decision. These findings are consistent with consumers searching less when the benefit of continuing their search is low because there is little new differentiation to be found. Since the search frictions are quite high for some categories, we expect to see beneficial outcomes from using pricing algorithms in these categories.

\subsection{Evidence of Negative Interaction between Search Costs and Algorithm Usage}

As we do not know whether a seller is using algorithms or not we need to infer it from the data. We construct an algorithm usage index derived from the correlation in pricing patterns, following \cite{chen2016empirical}. The intuition is that sellers using algorithmic pricing are likely to base their prices, at least partially, on prices set by other sellers. Our data contains prices at quite high frequency, and if a seller shows high correlation with competitor prices, it suggests that they likely use pricing algorithms to respond to competitors at fast pace. We compute the correlation of the price vector of a product, $\boldsymbol{\overrightarrow{p_j}}$, with the market average price, $\boldsymbol{\overrightarrow{p_{j^{\prime}}}}$. If the correlation is greater than a threshold $\rho$, we infer that the seller of product $j$ is using algorithms to adjust prices. We then average the algorithm usage indicator for all products in a market as the market's algorithm usage index:
\begin{equation}
  algo_k= \frac{1}{\left|J_k\right|} \sum_{j \in J_k}   \mathbf{1}\left\{\operatorname{corr}\left(\boldsymbol{\overrightarrow{p_j}}, \boldsymbol{\overrightarrow{p_{j^{\prime}}}}\right)\ge \bar{\rho}\right\} 
\end{equation}

\begin{figure}[!ht]
\begin{center}
\caption{Algorithm Usage Index}
\label{fig:by_category_algo} 
\includegraphics[width=0.75\textwidth]{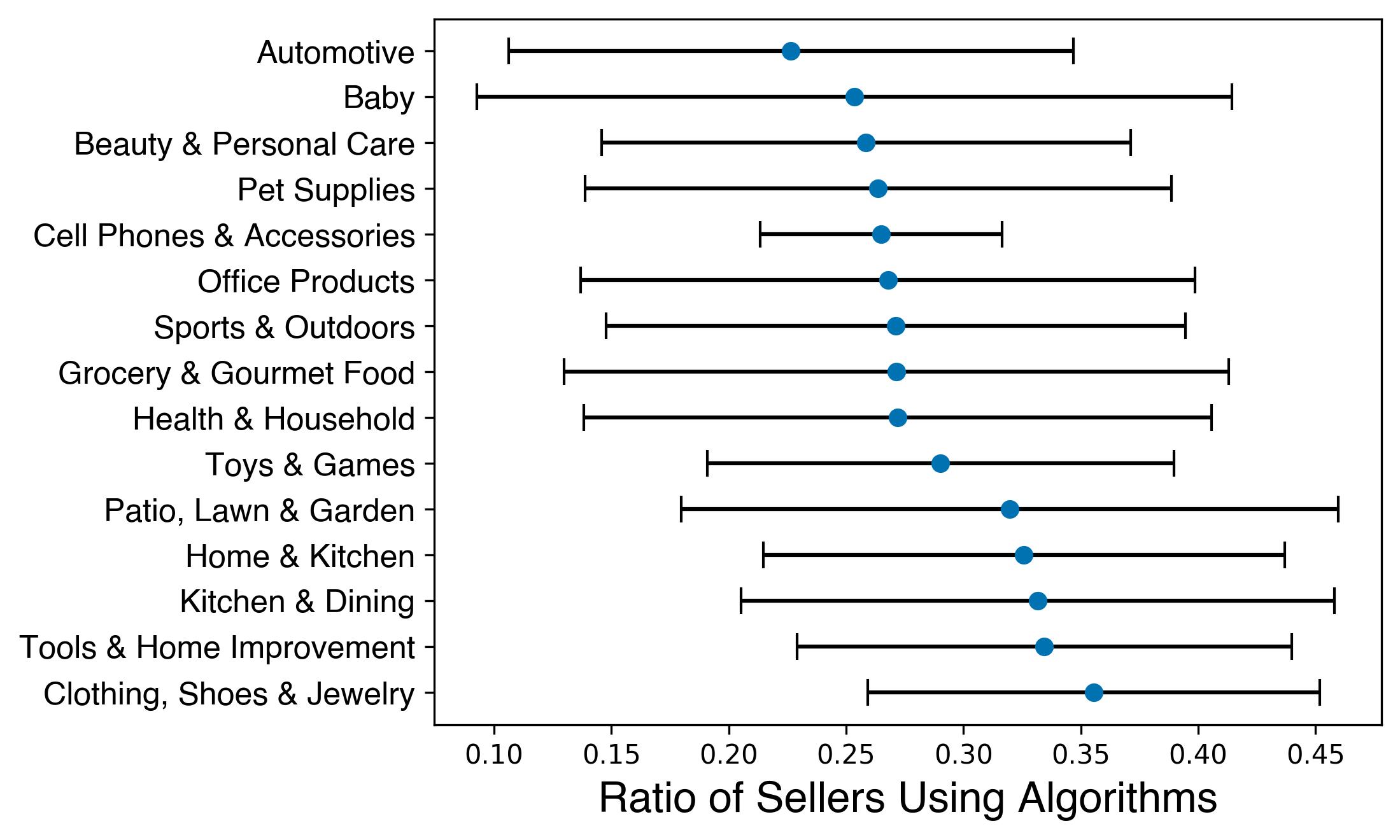}
\end{center}
Fraction of sellers using algorithms in each product keyword market. The mean and standard deviation within each product category are computed over all product-keywords in that category.
\end{figure}

Figure \ref{fig:by_category_algo} shows the fraction of sellers using algorithms in each product-keyword category.  When we set the threshold $\rho = 0.3$, the mean ratio of sellers using algorithms across all keywords is 31.1\%, with a standard deviation of 12\%. The empirical findings we provide is robust to different thresholds we test.

\begin{figure}[!ht]
\begin{center}
\caption{Interaction of Consumer Search Costs on Prices}
\label{fig:Interaction_effect_3} 
\includegraphics[width=0.6\textwidth]{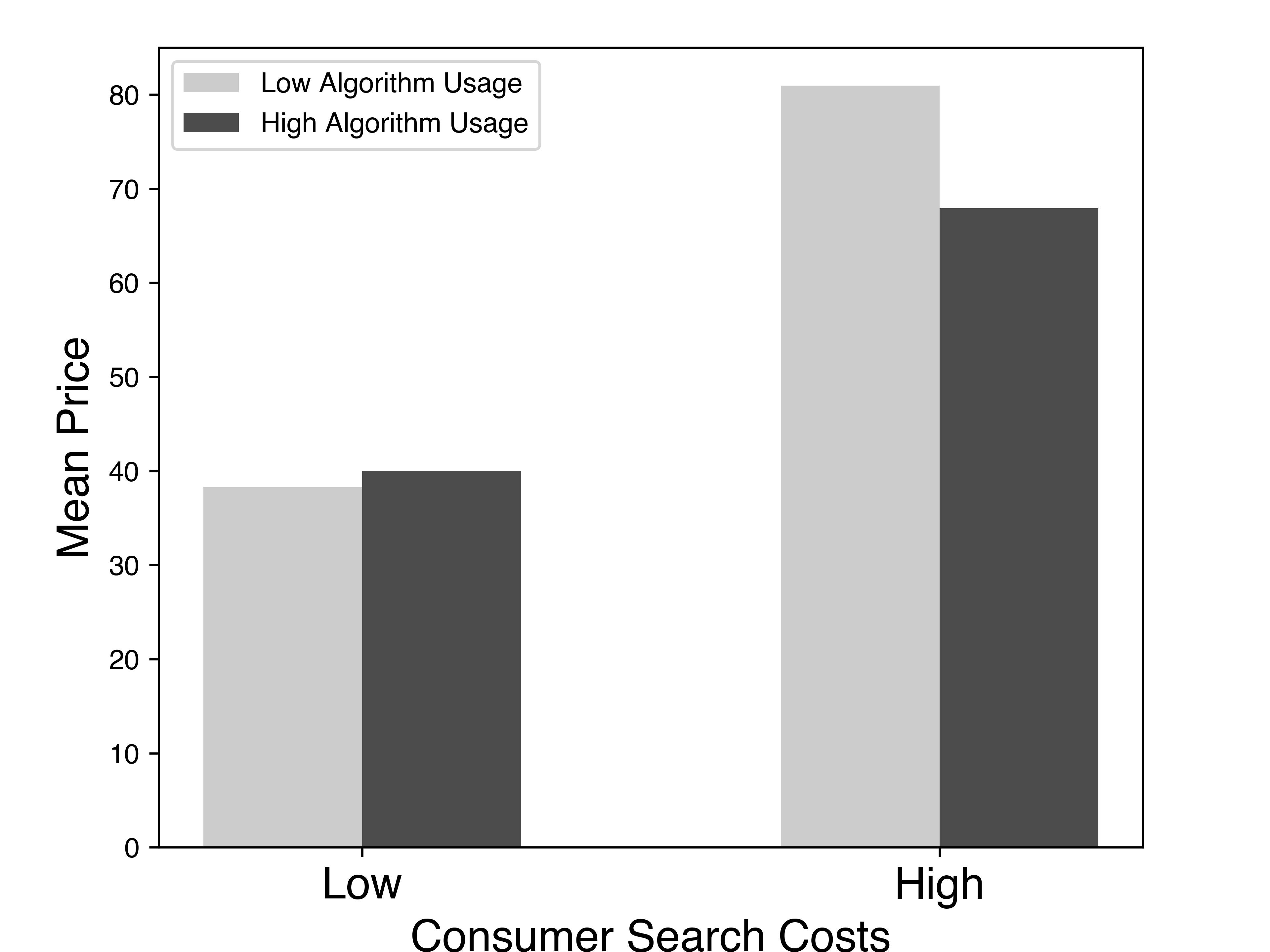}
\end{center}
The figure shows the mean prices in markets with low versus high consumer search costs and algorithm usage.
\end{figure}

Figure \ref{fig:Interaction_effect_3} compares the mean product prices when we split the data into low vs high consumer search costs and algorithm usage. With low consumer search costs, higher algorithm usage correlates with higher prices, while with higher consumer search costs, it correlates with lower prices. This negative interaction is consistent with the prediction from our model about pricing and advertising collusion. 

We also conducted a regression analysis to control for category fixed effects and rule out alternative explanations such as that products with higher search costs may also tend to be more expensive or have higher production costs. Table \ref{tab:algo_search_inter} in the web appendix shows that the results remain similar even after controlling for these effects.

\section{Discussion and Conclusion}
\label{sec:conclusion}
Our paper analyzes competing sellers who need to decide on product prices and ad-auction bids in an e-commerce setting. Many of these sellers opt to using algorithms to make these decisions, and past research raised concerns that the outcome might be harmful for consumers because of collusion on higher prices. Our findings show that when consumers have heterogeneous search costs, and enough of them have high costs, then employing reinforcement learning algorithms can surprisingly result in prices that are lower than competitive prices, because the algorithms learn to collude on lower advertising bids, which lower the marginal cost for pricing. Lower prices are beneficial to consumers, but can also be beneficial to the e-commerce platform because they increase the demand from consumers which compensates for lower advertising revenues. The sellers also benefit, but mostly from saving on advertising costs. 
Our analysis of a large-scale dataset of over 2 million products on Amazon.com showed that consumer search costs are high and heterogeneous for many categories and products. We found a negative interaction between consumer search costs and algorithm usage on pricing, which is consistent with our theory.

%Algorithmic collusion has consistently been a concern from a regulatory standpoint, primarily due to its propensity to increase prices and adversely affect consumer welfare. In this paper, we considered the critical role of sponsored product ads in a digital  platform setting where sellers competitively bid for these ads, aiming to capture the limited attention of consumers. To explore the broader impacts of algorithms usage when involved in multi-dimensional decision-making of pricing and bidding, we used a combination of theoretical modeling, stimulational experiments  and empirical exercise. 

%Our findings indicate that algorithms can facilitate outcomes beneficial for both sellers and consumers compared to the competition scenario, when consumer search cost is high. This results from the interesting interactions between the pricing and bidding of learning algorithms. The tendency of AI algorithms to soften competition and coordinate on lower bids in the bidding dimension reduces sellers' costs and, consequently, leads to lower prices in another dimension. The outcomes of the pricing dimension are more beneficial outcomes for consumers. We show that the algorithms can also benefit the platform, thus, the platform might not strategically respond to sellers adopting algorithms.

Platforms have multiple ways to strategically respond to sellers using such algorithms. Our analysis of incentive based strategies %One way is to limit the amount of information to sellers (e.g., not disclose competing bids) making it harder for algorithms to learn. This strategy turns out to not make much of a difference, leading us to focus on incentive based strategies: changing sales commissions or auction reserve prices. On reserve prices we interestingly find that counter to classic results, increasing the auction reserve price hurts the platform. Reserve prices do not make much difference when they are low, but they lower revenues dramatically when they are too high by causing algorithms to coordinate on even lower bids.
shows that adjusting sales commission rates is better for recoup some of the revenues the platform loses from lower advertising bids compared to changing reserve prices or disclosing bid information. Increased commissions do erode some of the benefits for sellers and consumers from lower prices achieved by algorithms, but these prices still remain lower than without these algorithms.

Our analysis also has a few limitations. First, we focused on two specific decisions: pricing and advertising, while in reality, sellers might need to learn to make others decisions. However, the principles that we uncovered regarding advertising should operate for most decisions that are realized as costs to the sellers, because sellers are better off by coordinating on lower costs if they do not affect demand. Second, we analyzed a specific reinforcement learning algorithm (Q-learning), and the results might not generalize to any learning algorithm. However, we can provide some insight in this direction. Even when initializing an algorithm at the competitive equilibrium prices and providing it with the true state-action value function (Q matrix), the algorithm will gradually shift to non-competitive equilibrium because of its exploration feature. We believe that in other cases where exploration is an inherent property of the algorithm, it will learn to converge on non-competitive outcomes.

Our paper also opens the door to a few additional research questions. For example, platforms might also act as a seller, selling their own private label products. This self-preferencing \citep{farronato2023self,lam2021platform} incentive will add another layer of complexity which is promising for analysis in future work. Another interesting direction is to consider targeting of keywords and allowing algorithms to learn who to target, and then coordinate to ``split the market'' creating inefficient bidding and matching. In this case, it might be possible that welfare is lowered for consumers.

One implication of our findings is to emphasize the impact of consumer search and heterogeneous search costs on learning algorithms. These costs create a unique interaction between pricing and advertising, that cause competition and collusion to yield unexpected outcomes. We believe that this finding makes a unique contribution to the growing literature on algorithmic collusion.

Another implication of our findings affects mostly platforms and consumers. There is currently an ongoing debate about the role of algorithmic decision making on platforms, and whether they are harmful or beneficial to consumers \citep{zhang2021frontiers,fu2022human,zhong2023platform}. Our findings provide a nuanced view on this question. Algorithms indeed create tacit collusion when they compete, but some level of collusion is not necessarily harmful for consumers (or platforms).

%Even if the platform does strategically respond, the beneficial outcomes for sellers and consumers are likely to remain. Whether the algorithms benefit consumers depends on the degree of consumer search costs. 

%Thus, our study helps mitigate the prevailing policy concerns that AI algorithms necessarily lead to detrimental collusion for consumers, especially in a realistic e-commerce digital platform context.

%Given many practical considerations faced by the platform and sellers, we do not claim to have completely resolved the concerns surrounding AI-facilitated collusion. For example, %our setting does not consider the entry and exit decisions of sellers. A less intense bidding competition that makes selling on the platform more profitable could attract more entrants. This dynamic aspect has not been explored in our study, leaving the long-term effects of AI usage in advertising open for future research. Finally, it's important to note that 

%Additionally, we only 

%Despite the limitation, our research significantly enriches our understanding by offering fresh insights into the impact of AI in marketplaces. It sheds light on the mechanism in which algorithms benefit sellers when used for two-dimensional learning. By decomposing the impacts in both dimensions, we find that the results in the pricing dimension can benefit consumers, alleviating the widespread policy concerns that AI collusion is necessarily harmful to consumers.

\section*{Funding and Competing Interests}
Partial financial support was received from (removed for blind review). All authors certify that they have no affiliations with or involvement in any organization or entity with any financial interest or non-financial interest in the subject matter or materials discussed in this manuscript.

\small
\singlespacing
\bibliography{references}
\onehalfspacing
\normalsize
\newcommand{\listappendicesname}{List of Appendices}
\newlistof{appendices}{app}{\listappendicesname}

% ==================================================
%   Regular Appendices (A.1, A.2, A.1.1, etc.)
% ==================================================
\newcounter{appsection}
\newcounter{appsubsection}[appsection]

\renewcommand{\theappsection}{A.\arabic{appsection}}
\renewcommand{\theappsubsection}{A.\arabic{appsection}.\arabic{appsubsection}}

\titleformat{\section}{\large\bfseries}{\theappsection}{1em}{}
\titleformat{\subsection}{\normalsize\bfseries}{\theappsubsection}{1em}{}

\newcommand{\appsection}[1]{%
  \refstepcounter{appsection}%
  \section*{Appendix~\theappsection.~#1}%
  \addcontentsline{app}{section}{Appendix~\theappsection.~#1}%
}

\newcommand{\appsubsection}[1]{%
  \refstepcounter{appsubsection}%
  \subsection*{\theappsubsection.~#1}%
  \addcontentsline{app}{subsection}{\theappsubsection.~#1}%
}

% ==================================================
%   Web Appendices (WA.1, WA.2, WA.1.1, etc.)
% ==================================================
\newcounter{webappsection}
\newcounter{webappsubsection}[webappsection]

\renewcommand{\thewebappsection}{WA.\arabic{webappsection}}
\renewcommand{\thewebappsubsection}{WA.\arabic{webappsection}.\arabic{webappsubsection}}

\titleformat{\section}{\large\bfseries}{\thewebappsection}{1em}{}
\titleformat{\subsection}{\normalsize\bfseries}{\thewebappsubsection}{1em}{}

\newcommand{\webappendicesstart}{%
  \clearpage
  \setcounter{page}{1}% restart counting
  \renewcommand{\thepage}{WA-\arabic{page}}% <-- show pages as WA-1, WA-2, ...
}

\newcommand{\webappsection}[1]{%
  \refstepcounter{webappsection}%
  \section*{Web Appendix~\thewebappsection.~#1}%
  \addcontentsline{app}{section}{Web Appendix~\thewebappsection.~#1}%
}

\newcommand{\webappsubsection}[1]{%
  \refstepcounter{webappsubsection}%
  \subsection*{\thewebappsubsection.~#1}%
  \addcontentsline{app}{subsection}{\thewebappsubsection.~#1}%
}

% ==================================================
%   List of Appendices
% ==================================================
% \makeatletter
% \newcommand{\listofappendices}{%
%   \section*{\listappendicesname}%
%   \addcontentsline{toc}{section}{\listappendicesname}%
%   \begin{center}
%     \begin{minipage}{0.8\textwidth}
%       \small
%       \@starttoc{app}
%     \end{minipage}
%   \end{center}
% }
\makeatother
% \clearpage
\appendix
\renewcommand{\thetable}{A\arabic{table}} % changes table numbering to A1, A2, ...
\setcounter{table}{0}  % reset table counter

\renewcommand{\thefigure}{A\arabic{figure}} % changes figure numbering to A1, A2, ...
\setcounter{figure}{0}  % reset figure counter

\renewcommand\thesection{A.\arabic{section}}

\section*{Appendix}
% \listofappendices  % Prints the list

% \clearpage

\appsection{ Details of Pricing and Bidding Model}
\label{subsec:appendix_model_detail}
For simplicity, we drop the time superscript in the one-shot game. 
The expected demand for seller $i$ is:
$D_i(\boldsymbol{p}, \boldsymbol{b}) 
=\theta\cdot  \Pr\{\tilde{b_i}>\tilde{b_j}\}\cdot s_i(p_i) + (1-\theta)\cdot s_i(p_i, p_j)
$.

To model the uncertainty in auction outcomes, we assume that when a seller submits a bid $b_i$, the realized bid $\tilde{b}_i$ is stochastic $\tilde{b}_i=\omega_i b_i$, and follows a log-normal distribution: $\log \left(\tilde{b}_i\right) \stackrel{\text { i.i.d. }}{\sim} \mathcal{N}\left(\log \left(b_i\right), \sigma^2\right)$. Hence, when sellers $i$ and $j$ submit bids $b_i$ and $b_j$, respectively, the realized bid of seller $i$, $\tilde{b_i}$,  is higher than that of seller $j$ with probability $\Pr\{\tilde{b_i}>\tilde{b_j}\}=\Pr\{\log(b_i) >\log(b_j)\}=\Phi\left(\frac{\log\left(\frac{b_i}{b_j}\right)}{\sqrt{2}\sigma}\right)$.

% Thus, the expected demand becomes
% $D_i(\boldsymbol{p}, \boldsymbol{b}) = \theta\cdot \Phi\left(\frac{\log\left(\frac{b_i}{b_j}\right)}{\sqrt{2}\sigma}\right)\cdot s_i(p_i) + (1-\theta)\cdot s_i(p_i, p_j)$.

Next, we derive the expected Cost Per Click (CPC) that seller $i$ needs to pay, given that he wins the auction, $\mathrm{E}[\tilde{b_i}|\tilde{b_i}>\tilde{b_j}]$. Let $u=\frac{\log \left(\omega_i\right)}{\sigma}   \sim \mathcal{N}(0,1)$ and $v=\frac{\log \left(\omega_j\right)}{\sigma}   \sim \mathcal{N}(0,1)$.
\begin{align}\nonumber
\mathrm{E} \left[\tilde{b_i}|\tilde{b_i}>\tilde{b_j}\right] & =\frac{1}{\Pr\{\tilde{b_i}>\tilde{b_j}\}} \int_{\frac{\log \left(b_i\right)} { \sigma}+u>\frac{\log \left(b_j\right) }{\sigma}+v} b_i \omega_i \phi(u) \phi(v) \mathrm{d} u \mathrm{d} v \\ \nonumber 
& =\frac{1}{\Pr\{\tilde{b_i}>\tilde{b_j}\}} \int_{\frac{\log \left(b_i\right)} { \sigma}+u>\frac{\log \left(b_j\right) }{\sigma}+v} b_i \exp (\sigma u) \phi(u) \phi(v) \mathrm{d} u \mathrm{d} v \\ \nonumber 
& =\frac{b_i}{\Pr\{\tilde{b_i}>\tilde{b_j}\}} \int_{-\infty}^{+\infty} \exp (\sigma u) \phi(u) \mathrm{d} u \int_{-\infty}^{\frac{\log \left(b_i\right)} { \sigma}+u-\frac{\log \left(b_j\right) }{\sigma}} \phi(v) \mathrm{d} v \\
& =\frac{b_i}{\Pr\{\tilde{b_i}>\tilde{b_j}\}} \int_{-\infty}^{+\infty}  \Phi\left(\frac{\log \left(\frac{b_i}{b_j} \right) }{\sigma} +u\right) \exp (\sigma u) \phi(u) \mathrm{d} u
\end{align}

Resulting in the expected seller profit: 
\begin{align}\nonumber
\pi_i(\boldsymbol{p}, \boldsymbol{b})=%&\theta \cdot  \Phi\left(\frac{\log\left(\frac{b_i}{b_j}\right)}{\sqrt{2}\sigma}\right)\left((1-\tau)\cdot p_i-c_i-\gamma_i\cdot  \mathrm{E} \left[\tilde{b_i}|\tilde{b_j}<\tilde{b_i}\right]\right)\cdot  s_i(p_i)+ (1-\theta)\cdot \left((1-\tau)\cdot p_i-c_i\right) \cdot s_i(p_i, p_j)\\\nonumber
%= \, 
&\theta \cdot  \Phi\left(\frac{\log\left(\frac{b_i}{b_j}\right)}{\sqrt{2}\sigma}\right)\cdot s_i(p_i)\left((1-\tau)\cdot p_i-c_i-\tilde{c_i}\right) + (1-\theta)\cdot((1-\tau)\cdot p_i-c_i) \cdot s_i(p_i, p_j)
\end{align}
where ${\tilde{c_i}}= {\gamma_i\cdot \frac{b_i}{\Phi\left(\frac{\log\left(\frac{b_i}{b_j}\right)}{\sqrt{2}\sigma}\right)} \int_{-\infty}^{+\infty} \Phi\left(\frac{\log \left(\frac{b_i}{b_j} \right) }{\sigma} +u\right) \exp (\sigma u) \phi(u) \mathrm{d} u}$.

% \clearpage
\appsection{Proof of Lemma \ref{obs:competitive_price}: Monotonicity and Comparison of Competitive Price and Benchmark Price }
\label{sec:appendix_competitive_price}

\appsubsection{Monotonicity of Equilibrium Price Under Price Only Competition}

We show that equilibrium prices under price only competition $p^{oN}$ increase with $\theta$ using the implicit function theorem.
The sellers' profit is: 
\[
\pi_i(p_i,p_j;\theta)=
\Big(\tfrac{\theta}{2}\,s_i(p_i)+(1-\theta)\,s_i(p_i,p_j)\Big)\,\Big((1-\tau)p_i-c\Big)
\equiv q(\theta)\,m,
\]
where we denote $m=(1-\tau)p_i-c$, and $s_i(p_i)=\frac{\exp\!\left(\frac{a_i-p_i}{\mu}\right)}{1+\exp\!\left(\frac{a_i-p_i}{\mu}\right)},
s_i(p_i,p_j)=\frac{\exp\!\left(\frac{a_i-p_i}{\mu}\right)}{1+\exp\!\left(\frac{a_i-p_i}{\mu}\right)+\exp\!\left(\frac{a_j-p_j}{\mu}\right)}$, $ q(\theta)=\tfrac{\theta}{2}s_i(p_i)+(1-\theta)s_i(p_i,p_j)$.

We first compute the first order condition. Assume $a_i=a_j=a$, $c_i=c_j=c$, we have $p_i=p_j=p$ in 
a symmetric equilibrium. Define $\mathrm{exp}\;\equiv\;\exp\!\left(\frac{a-p}{\mu}\right)>0,
a_1:=1+\mathrm{exp},\;\; b_1:=1+2\,\mathrm{exp}.
$
Then $s_i(p_i)=\frac{\mathrm{exp}}{a_1}, s_i(p_i,p_j)=\frac{\mathrm{exp}}{b_1}$. 

The derivatives of these expressions with respect to $p_i$ are $\frac{d\,s_i(p_i)}{dp_i}
=\frac{d}{dp}\!\left(\frac{\mathrm{exp}}{1+\mathrm{exp}}\right)
=-\frac{\mathrm{exp}}{\mu(1+\mathrm{exp})^2}
=-\frac{\mathrm{exp}}{\mu a_1^2}$, $\frac{\partial s_i(p_i,p_j)}{\partial p_i}|_{p_i=p_j=p}
=-\frac{\mathrm{exp}\,a_1}{\mu b_1^2}$, $
q_p(\theta)=\tfrac{\theta}{2}\frac{d s_i(p_i)}{dp}+(1-\theta)\frac{d s_i(p_i,p_j)}{dp}\ (<0)$.

\paragraph{First–order condition (FOC)}
Combining the above expressions yields:
\[
G(p,\theta):=\frac{\partial \pi_i}{\partial p_i}=q_p(\theta)\,m+q(\theta)(1-\tau)=0.
\]
Providing the equilibrium equation:
\begin{equation}
m=-\frac{q(\theta)}{q_p(\theta)}(1-\tau).
\label{eq:m_from_FOC}
\end{equation}

\paragraph{Comparative static with respect to $\theta$:}
$
G_\theta
=\Big(\tfrac12 s_i'(p)-s_i'(p,p)\Big)\,m
+\Big(\tfrac12 s_i(p)-s_i(p,p)\Big)(1-\tau).
$
Plugging in \eqref{eq:m_from_FOC}, we get
$
\frac{G_\theta}{1-\tau}
=\Big(\tfrac12 s_i(p)-s_i(p,p)\Big)
-\Big(\tfrac12 s_i'(p)-s_i'(p,p)\Big)\,
\frac{q(\theta)}{q_p(\theta)}.
$
We then simplify:
$$G_\theta
=(1-\tau)\,
\frac{\mathrm{exp}^{\,3}}{\ \theta(1+2\,\mathrm{exp})^2+2(1-\theta)(1+\mathrm{exp})^3\ }
=(1-\tau)\,
\frac{\mathrm{exp}^{\,3}}{\ 2(1+\mathrm{exp})^3-\theta\,(2\,\mathrm{exp}^3+2\,\mathrm{exp}^2+2\,\mathrm{exp}+1)\ }>0
$$
For all $\mathrm{exp}>0$ and $\theta\in[0,1]$, the denominator is strictly positive
(e.g.\ it is minimized at $\theta=1$ and equals $4\mathrm{exp}^2+4\mathrm{exp}+1>0$).
Hence $G_\theta>0$.

By the Implicit Function Theorem
$\frac{dp^*}{d\theta}=-\frac{G_\theta}{G_p}$. Since $G_\theta>0$ we only focus on the sign of $G_p=\partial^2\pi_i/\partial p_i^2$ at the interior optimum.

\paragraph{Second–order condition (SOC): $G_p$ at the optimum}

\[
\pi_i'(p)=q_p\,m+q(1-\tau),\qquad
\pi_i''(p)=q_{pp}\,m+2(1-\tau)\,q_p.
\]
At the interior optimum the FOC holds: $q_p\,m+q(1-\tau)=0$, and we can eliminate $m$ in $\pi_i''$,
\[
\pi_i''(p)
= (1-\tau)\Big(2q_p-\frac{q\,q_{pp}}{q_p}\Big)
=(1-\tau)\,\frac{2q_p^2-q\,q_{pp}}{q_p}.
\tag{$\dagger$}
\]

We compute
$
2q_p^2-q\,q_{pp}
=\frac{\mathrm{exp}^{\,2}}{4\,\mu^{2}\,a_1^{3}\,b_1^{3}}
\Big[
(4\mathrm{exp}^{4}+2\mathrm{exp}^{3}+2\mathrm{exp}^{2}+2\mathrm{exp}+1)\,\theta^{2}
-(8\mathrm{exp}^{4}+10\mathrm{exp}^{3}+14\mathrm{exp}^{2}+12\mathrm{exp}+4)\,\theta
+4(1+\mathrm{exp})^{4}
\Big]\;>0
\text{ for all }p\text{ (all }\mathrm{exp}>0\text{) and }\theta\in[0,1],
$

Together with $q_p<0$ this implies by \((\dagger)\) that $\,G_p<0\ \text{ at }p^*$, therefore,
$
\frac{dp^*}{d\theta}=-\frac{G_\theta}{G_p}>0
$.

\appsubsection{Price with pricing and bidding are higher than price only competition}

We show that $p_i^N \ge p_i^{oN}$. In a symmetric equilibrium $b_i = b_j$, yielding the sellers' profit:
$$
\pi_i(p_i,p_j;b_i, b_j)|_{b_i = b_j}=
\tfrac{\theta}{2}\,s_i(p_i)\,\Big((1-\tau)p_i-c - \tilde{c_i} \Big) + (1-\theta)\,s_i(p_i,p_j)\,\Big((1-\tau)p_i-c\Big)
$$
where $\tilde{c_i}=\gamma_i\cdot \frac{b_i}{\Phi\left(\frac{\log\left(\frac{b_i}{b_j}\right)}{\sqrt{2}\sigma}\right)} \int_{-\infty}^{+\infty} \Phi\left(\frac{\log \left(\frac{b_i}{b_j} \right) }{\sigma} +u\right) \exp (\sigma u) \phi(u) \mathrm{d} u\ge0$. Define $c_H = c +\tilde{c_i}$. 

When $c_H = c$, the setting is equivalent to the price only competition case. Thus, it suffices to show that the equilibrium prices increase with $c_H$.

We compute the first order condition (FOC):
{\small
\[
\begin{aligned}
G(p;\theta,c_H)
&= \frac{\partial \pi_i}{\partial p_i} = \tfrac{\theta}{2}s_i'(p)\big((1-\tau)p-c_H\big)
+ (1-\tau)\tfrac{\theta}{2}s_i(p) + (1-\theta)s_i'(p_i,p_j) m  + (1-\tau)(1-\theta)s_i(p_i,p_j)= 0.
\end{aligned}
\tag{FOC}
\]
}

The Implicit Function Theorem implies $
\frac{dp^*}{dc_H} = -\frac{G_{c_H}}{G_p}$.
With  
$G_{c_H}
=\frac{\partial G}{\partial c_H}
= -\tfrac{\theta}{2}s_i'(p) > 0$ and with

\[
G_p
= \tfrac{\theta}{2}s_i''(p_i)\big((1-\tau)p-c_H\big)
+ (1-\theta) s_i''(p_i, p_j)\big((1-\tau)p-c\big)
+ 2(1-\tau)\Big(\tfrac{\theta}{2}s_i'(p_i)+(1-\theta) s_i'(p_i, p_j)\Big)  < 0.
\]
we can combine the signs to receive:
$
\frac{dp^*}{dc_H} = -\frac{G_{c_H}}{G_p}
= -\frac{(+)}{(-)} > 0.
$

\appsection{Proof of Lemma \ref{obs:monopoly_price}: Monotonicity of  Collusive Prices}
\label{sec:appendix_theoretical_collusion}
We show that the equilibrium collusive prices decrease with $\theta$. The approach is similar to that of Lemma \ref{obs:competitive_price} and we relegate the details to Web Appendix \ref{webapp:lemma-monopoly}. The analysis using the implicit function theorem shows that because $G_\theta<0$ for all $\theta\in[0,1]$, and
because $G_p<0$ at the interior optimum, then 
$\frac{dp^*}{d\theta}=-\frac{G_\theta}{G_p}<0$
.

\appsection{Proof of Proposition \ref{prop_crossing}}
\label{appsec:prop_proof}

When $\theta = 1$, by Lemma \ref{obs:competitive_price}, $\mathbf{p}^{oN}$ is lower than $\mathbf{p}^{N}$ because ad costs increase prices $\mathbf{p}^{N} \ge \mathbf{p}^{oN}$, and by Lemma \ref{obs:limiting_case}, $\mathbf{p}^{oN} = \mathbf{p}^{M}$. Hence, \(\mathbf{p}^{N} > \mathbf{p}^{oN} = \mathbf{p}^{M}\).

When $\theta=0$, the collusive prices $\mathbf{p}^{M}$ are higher than the full competition prices $\mathbf{p}^{N}$ because competition leads to lower prices. The ads do not affect demand and cost as $\theta=0$, $\mathbf{p}^{oN} = \mathbf{p}^{N}$. So, \(\mathbf{p}^{M} > \mathbf{p}^{oN} = \mathbf{p}^{N}\). 

By Lemma \ref{obs:monopoly_price}, $\mathbf{p}^{M}$ decreases with $\theta$. Therefore, there always exists a point at which the collusive price $\mathbf{p}^{M}$ and the competitive price $\mathbf{p}^{N}$ intersect.

\appsection{Robustness to additional bidding information, differentiated products and multiple sellers}
\label{sec:appendix_robustness}

\begin{figure}[ht]
\caption{Complete State vs Partial Bid Information}
\label{fig:stateful_results} 
\begin{subfigure}{0.33\textwidth}
\caption{Sellers' Profit}
\centering
\includegraphics[width=\textwidth]{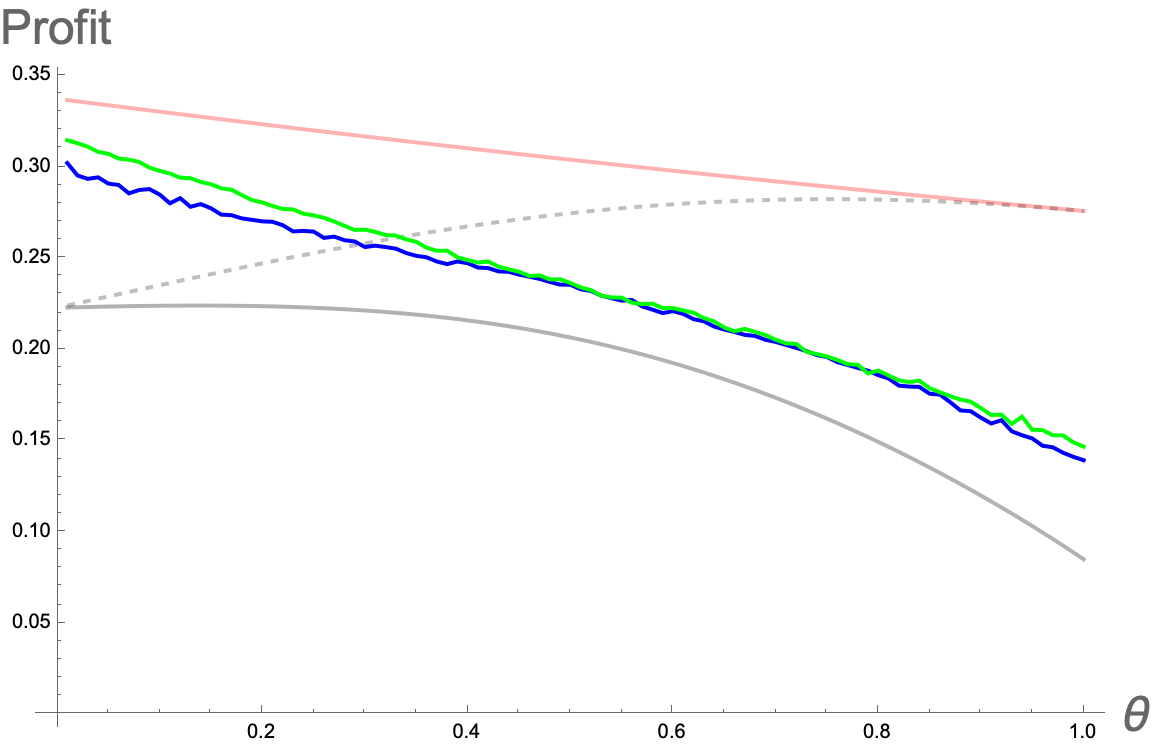}
\end{subfigure}~
\begin{subfigure}{0.33\textwidth}
\caption{Price}
\centering
\includegraphics[width=\textwidth]{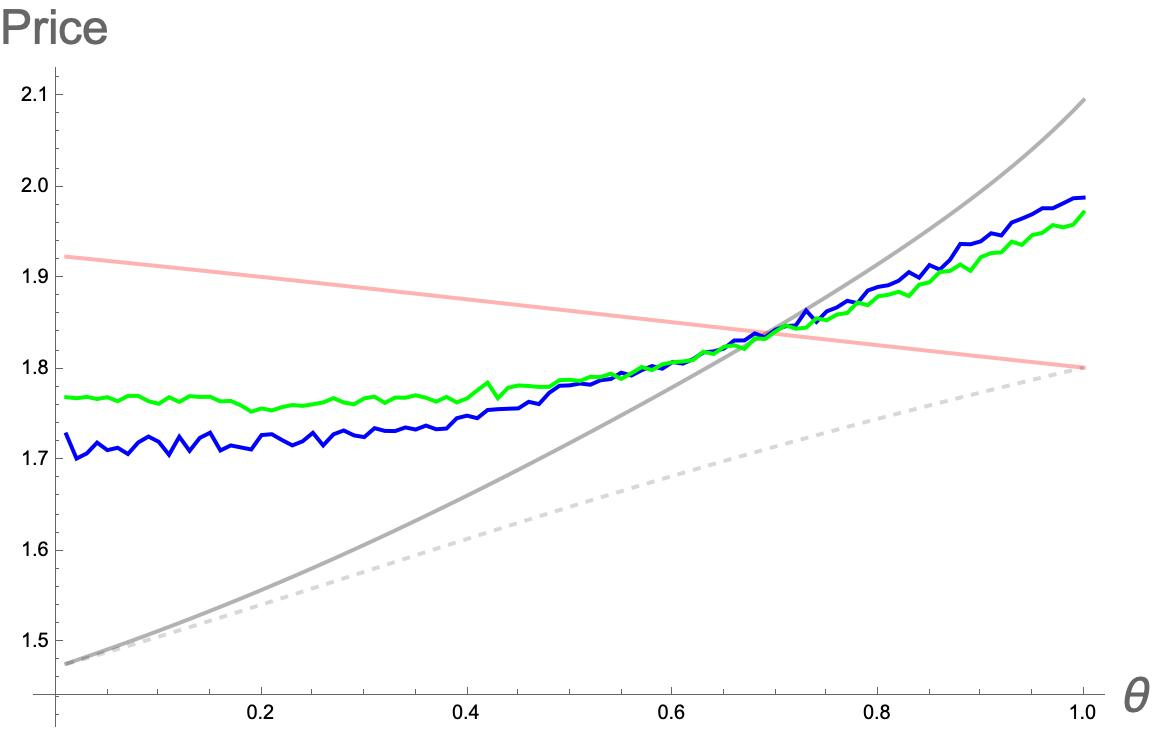}
\end{subfigure}
\begin{subfigure}{0.32\textwidth}
\caption{Bid}
\centering
\includegraphics[width=\textwidth]{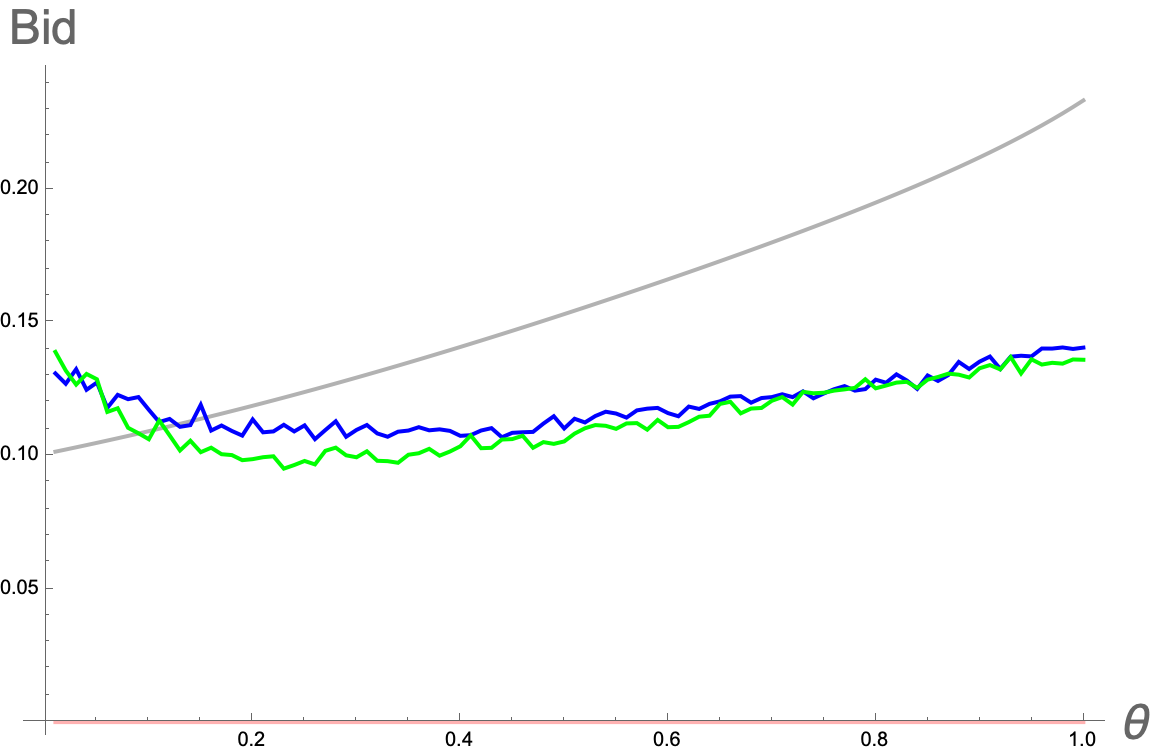}
\end{subfigure}
    Panels (a), (b), and (c) show the sellers' profit, price, and bid, respectively. The blue solid lines represent the benchmark scenario, while the green lines represent the \textit{complete state} scenario. Solid red indicates the fully collusive case, solid black the fully competitive, and dashed is fully competitive without advertising.
\end{figure}

\appsubsection{Alternative Bid Information in State Space}
\label{subsubsection:alternative_bid}
We consider the \textit{complete state} scenario where the state space is \( s_{it} = (p_{it-1}, p_{jt-1}, b_{it-1}, b_{jt-1} ) \) and sellers know competitor bids. Figure \ref{fig:stateful_results} compares the  \textit{complete state} scenario to our previous scenario. The \textit{complete state} condition yields outcomes closer to the fully collusive case, but still similar to what we found before. %That is, it generates higher profits for the sellers, bids closer to zero, and higher prices for small values of $\theta$, and lower prices otherwise, compared with the benchmark state space.

%We find that the outcomes facilitated by the algorithms, which are beneficial to both sellers and consumers, are robust to assumptions regarding the bid information available in the algorithm's state space. 

\begin{figure}[!ht]
\caption{Heterogeneous Sellers With Differentiated Quality}
\begin{center}
\begin{subfigure}{0.45\textwidth}
\caption{Sellers' Profit}
\centering
\includegraphics[width=\textwidth]{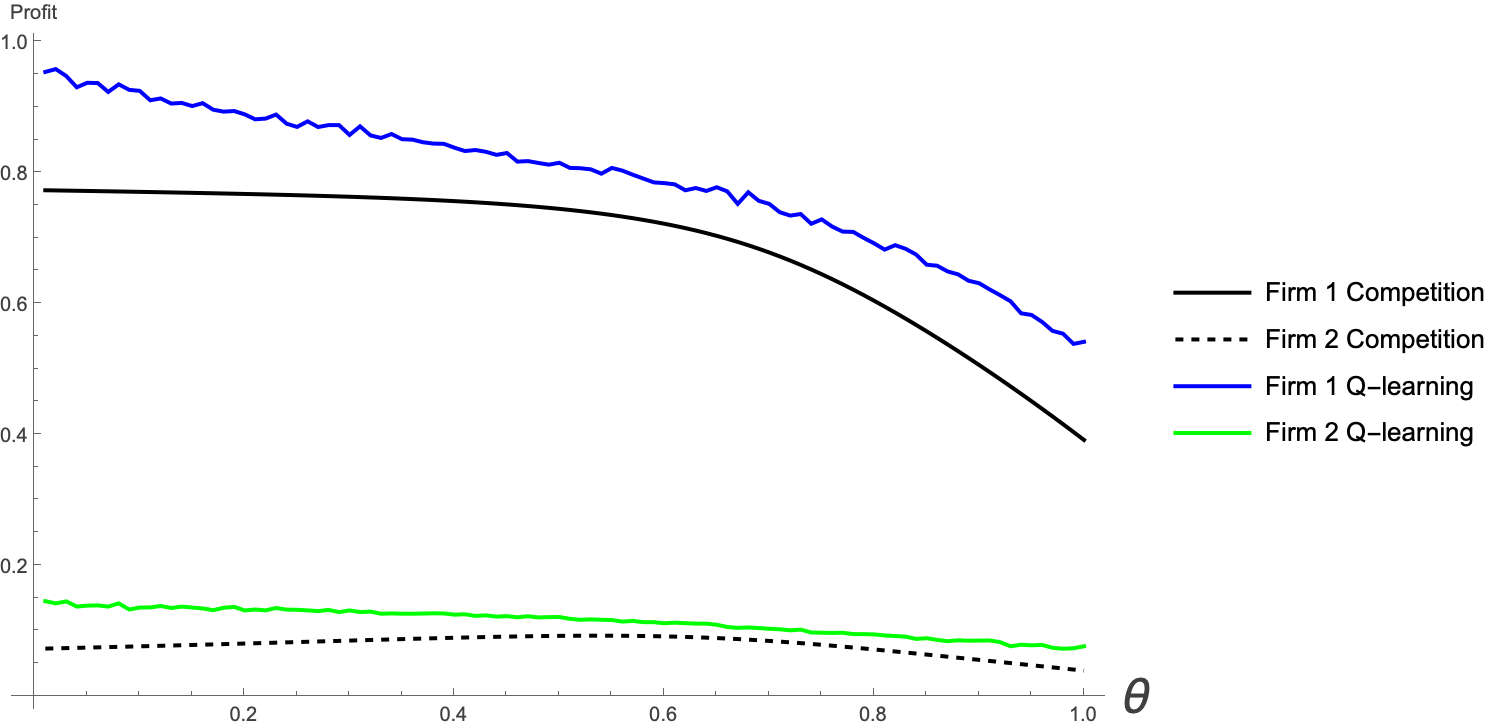}
\end{subfigure}~
\begin{subfigure}{0.45\textwidth}
\caption{Consumer Surplus}
\centering
\includegraphics[width=\textwidth]{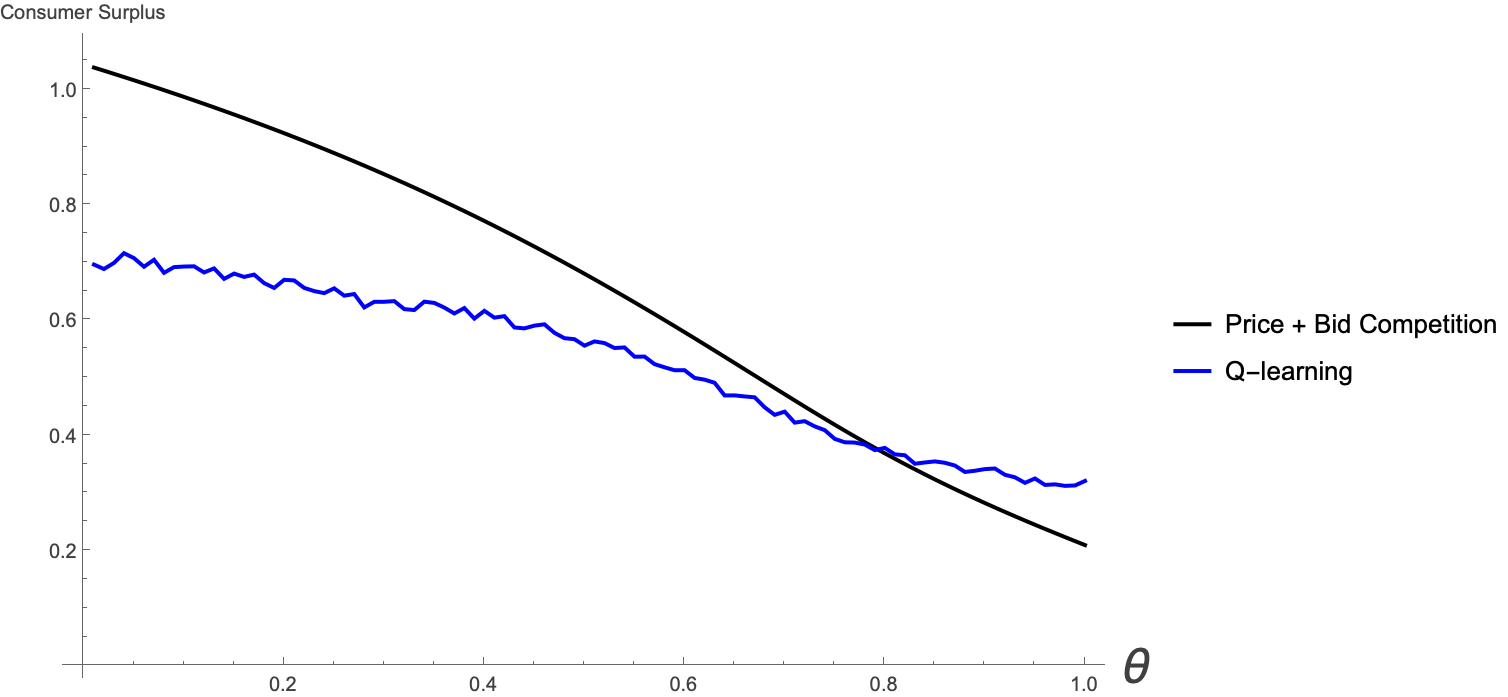}
\end{subfigure}~
\label{fig:heter_seller}
\end{center}
Panels (a) and (b) show the Q-learning and full competition  sellers' profits and consumer surplus as functions of $\theta$. We set 
$a_i=3$ and $a_j=2$, and keep other parameters the same as in Figure \ref{fig:theoretical-results}. In panel (a), the black line and the blue line represent seller 1's profits in the full competition and Q-learning cases, respectively; the dashed black line and the green line represent seller 2's profits in the full competition and Q-learning cases, respectively. In panel (b), the solid black line represents the consumer surplus in the full competition case, while the solid blue line denotes the Q-learning consumer surplus from our simulation experiments.\end{figure}

\appsubsection{Asymmetric Sellers}
\label{subsubsection:asymmetric_sellers}
We present an analysis of full competition and algorithmic pricing for sellers with differentiated quality. The sponsored ranking is determined by the bids of the two sellers using the same auction as before. Organic listings are ranked by the platform’s recommendation system, and the order depends probabilistically on their quality. For consumers , we assume that a fraction $\theta$ of them consider only the first listing, while the remaining consumers consider all listings. Their choice of products depends on the differentiated qualities. The seller's profit is:
{\small
\begin{align*}
\pi_i(\boldsymbol{p}, \boldsymbol{b}) =& \Pr\{\tilde{b_i}> \tilde{b_j}\}\cdot \left[\theta\cdot s_i(p_i,a_i)\left((1-\tau)p_i-c_i -\gamma_i \mathrm{E} \left[\tilde{b_i}|\tilde{b_i}>\tilde{b_j}\right]\right) \right]+ (1-\theta)\cdot s_i(p_i, p_{j}, a_i,a_j) ((1-\tau)p_i-c_i)
\end{align*}
% \begin{align*}
% \pi_i(\boldsymbol{p}, \boldsymbol{b}) =& \Pr\{\tilde{b_i}> \tilde{b_j}\}\cdot \left[\theta\cdot s_i(p_i)\left((1-\tau)p_i-c_i -\gamma_i \mathrm{E} \left[\tilde{b_i}|\tilde{b_i}>\tilde{b_j}\right]\right) + \left(\frac{a_i}{a_i+a_j}\left(\theta(1-\theta)\cdot s_i(p_i)+(1-\theta)^2\cdot s_i(p_i, p_{j})\right)\right.\right.\\
% &+\left.\left.\frac{a_j}{a_i+a_j}(1-\theta)\cdot s_i(p_i, p_{j})\right) ((1-\tau)p_i-c_i)\right]\\
% &+\Pr\{\tilde{b_i}< \tilde{b_j}\}\cdot \left[ \left(\frac{a_i}{a_i+a_j}(1-\theta)\cdot s_i(p_i, p_{j})+\frac{a_j}{a_i+a_j}(1-\theta)^2\cdot s_i(p_i, p_{j})\right) ((1-\tau)p_i-c_i)\right]
% \end{align*}
}
Figure \ref{fig:heter_seller} shows the sellers' profit and consumer surplus. Similar to our previous results, algorithms benefit both sellers. When consumer search costs are high, algorithmic pricing can benefit consumers, generating a higher consumer surplus.

\appsubsection{Multiple Sellers}
\label{subsubsection:multiple_sellers}
We now consider the scenario where there $n>2$ sellers competing for the sponsored position. In the price and ads competition case, when sellers are homogeneous, a sellers' profit is:
\begin{align*}
&\pi_i(\boldsymbol{p}, \boldsymbol{b}) \\
=&\theta\cdot  \Pr\{\tilde{b_i}> \max_{j\neq i}\{\tilde{b_j}\}\}\cdot s_i(p_i)\left((1-\tau)p_i-c_i -\gamma_i \mathrm{E} \left[\tilde{b_i}|\tilde{b_i}>\max_{j\neq i} \tilde{b_j}\right]\right) + (1-\theta)\cdot s_i(p_i, \boldsymbol{p_{j}}) ((1-\tau)p_i-c_i)
\end{align*}

And in the price only competition case, the sellers' profit is 
\begin{align*}
&\pi_i(\boldsymbol{p}) =\frac{\theta}{n}\cdot s_i(p_i)\left((1-\tau)p_i-c_i\right) + (1-\theta)\cdot s_i(p_i, \boldsymbol{p_{j}}) ((1-\tau)p_i-c_i)
\end{align*}

\begin{figure}[!ht]
    \caption{Multiple sellers}
    \label{fig:multi_sellers}
\begin{center}
    \includegraphics[width=0.75\textwidth]{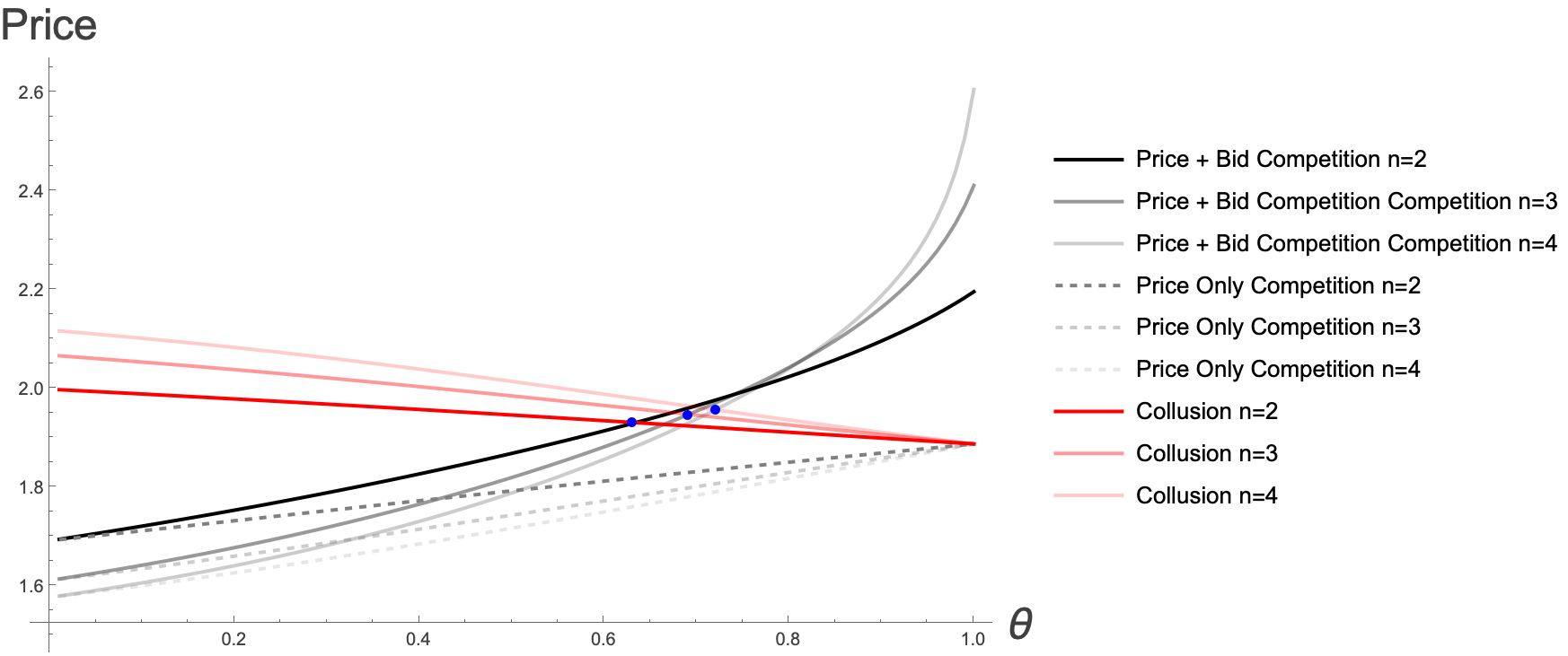}
\end{center}    
Equilibrium prices as a function of $\theta$ when there are multiple sellers. The dashed gray, black solid and red solid line represent the benchmark equilibrium price, the Nash-Bertrand equilibrium price and collusive price, respectively. Seller-related and ad-related parameters are set the same as in our default specifications, as in Figure \ref{fig:theoretical-results}.
\end{figure}

Figure \ref{fig:multi_sellers} shows the equilibrium prices under full competition, price-only competition, and monopoly for the cases where there are more than two sellers. In the multiple-seller case, the following results still hold:Lemma \ref{obs:competitive_price} (ads increase the equilibrium prices), Lemma \ref{obs:monopoly_price} (collusive prices decrease with $\theta$), Lemma \ref{obs:limiting_case} (limiting case, when $\theta = 1$, the collusive prices and the equilibrium prices under price-only competition are the same across different numbers of sellers, since consumers only consider the top product when $\theta = 1$), and Proposition \ref{prop_crossing}. Thus, even with more than two sellers, the collusive price can still be lower than the competitive prices, and we can also expect the algorithmic prices to fall below the competitive prices.

\appsubsection{Separate learning algorithms for pricing and bidding}
\label{subsubsection:separate_algos}

\begin{figure}[!ht]
\caption{Separate learning algorithms for pricing and bidding}
\begin{center}
\begin{subfigure}{0.49\textwidth}
\caption{Price}
\centering
\includegraphics[width=\textwidth]{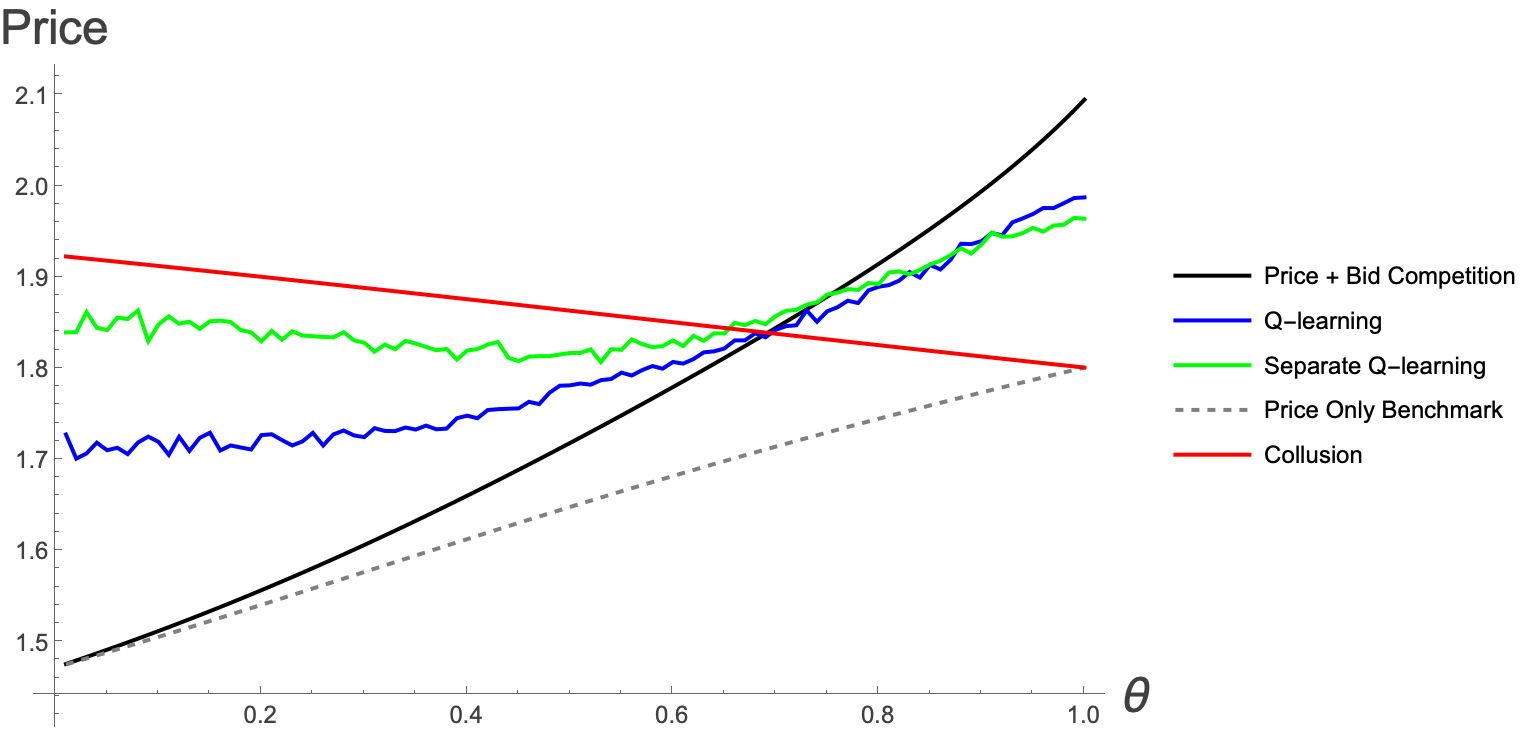}
\end{subfigure}~
\begin{subfigure}{0.49\textwidth}
\caption{Bid}
\centering
\includegraphics[width=\textwidth]{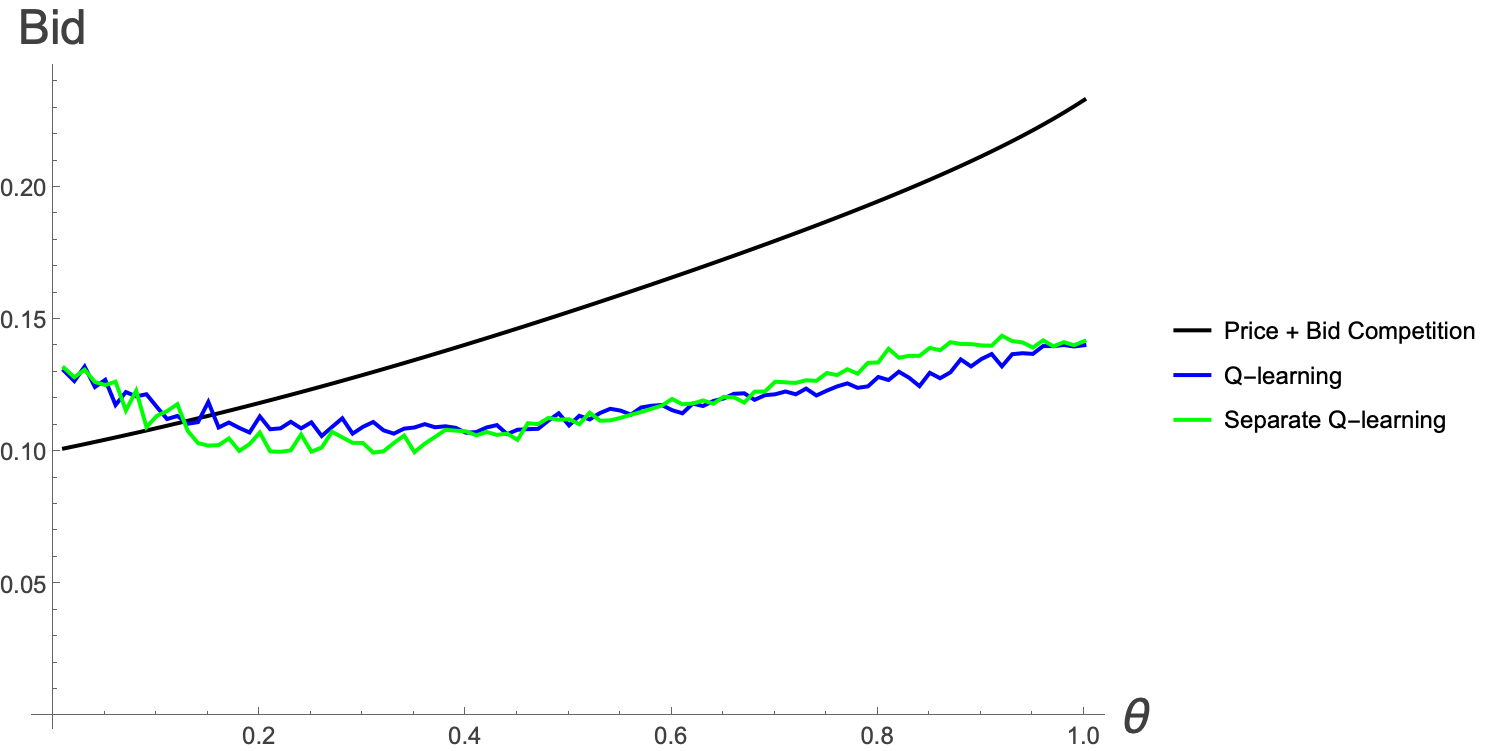}
\end{subfigure}
\label{fig:separate_q}
\end{center}
Panels (a) and (b) show the Q-learning and full competition prices and bids as functions of $\theta$. The black, blue, green, and red solid lines represent the equilibrium prices in the full competition, single Q-learning (for both pricing and bidding), separate Q-learning (for pricing and bidding), and collusive cases, respectively.
\end{figure}

We consider the scenario where sellers use two Q-learning algorithms separately for pricing and bidding. The pricing algorithm for seller $i$ learns to set the price $p_{it}$ with a state space of $(p_{it-1}, p_{jt-1})$, while the bidding algorithm learns to set $b_{it}$ with a state space of $(b_{it-1})$. The single-period profit remains the same as before, defined in Equation \eqref{equ:seller_profit}. The learning and updating procedures are the same as before, with the main idea outlined in Equation \eqref{equ:Q_learning}.  Figure \ref{fig:separate_q} presents the equilibrium pricing and bidding results and shows that algorithmic prices still yield lower prices. 

\appsection{Platform Strategic Response}
\appsubsection{Adjusting Commission Rate}
\label{subsec:commission_rate}

% The total impact of sellers using algorithms on the platform's own profit is unclear because advertising revenue might be lower than in competitive scenarios with sellers submitting lower bids when consumer search costs are high. At the same time, the platform's commission revenue might be higher if Q-learning prices are lower and the demand-enlarging effect dominates. 

\begin{figure}[!ht]
\caption{Platform Strategic Response: Commission Rate}
\label{fig:commission_platform} 
\begin{center}
\begin{subfigure}{0.45\textwidth}
\caption{Platform's Profit}
\label{subfig:commission_platform_profit}
\centering
\includegraphics[width=\textwidth]{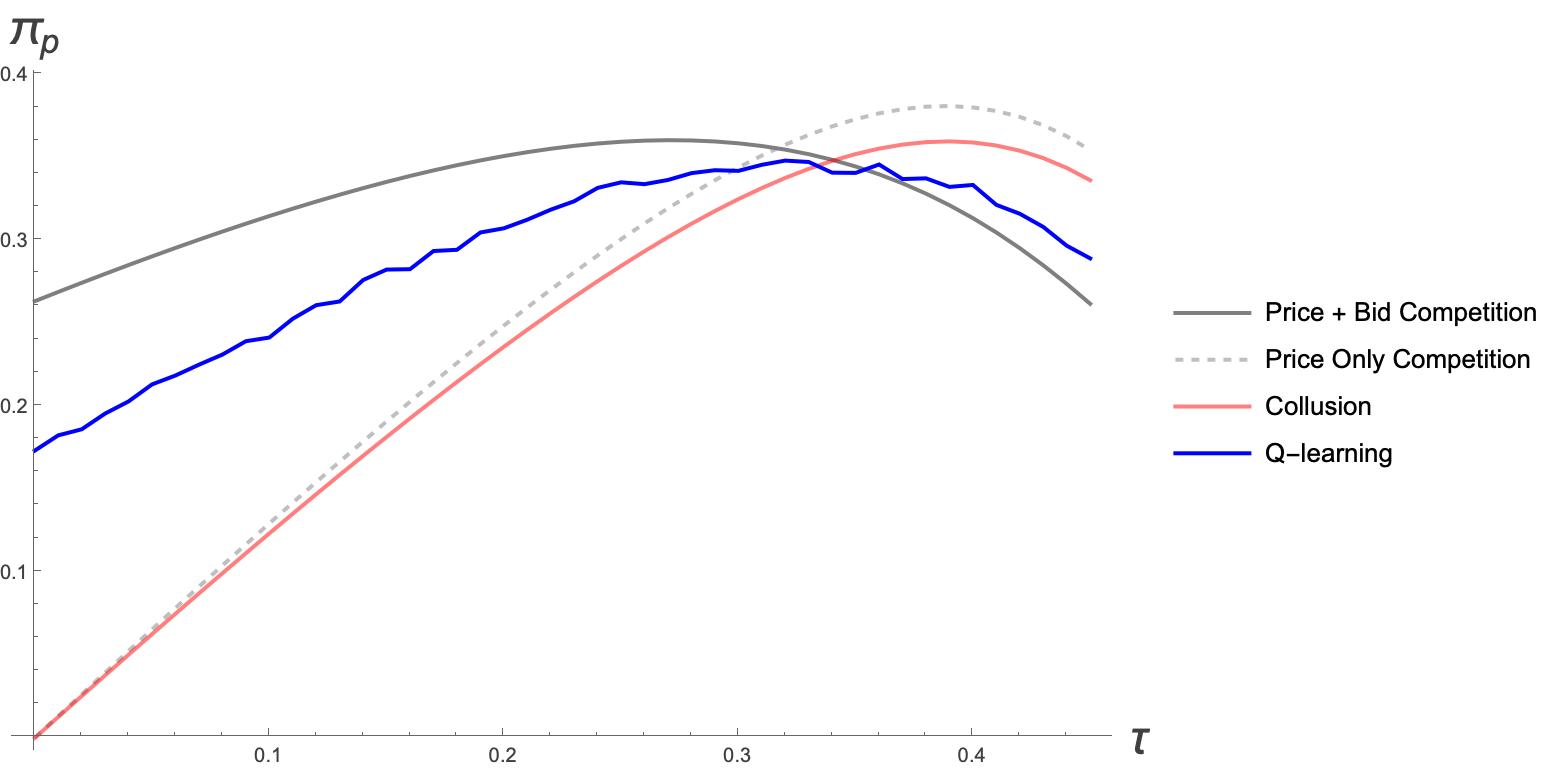}
\end{subfigure}~
\begin{subfigure}{0.45\textwidth}
\caption{Commission Revenue}
\label{subfig:commission_platform_comm}
\centering
\includegraphics[width=\textwidth]{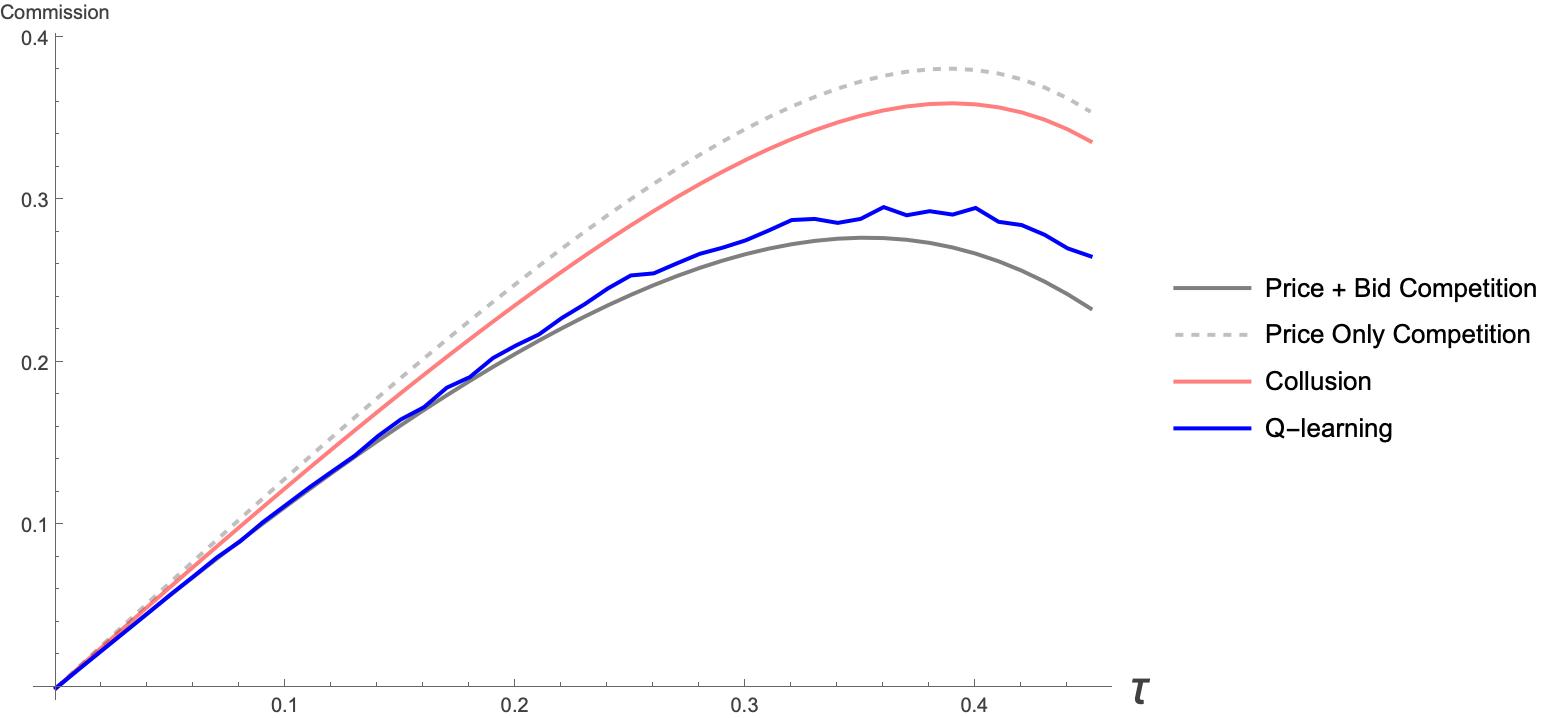}
\end{subfigure}
\end{center}
Panels (a) and (b) show the platform' profit and commission fees as a function of $\tau$ when $\theta = 0.8$. The platform's Q-learning  simulation profits and fees are in solid blue. Solid red are fully collusive, and solid black are fully competitive. Dashed are from pricing without advertising.
\end{figure}

Figure \ref{fig:commission_platform} presents results for adjusting the commission rate with high consumer search costs. Increasing the commission rate can increase the platform's profit when sellers use algorithms. Decomposing the platform's profit, \ref{fig:commission_platform}(b) shows that most of the increase is due to increased commissions revenue from increased commission rates. However, advertising revenues generally decrease as the bid coordination effect becomes more dominant (Figure \ref{fig:commission_platform_additional}(c) in the web appendix).

The platform might benefit from sellers using algorithmic pricing, thus not requiring any adjustment of commission rates. If the platform increases the commission rate, algorithms can still generate higher consumer surplus and seller profits than under full competition at the adjusted commission rate.\footnote{This is because the same consumer search cost will still fall within the region where algorithms yield beneficial outcomes for both sellers and consumers at a higher commission rate.}

\begin{figure}[!ht]
\caption{Platform Strategic Response: Reserve Price}
\label{fig:reserve_price_platform} 
\begin{center}
\begin{subfigure}{0.45\textwidth}
\caption{Platform's Profit}
\label{subfig:reserve_price_platform_profit}
\centering
\includegraphics[width=\textwidth]{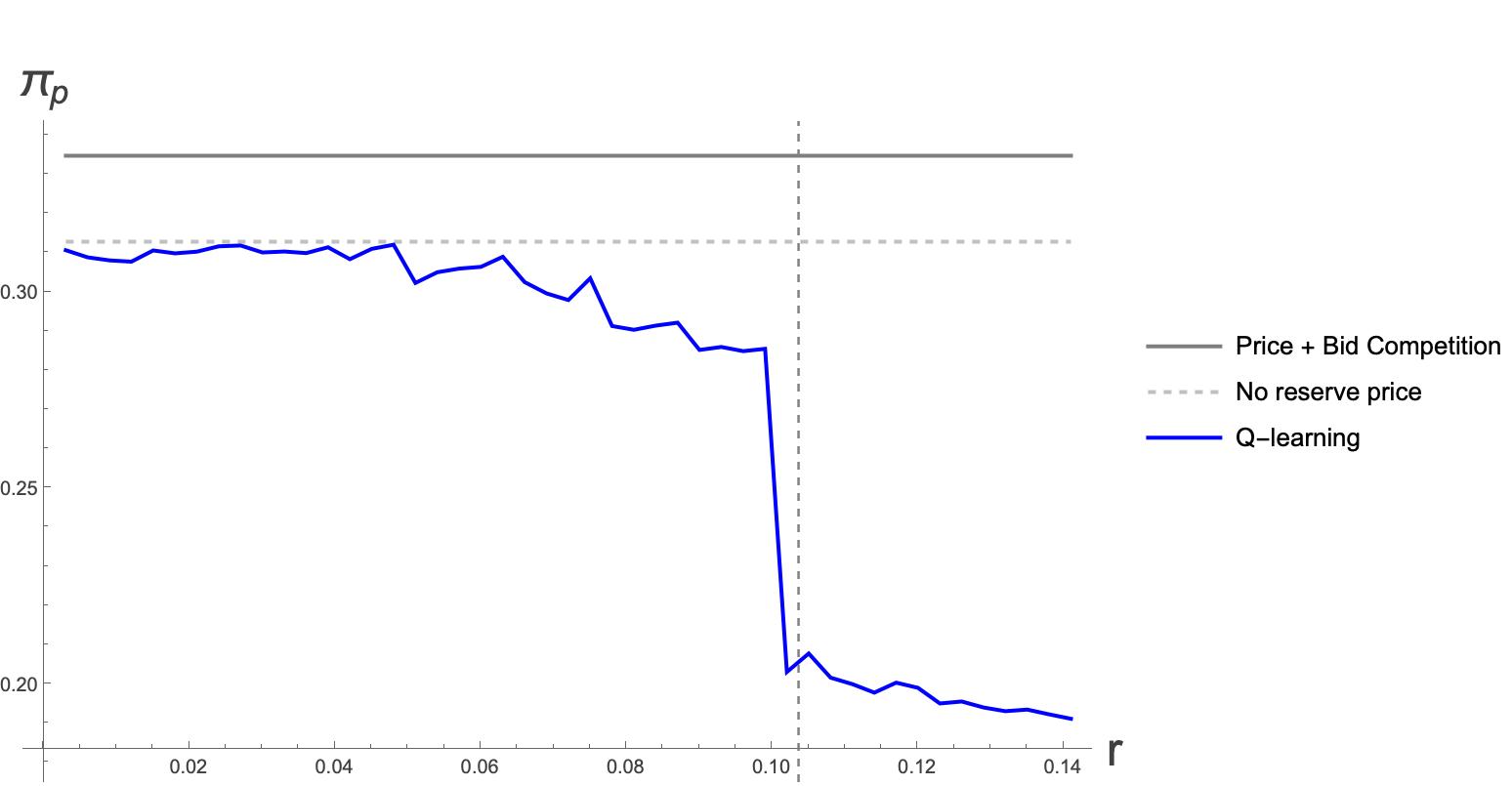}
\end{subfigure}~
\begin{subfigure}{0.45\textwidth}
\caption{Bids}
\label{subfig:reserve_price_bids}
\centering
\includegraphics[width=\textwidth]{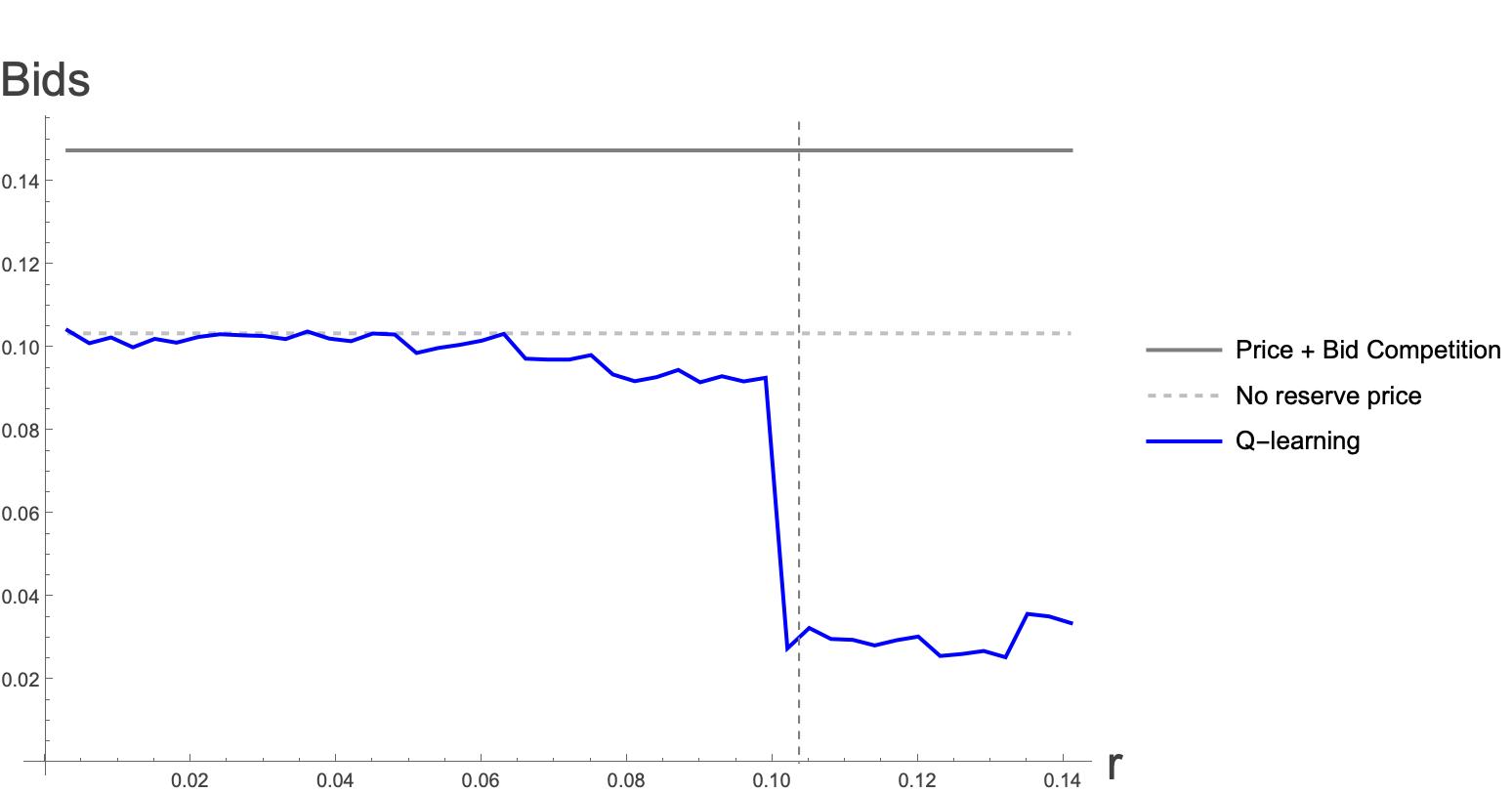}
\end{subfigure}
\end{center}
Panels (a) and (b) show the platform's profit and sellers' bids as functions of the reserve price when $\theta = 0.8$ and $\tau = 0.15$, respectively. The blue solid line denotes the Q-learning platform profits from our simulation experiments. The representation of other lines and the parameter specifications are the same as in Figure \ref{fig:theoretical-results}. The dashed vertical lines represent the Q-learning bids in the algorithmic pricing scenario.
\end{figure}

\appsubsection{Adjusting Ad Auction Reserve Price}
\label{subsec:reserve_price}
We also conduct simulation analyses where we allow the platform to gradually increase the reserve price from 0. We assume that if the bids submitted by all sellers are lower than the reserve price, the platform will need to randomly display one of the products in the top position without advertising costs for the sellers.\footnote{The sellers' profit function becomes
\begin{align}\nonumber
\pi_i^t(\boldsymbol{p^t}, \boldsymbol{b^t}) = & \theta \cdot \mathbf{1}\left(\tilde{b_i^t} \ge r\right)\cdot \operatorname{Pr}((\tilde{b_i^t} > \tilde{b_j^t})  ) \cdot s_i\left(p_i^t\right) \cdot \left(\left((1-\tau) \cdot p_i^t - c_i\right) - \gamma_i \cdot \mathbb{E}\left[\tilde{b_i^t} \mid \tilde{b_i^t} > \tilde{b_j^t}\right]\right) \\
& +\left(\frac{1}{2}\cdot \theta \cdot \mathbf{1}\left(\left(\tilde{b_i^t} <r\right)\land \left(\tilde{b_j^t} <r\right)\right) \cdot s_i\left(p_i^t\right) + (1-\theta) \cdot s_i\left(p_i^t, p_j^t\right)\right)\cdot \left((1-\tau) \cdot p_i^t - c_i\right) 
\end{align}
where $r$ denotes the platform's reserve price. 
}
Figure \ref{fig:reserve_price_platform} shows that interestingly, the reserve price might not be effective, as the platform raises the reserve price above the Q-learning equilibrium bid level, there is a large discontinuity point in the platform's revenue. When the reserve price is below the Q-learning equilibrium bid, increasing the reserve price has very little effect. When it reaches the equilibrium bid, the algorithms need to coordinate on a new equilibrium bid, and they learn to coordinate on lower bids below the new reserve price, implying that they will pay even lower advertising costs (figure \ref{fig:reserve_price_platform}(b)). %Effectively, increasing the reserve price allows the seller to reach outcomes that are fully collusive, since deviating to higher bids becomes not profitable. %This coordination further increases the sellers' profits (as shown in Figure \ref{fig:reserve_price_platform_additional}(c) in the appendix), since the sellers are still displayed in the sponsored position without any ad costs if the bids submitted by all sellers are lower than the reserve price. This cost-saving effect further increases consumer surplus, as shown in Figure \ref{fig:reserve_price_platform_additional}(d).\footnote{Although it might seem that the algorithms can profitably deviate by bidding above the reserve price to win the sponsored positions, the competitor will respond by bidding above the reserve price as well, which causes lower profits for both sellers in the long-run. Because of the experimentation property of learning algorithms, the algorithms gradually discover that coordinating on lower bids is more profitable, eventually converging to an equilibrium of lower bidding.}

% Now, for the benchmark case without ads, we assume that the platform will display the products in descending order of quality. Thus, the profits of the two firms in the benchmark case become
% \begin{align}\nonumber
% &\pi_i(\boldsymbol{p})=\left(\theta  \cdot s_i\left(p_i\right)+(1-\theta) \cdot s_i\left(p_i, p_j\right)\right) \cdot \left((1-\tau) \cdot p_i-c_i\right)\\
% &\pi_j(\boldsymbol{p})=\left((1-\theta) \cdot s_i\left(p_i, p_j\right)\right) \cdot \left((1-\tau) \cdot p_j-c_j\right)
% \end{align}

\webappendicesstart

\renewcommand\thesubsection{WA.\arabic{subsection}}
\clearpage
\setcounter{page}{1}

\section*{Web appendix}

\renewcommand{\thetable}{WA\arabic{table}} % changes table numbering to A1, A2, ...
\setcounter{table}{0}  % reset table counter

\renewcommand{\thefigure}{WA\arabic{figure}} 
\setcounter{figure}{0}  % reset table counter

%\webappsection{Example of E-commerce Software Considering Both Pricing and Bidding}

%Figure \ref{fig:ecommerce_software} below shows screenshots of e-commerce software websites that state they use AI to jointly optimize pricing and bidding decisions.

%\begin{figure}[!ht]
%    \centering
%    \caption{E-commerce Software Considering Both Pricing and Bidding}
%     \begin{subfigure}[b]{0.75\linewidth}
%         \centering
%         \caption{Feedvisor}
% \includegraphics[width=\linewidth]{Figure/Feedvisor.png}
%         \label{fig:feedvisor}
%     \end{subfigure}
%     \begin{subfigure}[b]{0.75\linewidth}
%         \centering
%         \caption{Profasee}
% \includegraphics[width=\linewidth]{Figure/Profasee.png}
%         \label{fig:profasee}
%     \end{subfigure}
%     \label{fig:ecommerce_software}
% \end{figure}

\webappsection{Sponsored Ads Examples}

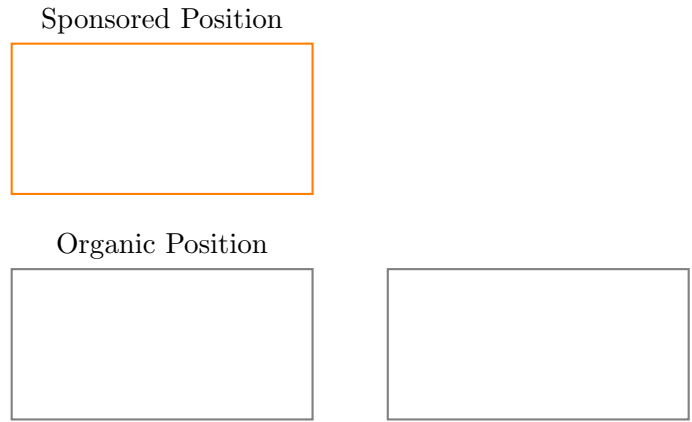
\begin{figure}[!ht]
\begin{center}
    \caption{Graphical Representation of the Theoretical Model}
    \begin{tikzpicture}
    % Draw the first rectangle with text "Sponsored" above
    \draw[orange, thick] (0,0) rectangle (4,2);
    \node[above] at (2,2) {Sponsored Position};

    % Draw the second rectangle of the same size below the first one
    \draw[gray, thick] (0,-1) rectangle (4,-3);
    \node[above] at (2,-1) {Organic Position};
    % Draw the third rectangle of the same size to the right of the second one
    \draw[gray, thick] (5,-1) rectangle (9,-3);
\end{tikzpicture}
\label{fig:Graphical_Representation_Theoretical_Model}
\end{center}    
    
    This figure provides a graphical representation of the search result page layout. There is a sponsored position on the top, followed by organic positions. 
\end{figure}

\begin{figure}[!ht]
    \caption{Example of Amazon Search Result Page With 22 Products}
    \label{fig:Amazon_sponsored_product_example_22}
\begin{center}
    \includegraphics[width=0.75\textwidth]{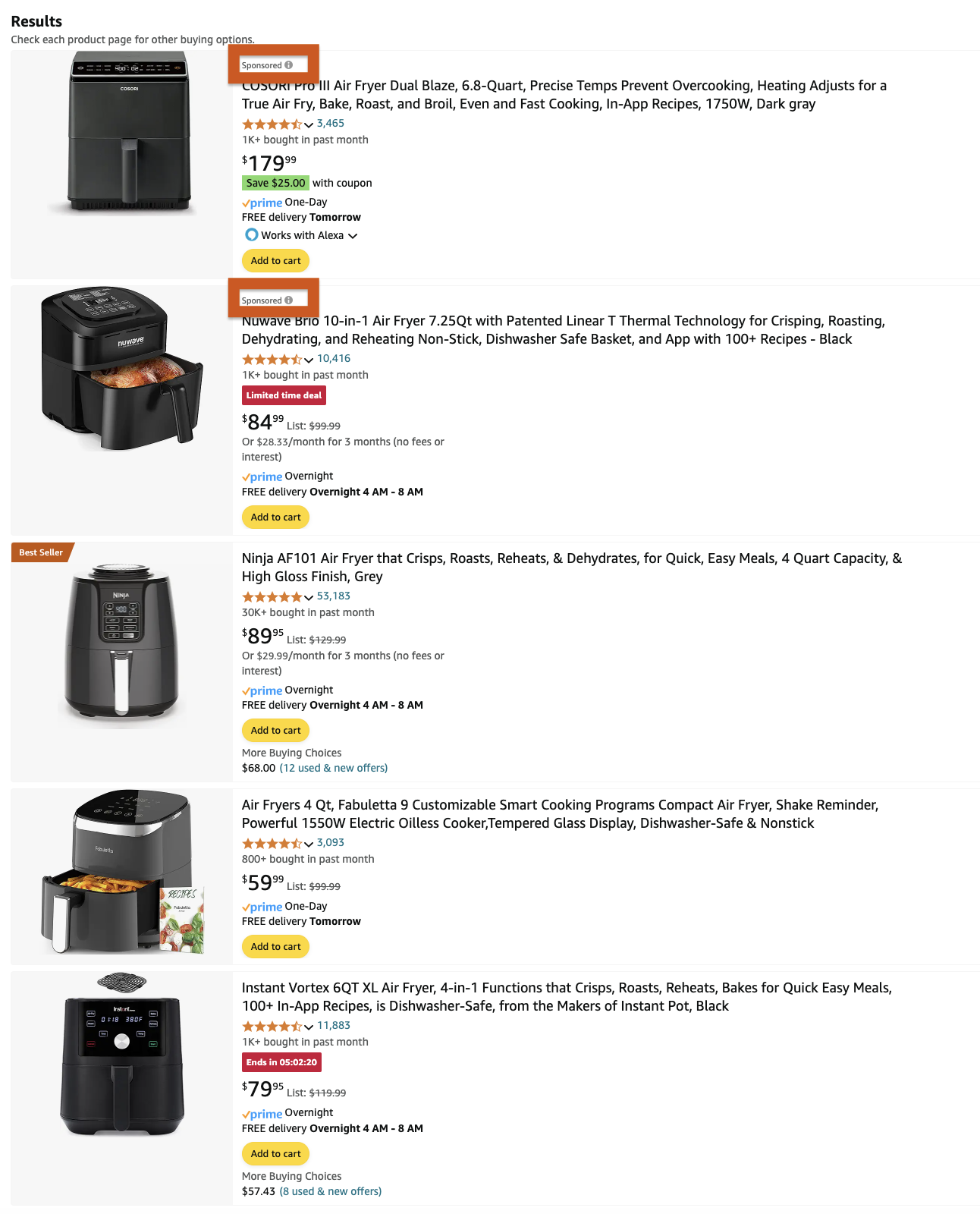}
\end{center}    
First five rows of the search results for ``air fryer.'' Orange rectangles emphasize sponsored listings that link to products by sellers who won the ad position in an auction.
\end{figure}

Figure \ref{fig:Graphical_Representation_Theoretical_Model} below provides an
illustration of the results page for the simplified model. Figure \ref{fig:Amazon_sponsored_product_example_22} shows an example of an Amazon search results page with 22 products as viewed on a desktop or laptop browser, which contains a mix of both sponsored and organic products. Key information presented on the search results page includes the product's image, price, sales, customer ratings, and delivery options. The mobile app experience is similar, displaying a mix of sponsored and organic products vertically that consumers can scroll down or click on for more details.

\webappsection{Details of Proof of Lemma \ref{obs:monopoly_price}}
\label{webapp:lemma-monopoly}
As before,  we show this result using the Implicit Function Theorem.

The monopolist's objective function is 
\begin{equation}
    \max _{p_i, p_j, b_i, b_j} \pi_M (\boldsymbol{p}, \boldsymbol{b}) = \pi_i (\boldsymbol{p}, \boldsymbol{b}) +\pi_j (\boldsymbol{p}, \boldsymbol{b})
\end{equation}

The two sellers would agree to set their bids to the minimum, \( b_i = b_j \rightarrow 0 \), because it minimizes advertising costs while maintaining the same probability of capturing the demand from consumers who consider only the first position. The monopolist objective becomes:
\begin{align}\nonumber
    \max_{p_i,  p_j=p_i} \qquad \pi_M(\boldsymbol{p},\boldsymbol{0})
%    &=\left( \frac{\theta}{2} \cdot\left( s_i\left(p_i\right)+s_j\left(p_j\right)\right)+(1-\theta) \cdot \left(s_i\left(p_i, p_j\right)+s_j\left(p_i, p_j\right)\right)\right) \cdot\left((1-\tau) \cdot p_i-c_i\right)\\
  &  \underset{p_j = p_i}{=}\left( \theta \cdot  s_i\left(p_i\right)+2 (1-\theta) \cdot s_i\left(p_i, p_i\right)\right) \cdot\left((1-\tau) \cdot p_i-c_i\right)
\end{align}

We first compute the first order condition. Assume $\mathrm{exp}:=\exp\!\Big(\frac{a-p}{\mu}\Big)>0,
a_1:=1+\mathrm{exp}, b_1:=1+2\,\mathrm{exp}, p_i=p_j=p$. Then, $s_i(p)=\frac{\mathrm{exp}}{a_1},
s_i(p,p)=\frac{\mathrm{exp}}{b_1}$, $\frac{d\,s_i(p)}{dp}=-\frac{\mathrm{exp}}{\mu\,a_1^2},
\;\frac{d\,s_i(p,p)}{dp}
=\frac{d}{dp}\!\left(\frac{\mathrm{exp}}{1+2\,\mathrm{exp}}\right)
=-\frac{\mathrm{exp}}{\mu\,b_1^2}$, $q(\theta)=\theta\,s_i(p)+2(1-\theta)\,s_i(p,p)$.

\paragraph{FOC.} Combining the above expressions yields:
\[
G(p,\theta):=\frac{\partial \pi_M}{\partial p}=q_p(\theta)\,m+q(\theta)\,(1-\tau)=0
\quad\Longrightarrow\quad
m=-\frac{q(\theta)}{q_p(\theta)}(1-\tau).
\tag{FOC}
\]

\paragraph{Comparative static with respect to $\theta$.}
Differentiate the FOC in $\theta$:
\[
G_\theta
=\Big(s_i'(p)-2\,\tfrac{d\,s_i(p,p)}{dp}\Big)\,m
+\Big(s_i(p)-2\,s_i(p,p)\Big)\,(1-\tau).
\]
We then simplify:
\[
G_\theta
=-(1-\tau)\,
\frac{2\,\mathrm{exp}^{\,2}}
{\ \theta(1+2\,\mathrm{exp})^{2}+2(1-\theta)(1+\mathrm{exp})^{2}\ }\;<\;0
\]
(Every denominator term is strictly positive for all $\mathrm{exp}>0$ and
$\theta\in[0,1]$.)

\subsection*{Second–order condition (SOC): $G_p$ at the optimum}

We also need to know the sign of $G_p=\partial^2\pi_M/\partial p^2$ at the interior optimum.
Note
\[
\pi_M'(p)=q_p\,m+q(1-\tau),\qquad
\pi_M''(p)=q_{pp}\,m+2(1-\tau)\,q_p
=(1-\tau)\Big(2q_p-\frac{q\,q_{pp}}{q_p}\Big)
=(1-\tau)\,\frac{2q_p^2-q\,q_{pp}}{q_p}.
\tag{$\star$}
\]

Plugging in 
$\frac{d^2 s_i(p)}{dp^2}
=\frac{\mathrm{exp}}{\mu^2}\left(\frac{1}{a_1^2}-\frac{2\mathrm{exp}}{a_1^3}\right),
\frac{d^2 s_i(p,p)}{dp^2}
=\frac{\mathrm{exp}}{\mu^2}\left(\frac{1}{b_1^2}-\frac{4\mathrm{exp}}{b_1^3}\right)$, and simplifying, we get 
\[
2q_p^2 - q\,q_{pp}
= \frac{\mathrm{exp}^{\,2}}{\mu^{2}\,a_1^{3}\,b_1^{3}}
\Big[
(2-\theta)^{2}
+ 6\,\mathrm{exp}\,(2-\theta)
+ 2\,\mathrm{exp}^{2}\,(6-\theta+\theta^{2})
+ 4\,\mathrm{exp}^{3}\,(1+\theta)
\Big]
> 0.
\]
All denominators are positive; each bracketed term is non-negative for
$\mathrm{exp}>0$ and $\theta\in[0,1]$ (indeed strictly positive). Since
$q_p<0$, \((\star)\) gives $G_p<0$.

\webappsection{Discussion of the modeling choice of   \texorpdfstring{$\theta$}{theta}}
\label{subsec:micro-foundation}

\webappsubsection{Endogenizing $\theta$}
\label{subsubsection:endogenizing_theta}

At each position $n$, consumers will form an expectation of the incremental utility of continuing their search to the next position, which is $\log\left(1+e^{\delta_1}+e^{\delta_2}+\cdots + E \left[\hat{e^{\delta_{n+1}}}|\delta_1, \delta_2, \cdots, \delta_{n}\right]\right)- \log \left(1+e^{\delta_1}+e^{\delta_2}+\cdots +e^{\delta_n}\right)$, and compare with the cost $s$ to search one more product. If the cost is lower than the expected incremental utility, they will continue to search for the product in position $n+1$; otherwise, they will stop at position $n$.

% Let $\mathcal{J}_n=\left\{j_1, \ldots, j_n\right\}$ denote the set of the top $n$ products. Consumer $i$ with a consideration set $\mathcal{J}_{n_1}$ will add new products to her consideration set to form a new consideration set $\mathcal{J}_{n_2} \supset \mathcal{J}_{n_1}$ (i.e., $\left.n_2>n_1\right)$, incurring a search cost of $\left(n_2-n_1\right) s_i$, if and only if the following condition holds:
% $$
% \mathbb{E}_\epsilon\left[\max \left\{\delta_{j_n}+\epsilon_{i j_n}\right\}_{n \leq n_1} \cup\left\{\tilde{\delta}_n+\epsilon_{i j_n}\right\}_{n_1<n \leq n_2}\right]-\mathbb{E}_\epsilon\left[\max \left\{\delta_{j_n}+\epsilon_{i j_n}\right\}_{n \leq n_1}\right]>\left(n_2-n_1\right) s_i .
% $$

% Stopping rule
% $$s\left(n_1\right)=\max _{n_2>n_1} \frac{\mathbb{E}_\epsilon\left[\max \left\{\delta_{j_n}+\epsilon_{i j_n}\right\}_{n \leq n_1} \cup\left\{\tilde{\delta}_n+\epsilon_{i j_n}\right\}_{n_1<n \leq n_2}\right]-\mathbb{E}_\epsilon\left[\max \left\{\delta_{j_n}+\epsilon_{i j_n}\right\}_{n \leq n_1}\right]}{n_2-n_1}$$

% $$\lambda_n=\sum_{n^{\prime}=n}^N \int \mathbf{1}\left(s_i \in\left(s(n), \min _{n^{\prime}<n} s\left(n^{\prime}\right)\right]\right) \mathrm{d} F\left(s_i\right)$$

For the two-product example, suppose that the cumulative distribution function of the cost is $F_s$.
 Suppose that every consumer search at least one product, then $\log \left(1+e^{\delta_1}\right)-0>s$
would always hold.
 For consumers to search the first position but not the second, we have $\log \left(1+e^{\delta_1}\right)-0>s>\log \left(1+e^{\delta_1}+e^{\delta_2}\right)-\log \left(1+e^{\delta_1}\right)$.
 For consumers to search the second, we have
 $\log \left(1+e^{\delta_1}+e^{\delta_2}\right)-\log \left(1+e^{\delta_1}\right)>s$. Therefore, $\theta$ and $1-\theta$ can be expressed as 
    $\theta=\frac{ F_s\left(\log \left(1+e^{\delta_1}\right)\right)-F_s\left(\log \left(1+e^{\delta_1}+e^{\delta_2}\right)-\log \left(1+e^{\delta_1}\right)\right)}{F_s\left(\log \left(1+e^{\delta_1}\right)\right)}$ and $1-\theta=\frac{ F_s\left(\log \left(1+e^{\delta_1}+e^{\delta_2}\right)-\log \left(1+e^{\delta_1}\right)\right)}{F_s\left(\log \left(1+e^{\delta_1}\right)\right)}$.

If we endogenize $\theta$, the crossing results in Proposition \ref{prop_crossing} still hold, since the limiting behavior and the fact that ads increase prices remain the same, although the exact point at which the competitive and collusive prices intersect would change slightly.

\webappsubsection{Sensitivity check: A fraction of consumers who consider all products would also click the sponsored links}
\label{websec:sensitivity_check}

\begin{figure}[ht]
\begin{center}
    \caption{Sensitivity check: fraction of consumers who consider all products also click ads}
        \includegraphics[width=0.7\textwidth]{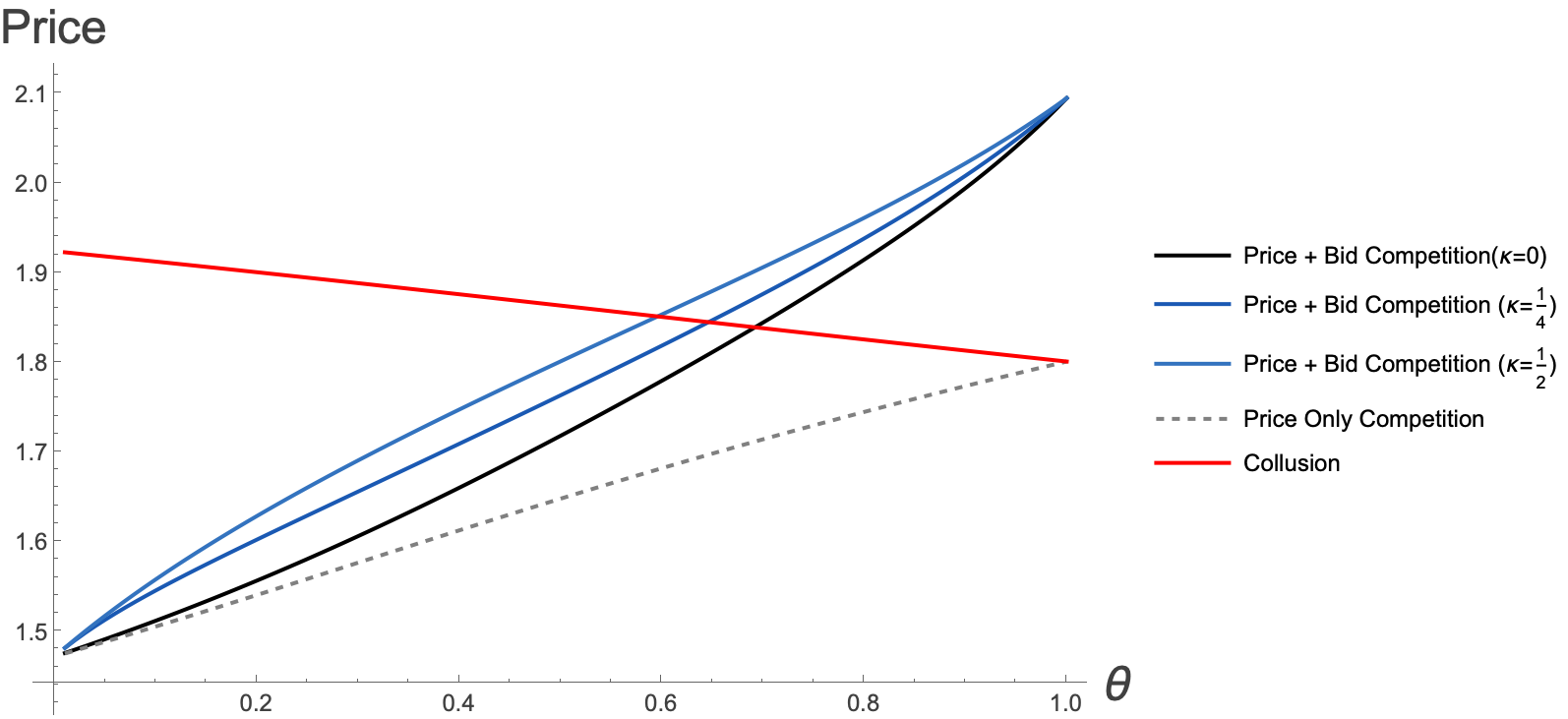}
\label{fig:sensitivity_check}
\end{center}    
    The figure shows the full competition prices (with advertising - black, dark blue and blue solid) as a function of search costs $\theta$ and under different values of $\kappa$. 
\end{figure}

We consider a sensitivity check in which a fraction $\kappa$ of consumers who consider all products also click the sponsored links. The seller’s profit becomes:

\begin{align}\nonumber
\pi_i^t(\boldsymbol{p^t}, \boldsymbol{b^t}) = & \theta \cdot \operatorname{Pr}(\tilde{b_i^t} > \tilde{b_j^t}) \cdot s_i\left(p_i^t\right) \cdot \left(\left((1-\tau) \cdot p_i^t - c_i\right) - \gamma_{i1} \cdot \mathbb{E}\left[\tilde{b_i^t} \mid \tilde{b_i^t} > \tilde{b_j^t}\right]\right) \\\nonumber
& + (1-\theta) \cdot \kappa \cdot s_i\left(p_i^t, p_j^t\right) \cdot \left(\left((1-\tau) \cdot p_i^t - c_i\right)- \gamma_{i2} \cdot \mathbb{E}\left[\tilde{b_i^t} \mid \tilde{b_i^t} > \tilde{b_j^t}\right]\right)\\\nonumber
& + (1-\theta) \cdot (1-\kappa) \cdot s_i\left(p_i^t, p_j^t\right) \cdot \left((1-\tau) \cdot p_i^t - c_i\right)
\end{align}
Figure \ref{fig:sensitivity_check} presents the equilibrium prices for different values of $\kappa$. When $\kappa = 0$, it corresponds to our baseline model. The crossing results of competition prices and collusive prices still hold.

% \begin{comment}

\webappsubsection{Alternative Characterization of Consumer Search Costs}
\label{subsubsection:alternative_characterization_of_search_costs}

% In \cite{reimers2023framework}, the authors consider an alternative, more reduced-form characterization of consumer search costs. We present the details of that model and demonstrate the correspondence between their model and ours. 

We consider an alternative way of modeling consumer search costs and show that the details of modeling consumer search costs do not change our main findings that collusive prices can fall below competitive prices. Following \cite{reimers2023framework}, we consider a reduced-form characterization of consumer search costs. Specifically, consumer's utility for product $j$ when ranked at $r_j$ is given by:
$$
u_{i j}=X_j' \beta - \alpha p_j+\zeta r_j+\zeta^{\prime} r_j^0+\xi_j+\epsilon_{i j}
$$
where the outside good has utility 0, $\xi_j$ represents unobserved product quality, $r_j^0$ is the initial rank of product $j$, and $\epsilon_{i j}$ is an extreme value error. Here, $\zeta$ is the causal effect of rank on sales, while $\zeta^{\prime}$ reflects the systematic product quality variation across ranks that is not accounted for by $x_j$. 

The mean expected utility of product $j$ when ranked at $r$ is $\delta_j(r)=X_j' \beta-\alpha p_j+\zeta^{\prime} r_j^0+\zeta r$. Even if product $j$ were moved to a different rank $r_j$, the part of utility reflected by $\zeta^{\prime} r_j^0$ would remain, while the part reflected by $\zeta r$ would change. The ``rank-independent expected mean utility" is
$$
\delta_j^0 \equiv X_j' \beta-\alpha p_j+\zeta^{\prime} r_j^0=\delta_j(r)-\zeta r .
$$

Product $j$ 's market share when ranked $r^{\text {th }}$ is given by $s_j(r)=\frac{e^{\delta_j(r)}}{1+\sum e^{\delta_j(r)}}$. The way in which the products are ranked affects both consumer well-being and the propensity for consumers to purchase.

Then assume that the ad auction part is the same as the one we consider in our benchmark model, while the consumer choice is from this model. In the full competition scenario, where sellers compete on both prices and bids, the profit function becomes:
\begin{align}\nonumber
\pi_i^t(\boldsymbol{p^t}, \boldsymbol{b^t}) = & \operatorname{Pr}(\tilde{b_i^t} > \tilde{b_j^t}) \cdot \frac{e^{\frac{a_i- p_i^t}{\mu}}}{e^{\frac{a_i- p_i^t}{\mu}}+ e^{\frac{a_j- p_j^t-\zeta}{\mu}}+1} \cdot \left(\left((1-\tau) \cdot p_i^t - c_i\right) - \gamma_i \cdot \mathbb{E}\left[\tilde{b_i^t} \mid \tilde{b_i^t} > \tilde{b_j^t}\right]\right) \\
& + \operatorname{Pr}(\tilde{b_i^t} < \tilde{b_j^t}) \cdot\frac{e^{\frac{a_i- p_i^t- \zeta}{\mu}}}{e^{\frac{a_i- p_i^t- \zeta }{\mu}}+ e^{\frac{a_j- p_j^t}{\mu}}+1}  \cdot \left((1-\tau) \cdot p_i^t - c_i\right)
\label{equ:rank_seller_profit_full_competition}
\end{align}

In the scenario where there are no sponsored ads and the platform randomly displays one of the products in the top position, the sellers' profit function is:
\begin{align}\nonumber
\pi_i^t(\boldsymbol{p^t}, \boldsymbol{b^t}) = & \frac{1}{2} \cdot \left( \frac{e^{\frac{a_i- p_i^t}{\mu}}}{e^{\frac{a_i- p_i^t}{\mu}}+ e^{\frac{a_j- p_j^t-\zeta}{\mu}}+1} + \frac{e^{\frac{a_i- p_i^t- \zeta}{\mu}}}{e^{\frac{a_i- p_i^t- \zeta }{\mu}}+ e^{\frac{a_j- p_j^t}{\mu}}+1} \right) \cdot \left(\left((1-\tau) \cdot p_i^t - c_i\right) \right) 
\label{equ:rank_seller_profit_random_display}
\end{align}

% In this model, the consumer surplus can be expressed as 
% \begin{align}\nonumber
% U\left(\boldsymbol{p^t}\right)&=\operatorname{Pr}(\tilde{b_i^t} > \tilde{b_j^t}) \cdot \mu \cdot \log \left[ e^{\frac{a_i- p_i^t}{\mu}}+ e^{\frac{a_j- p_j^t-\zeta}{\mu}}+1\right]+ \operatorname{Pr}(\tilde{b_i^t} < \tilde{b_j^t}) \cdot \mu \cdot \log \left[e^{\frac{a_i- p_i^t- \zeta }{\mu}}+ e^{\frac{a_j- p_j^t}{\mu}}+1\right]\end{align}

\begin{figure}[!ht]
\caption{Casual Effect of Rank Model}
\label{fig:rank_casual} 
\begin{center}
\begin{subfigure}{0.49\textwidth}
\caption{Prices}
\label{subfig:rank_casual_price}
\centering
\includegraphics[width=\textwidth]{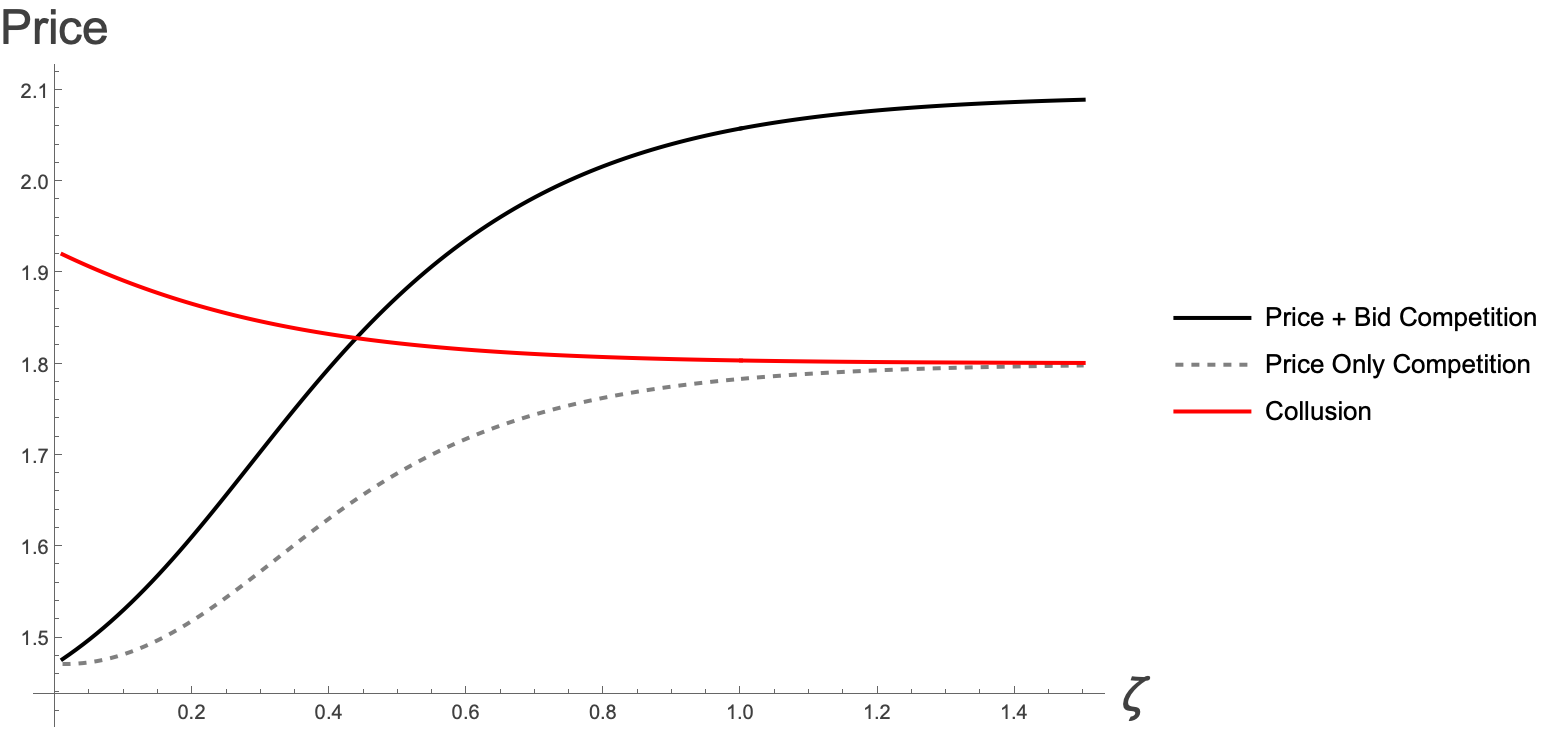}
\end{subfigure}~
\begin{subfigure}
{0.49\textwidth}
\centering
\caption{Bids}
\includegraphics[width=\textwidth]{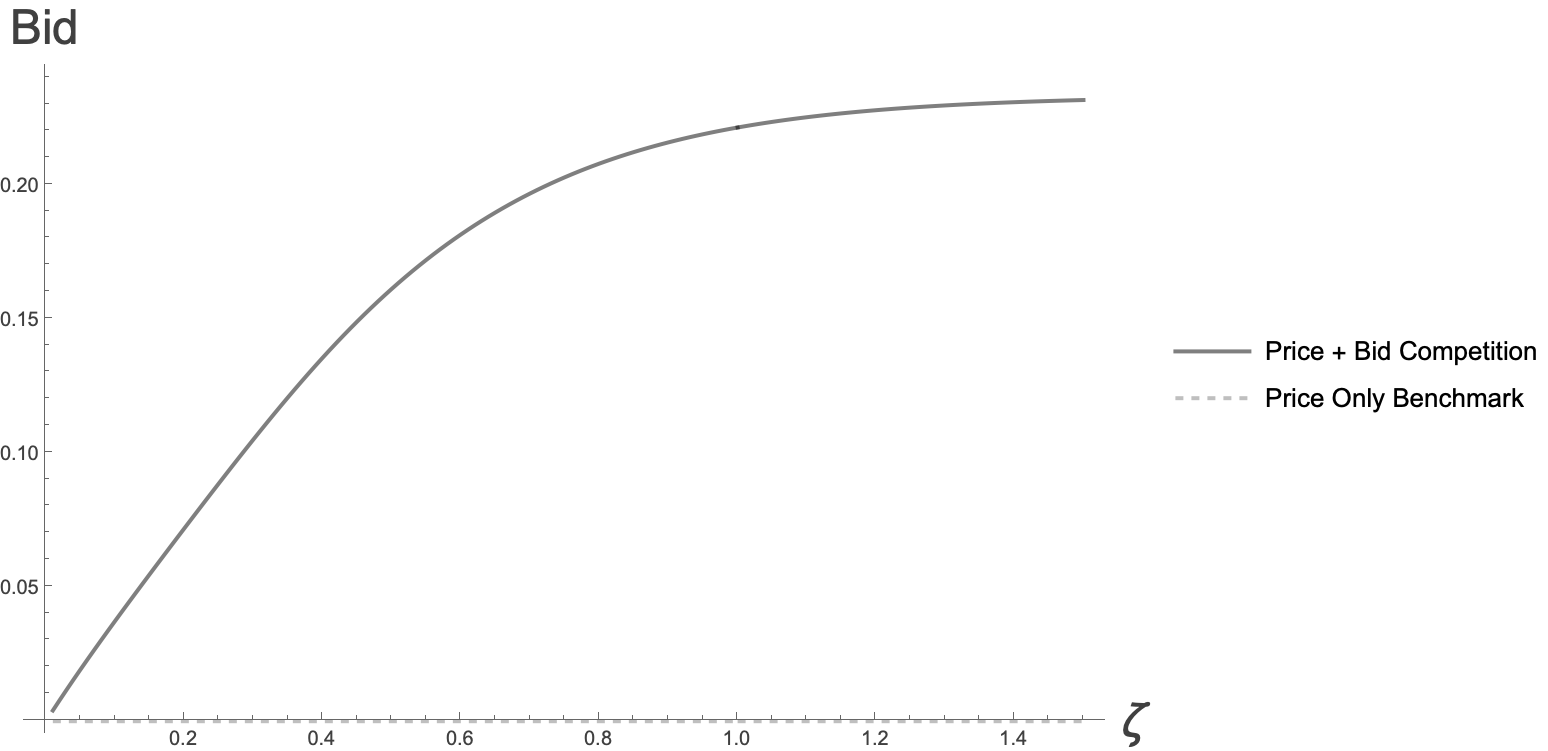}
\label{subfig:rank_casual_bid}
\end{subfigure}
% \begin{subfigure}{0.49\textwidth}
% \caption{Seller Profit}
% \label{subfig:rank_casual_profit}
% \centering
% \includegraphics[width=\textwidth]{Figure/rank_casual_profit.png}
% \end{subfigure}
% \begin{subfigure}{0.49\textwidth}
% \caption{Consumer Surplus}
% \label{subfig:rank_casual_cs}
% \centering
% \includegraphics[width=\textwidth]{Figure/rank_casual_cs.png}
% \end{subfigure}
\end{center}
Panels (a) and (b) show the equilibrium prices and bids as functions of the casual effect on sales parameter $\zeta$, respectively. The representation of other lines and the parameter specifications are the same as in Figure \ref{fig:theoretical-results}. 
\end{figure}

Figure \ref{fig:rank_casual} presents the equilibrium results. The results are robust and similar to those in Figure \ref{fig:theoretical-results}.\footnote{In the limiting case, when the causal effect of rank on sales, denoted by $\zeta$, approaches infinity ($\zeta \rightarrow \infty$) or zero ($\zeta \rightarrow 0$), it is equivalent to the case where the fraction of consumers who only care about the first position approaches one ($\theta \rightarrow 1$) or zero ($\theta \rightarrow 0$), respectively.} The main intuitions from Lemma \ref{obs:competitive_price} (advertising increases prices), from Lemma \ref{obs:monopoly_price} (monopoly prices decrease with the rank effect), as well as the limiting behavior as the rank effect approaches infinity, remain robust, and collusive prices can still be lower than competitive prices when consumer search costs are high.

% \end{comment}

\webappsection{Platform Strategic Response Additional Results}

\webappsubsection{Platform Strategic Response: Commission Rate Additional Results}

\begin{figure}[!ht]
\caption{Platform Strategic Response: Commission Rate Additional Results}
\label{fig:commission_platform_additional} 
\begin{center}
\begin{subfigure}
{0.45\textwidth}
\centering
\caption{Ad Revenue}
\includegraphics[width=\textwidth]{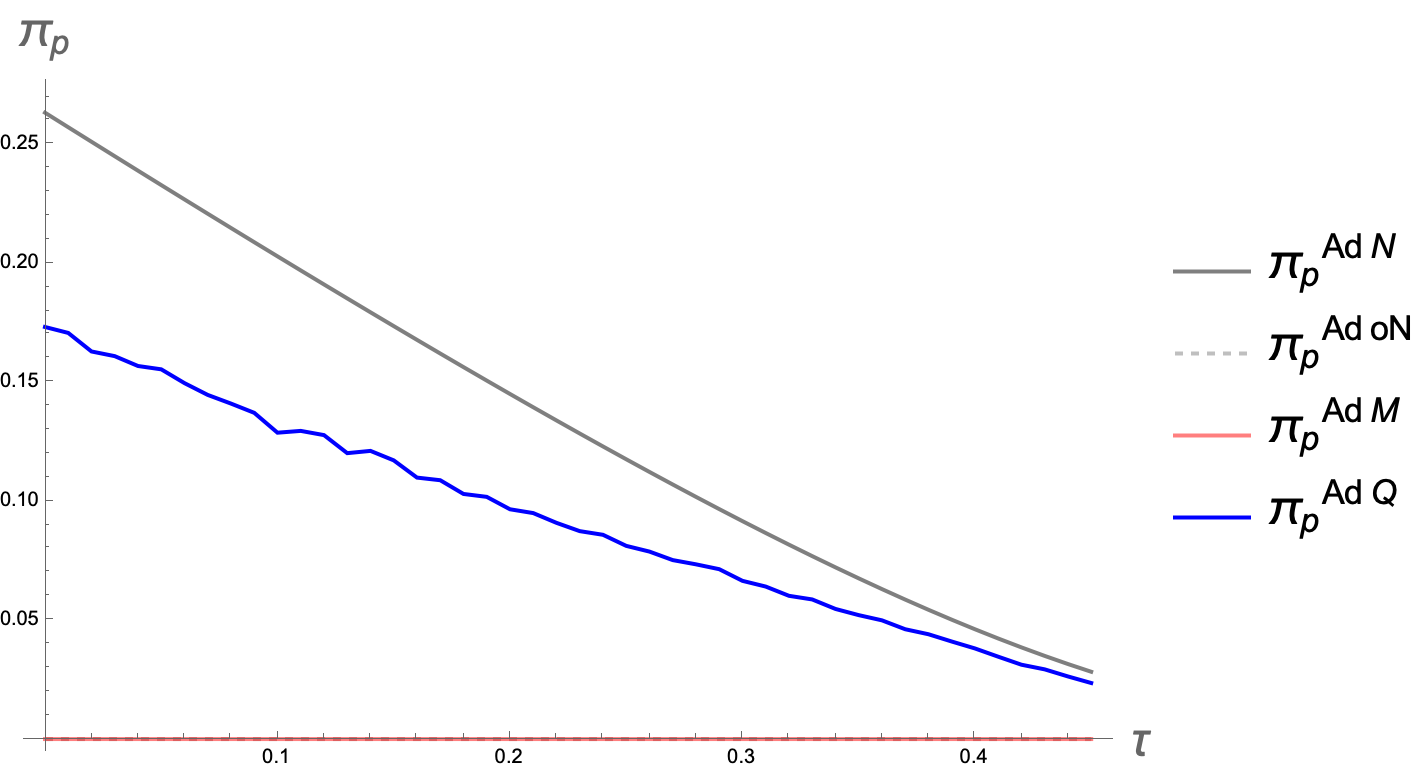}
\label{subfig:commission_platform_ads}
\end{subfigure}
\begin{subfigure}{0.45\textwidth}
\caption{Sellers' Profit}
\label{subfig:commission_seller_profit}
\centering
\includegraphics[width=\textwidth]{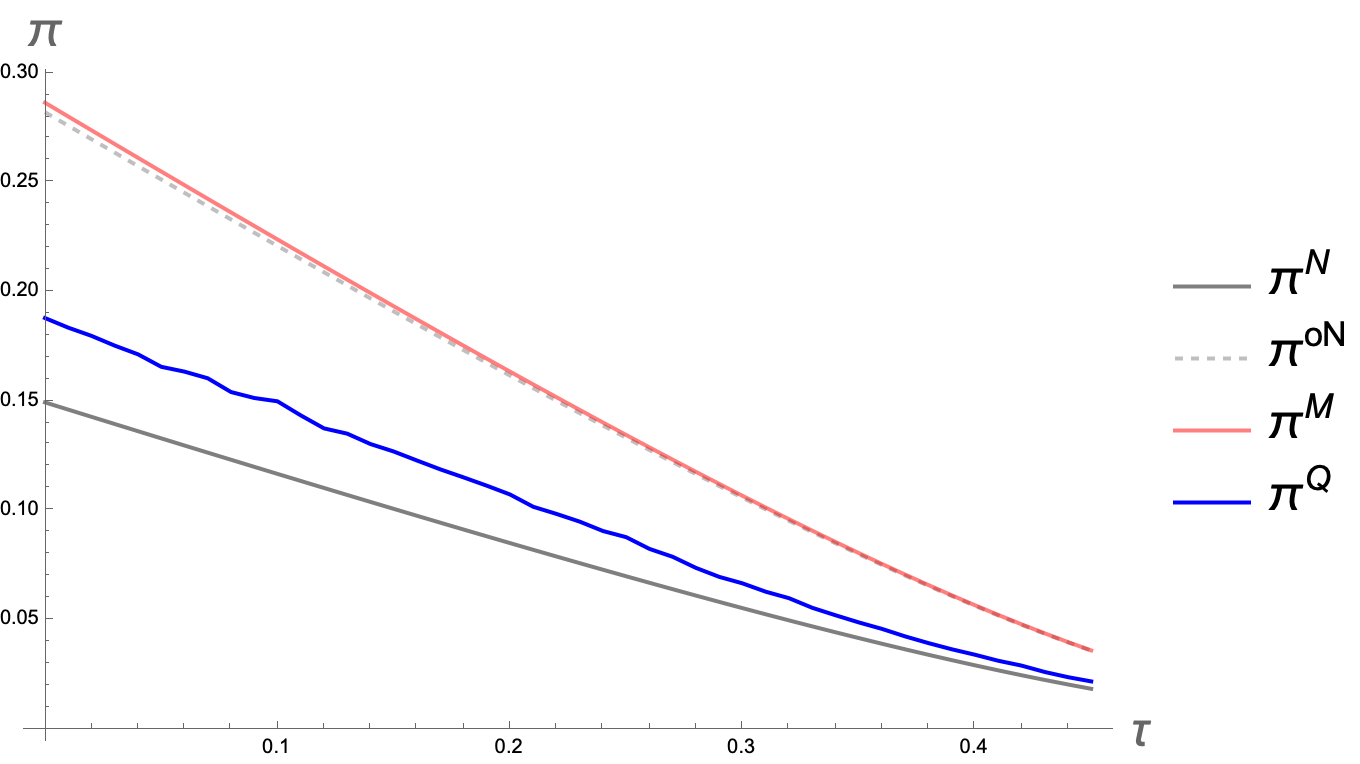}
\end{subfigure}
\begin{subfigure}
{0.45\textwidth}
\centering
\caption{Prices}
\includegraphics[width=\textwidth]{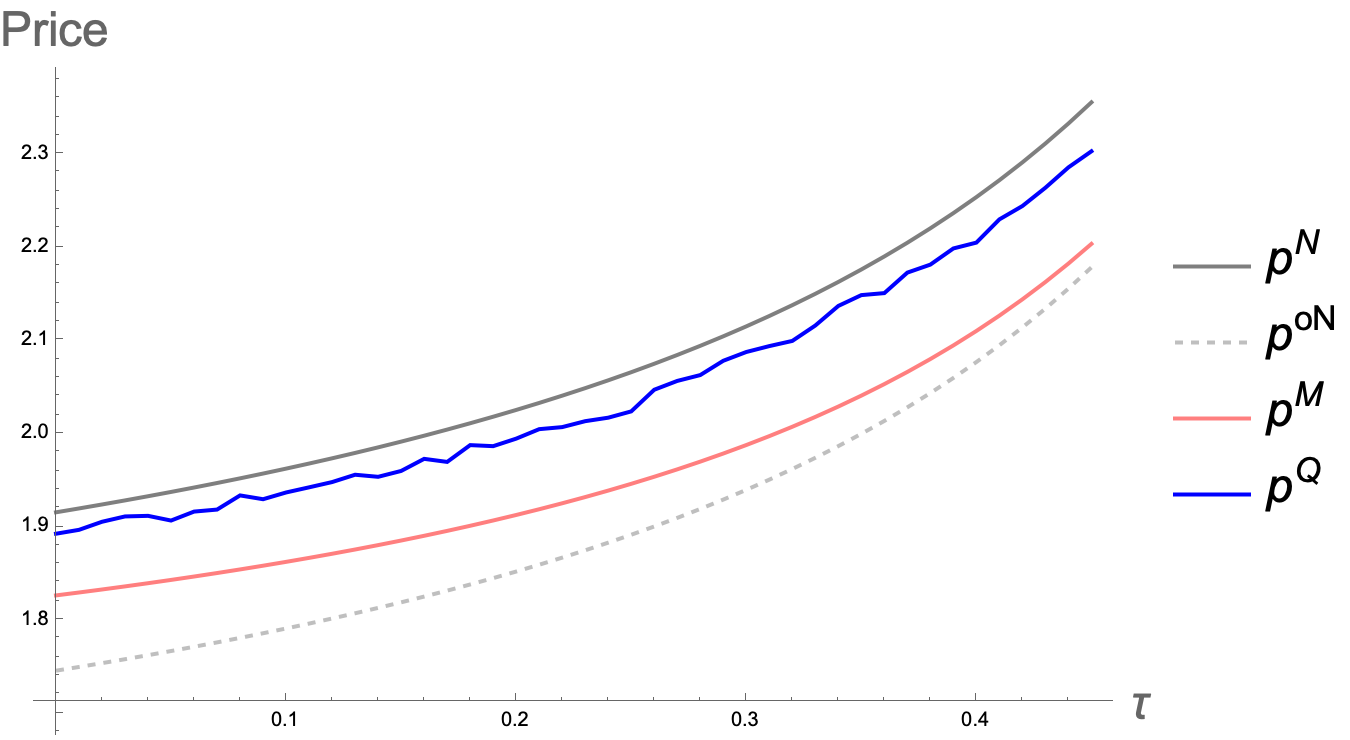}
\label{subfig:commission_market_price}
\end{subfigure}
\begin{subfigure}{0.45\textwidth}
\caption{Consumer Surplus}
\label{subfig:commission_CS}
\centering
\includegraphics[width=\textwidth]{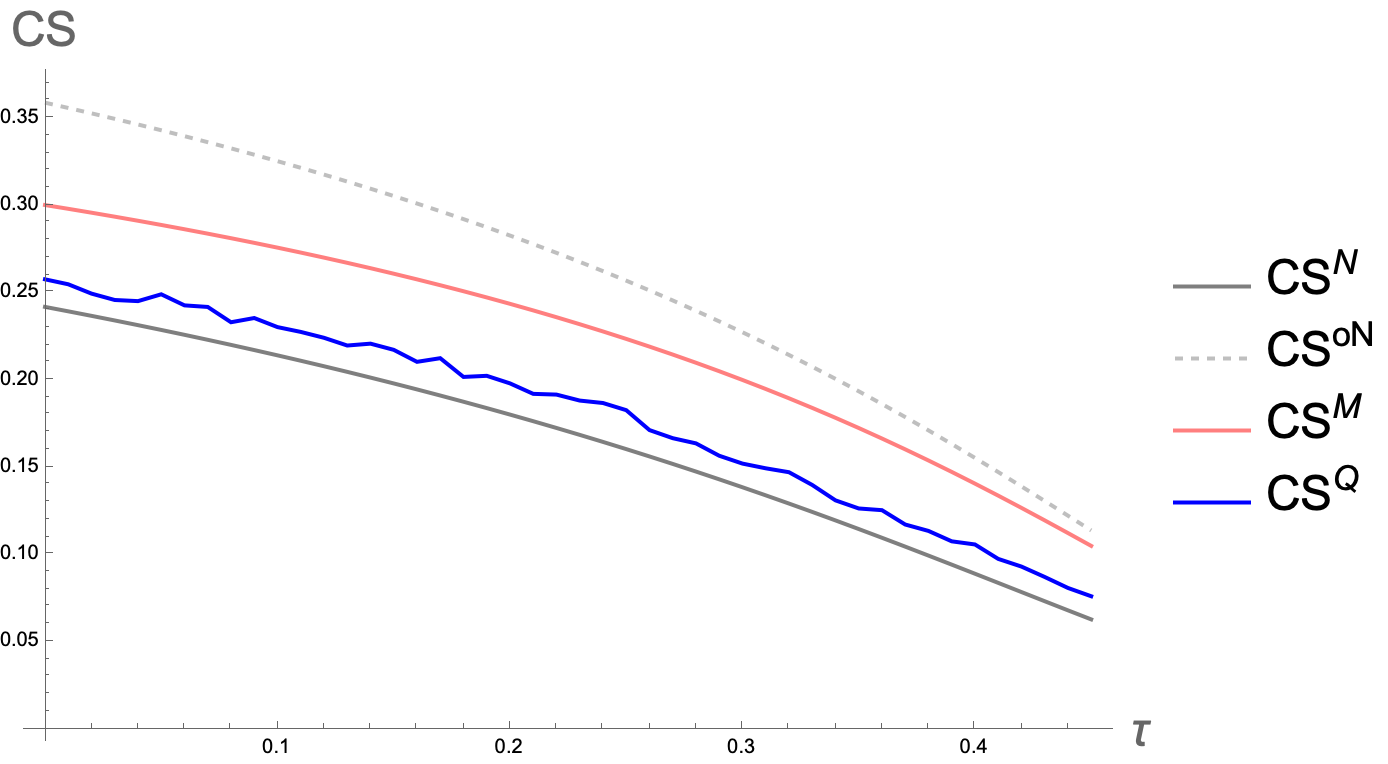}
\end{subfigure}~
\end{center}
Panels (a) to (d) show the platform' ads revenue, sellers' profit, market price and consumer surplus as a function of $\tau$ when $\theta = 0.8$, respectively. The blue solid line denotes the Q-learning platform profits from our simulation experiments. The representation of other lines and the parameter specifications are the same as in Figure \ref{fig:theoretical-results}.  
\end{figure}

Figure \ref{fig:commission_platform_additional} presents additional results when the platform adjusts the commission rate.

\webappsubsection{Platform Strategic Response: Reserve Price Additional Results}

\begin{figure}[!ht]
\caption{Platform Strategic Response: Reserve Price}
\label{fig:reserve_price_platform_additional} 
\begin{center}
\begin{subfigure}
{0.49\textwidth}
\centering
\caption{Commission}
\includegraphics[width=\textwidth]{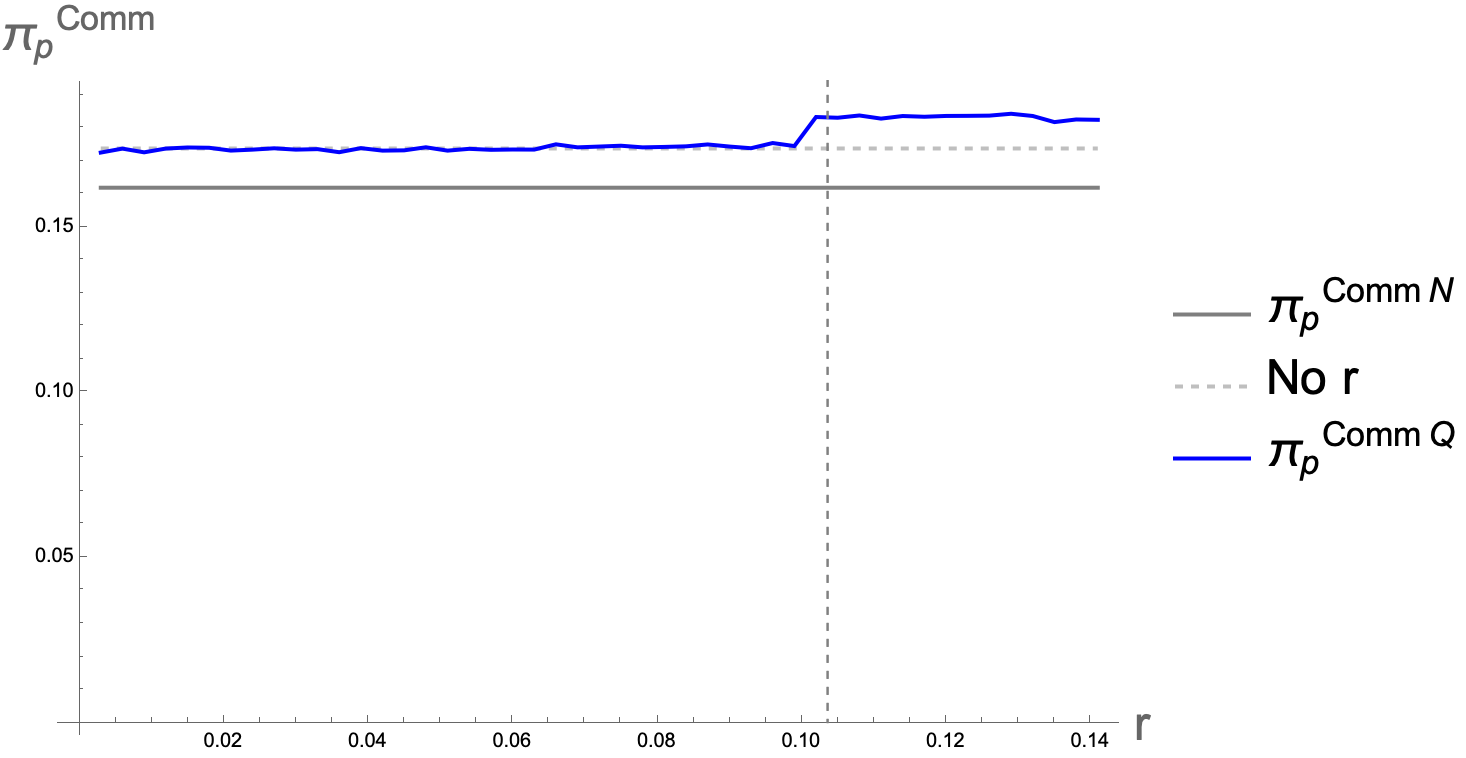}
\label{subfig:reserve_price_platform_comm}
\end{subfigure}
\begin{subfigure}{0.49\textwidth}
\caption{Ads Revenue}
\label{subfig:reserve_price_platform_ads}
\centering
\includegraphics[width=\textwidth]{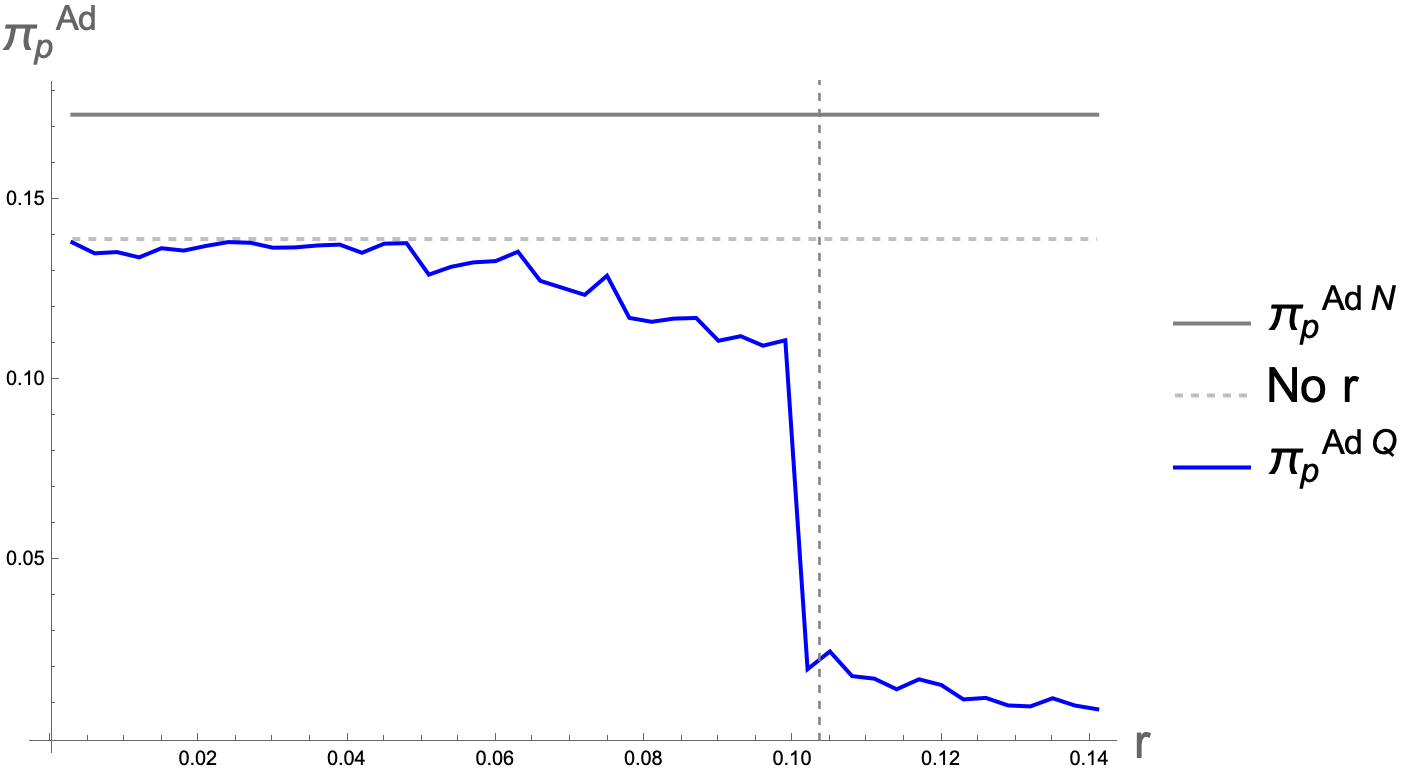}
\end{subfigure}
\begin{subfigure}{0.49\textwidth}
\caption{Seller Profit}
\label{subfig:reserve_price_seller_profit}
\centering
\includegraphics[width=\textwidth]{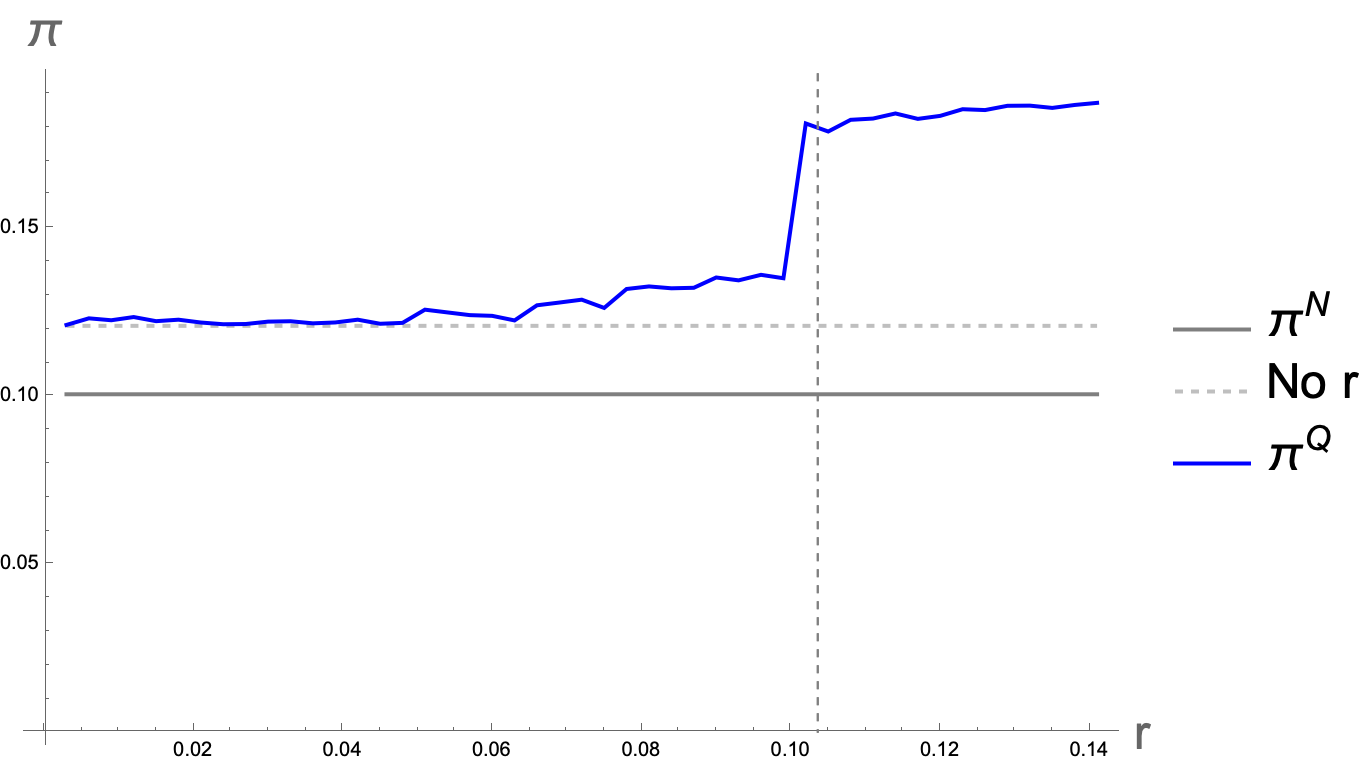}
\end{subfigure}
\begin{subfigure}{0.49\textwidth}
\caption{Consumer Surplus}
\label{subfig:reserve_price_cs}
\centering
\includegraphics[width=\textwidth]{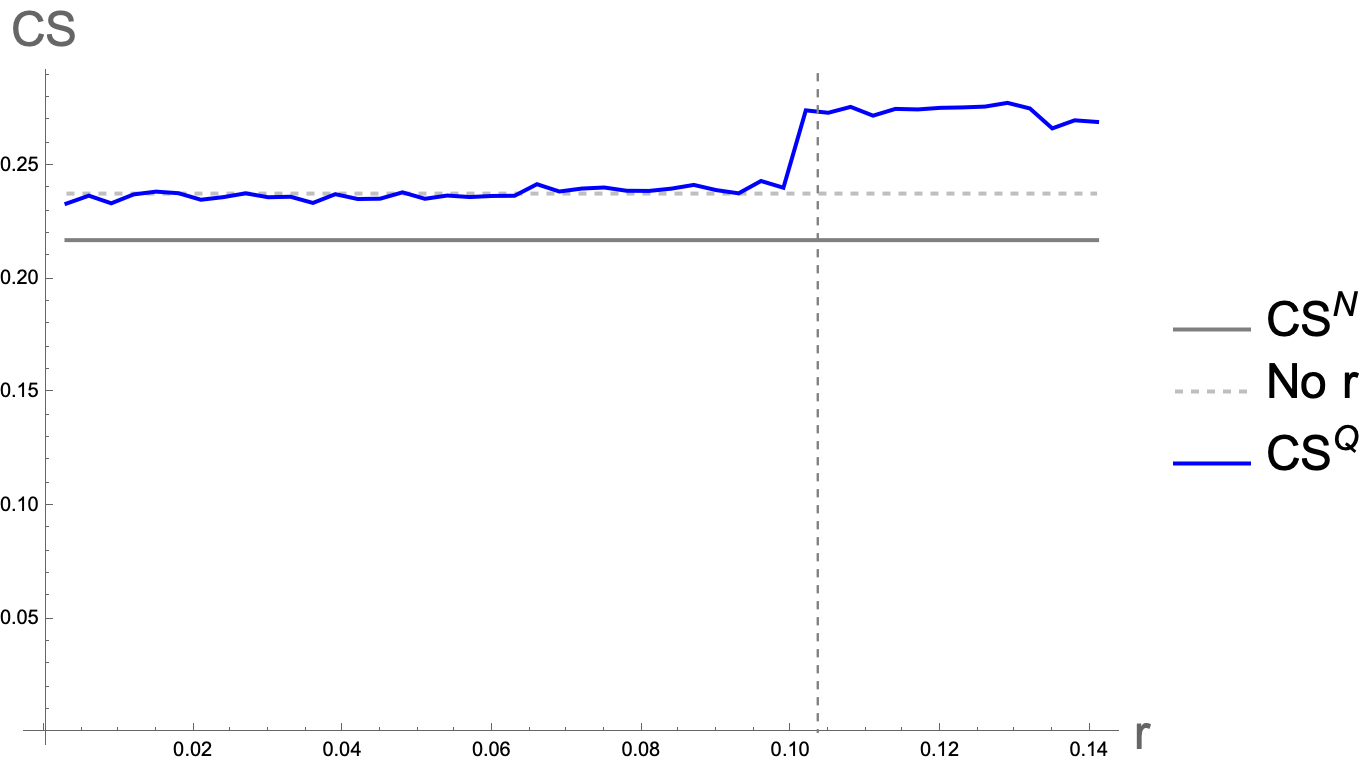}
\end{subfigure}
\end{center}
Panels (a) - (d) show the platform's commission revenue, ads revenue, sellers' profit, and consumer surplus as functions of the reserve price when $\theta = 0.8$ and $\tau = 0.15$, respectively. The blue solid line denotes the Q-learning platform profits from our simulation experiments. The representation of other lines and the parameter specifications are the same as in Figure \ref{fig:theoretical-results}. The dashed vertical lines represent the Q-learning bids in the algorithmic pricing scenario.
\end{figure}

Figure \ref{fig:reserve_price_platform_additional} presents additional results when the platform adjusts the reserve price.
 
\webappsubsection{Platform's Alternative Objective: Weighted Average of Total Surplus}
\label{subsec:total_surplus}

Given competition between platforms, a platform might also consider consumer surplus and seller profits to encourage entry for long-term business growth. Hence, we consider a scenario where the platform is maximizing a weighted average of its own profit, sellers' profit, and consumer surplus, with weights $\omega$ and $1-\omega$, respectively. The platform's objective can be written as follows:
{\small $$ \omega \cdot\pi_p(\boldsymbol{p}, \boldsymbol{b})+ (1-\omega)\cdot\left(\sum_j \pi_j(\boldsymbol{p}, \boldsymbol{b}) + CS(\boldsymbol{p}) \right) = \omega \cdot \left(\pi_p^{Ad }(\boldsymbol{p}, \boldsymbol{b}) +\pi_p^{Com }(\boldsymbol{p}, \boldsymbol{b})\right) + (1-\omega)\cdot \left(\sum_j \pi_j(\boldsymbol{p}, \boldsymbol{b})  + CS(\boldsymbol{p}) \right)
$$}
where $\pi_p(\boldsymbol{p}, \boldsymbol{b})$ is the platform's own-profit, $\pi_j(\boldsymbol{p}, \boldsymbol{b})$ is seller $j$'s profit as defined in \eqref{equ:seller_profit}, and $CS(\boldsymbol{p})$ is the consumer surplus as defined in \eqref{equ:consumer_surplus}.

When $\omega = \frac{1}{2}$, i.e., the platform puts equal weight on its own profit versus the sum of sellers' profits and consumer surplus, then its objective is equivalent to maximizing the total surplus.\footnote{Since the commission fee and advertising costs are transfers between sellers and consumers, they cancel out when $\omega=\frac{1}{2}$. Therefore, the total surplus in this scenario is also equivalent to the duopoly case where there is no platform, no commission fee, and no advertising costs.} The platform's objective becomes $\sum_j \left( \theta \cdot \operatorname{Pr}(\tilde{b_j} > \tilde{b_{-j}})  \cdot s_j\left(p_j\right) +(1-\theta) \cdot  s_j\left(p_j, p_{-j}\right)   \right)\cdot  (p_j- c_j)
- \theta \cdot \sum_j \mathbf{1}\{j \in \mathcal{J}_1(\boldsymbol{\Gamma})\} \cdot \mu \cdot \log \left( 1-s_j\left(p_j\right) \right)- (1- \theta) \cdot \mu \cdot \log \left( 1- \sum_{j} s_j\left(\boldsymbol{p}\right)\right)$.

\begin{figure}[!ht]
\caption{Platform Strategic Response: Total Surplus}
\label{fig:total_surplus} 
\begin{center}
\begin{subfigure}{0.49\textwidth}
\caption{Total Surplus vs $\theta$}
\label{subfig:search_total_surplus}
\centering
\includegraphics[width=\textwidth]{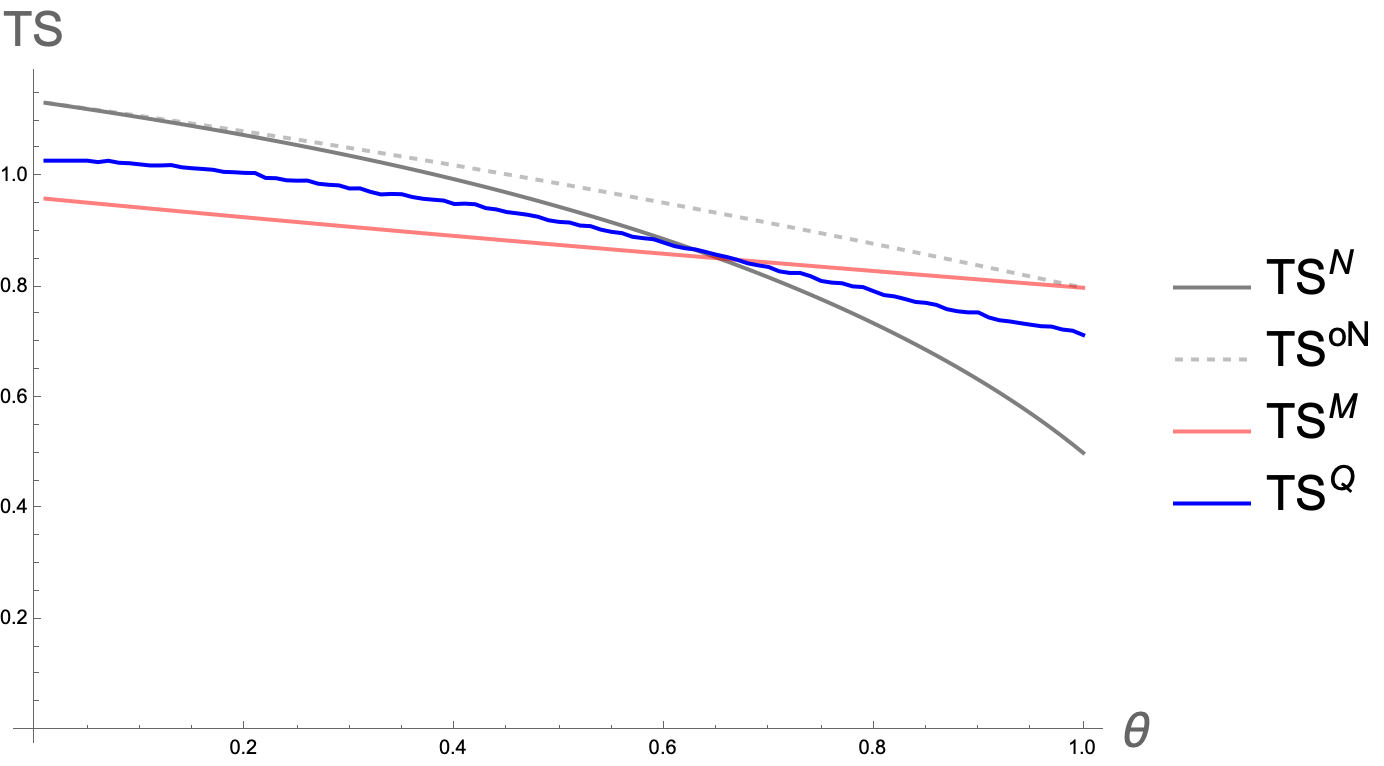}
\end{subfigure}~
\begin{subfigure}{0.49\textwidth}
\caption{Total Surplus vs $\tau$}
\label{subfig:commission_total_surplus}
\centering
\includegraphics[width=\textwidth]{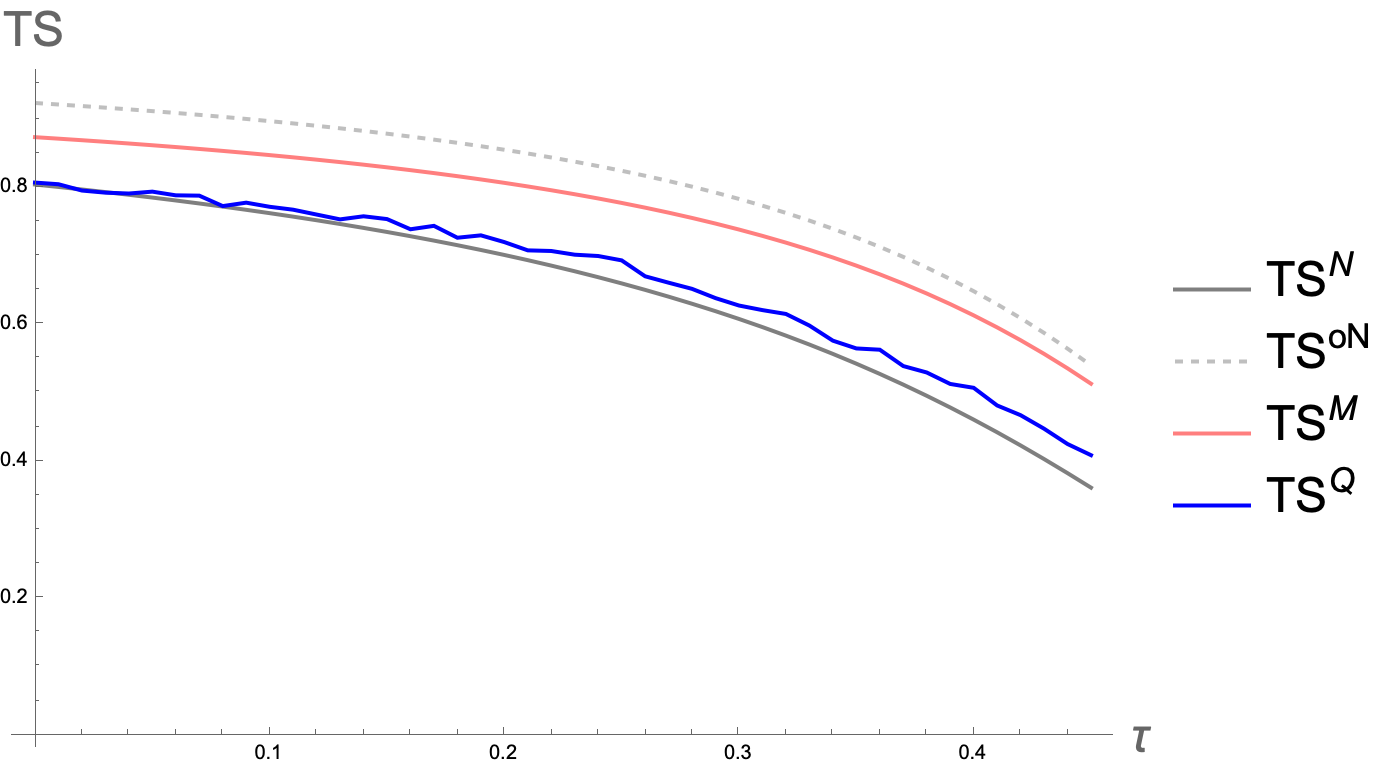}
\end{subfigure}~
\end{center}

Panel (a) shows the total surplus as a function of consumer search friction when $\tau = 0.15$, and (b) shows the total surplus as a function of the reserve price when 
$\theta = 0.8$. The blue solid line denotes the Q-learning platform profits from our simulation experiments. The representation of other lines and the parameter specifications are the same as in Figure \ref{fig:theoretical-results}. 
\end{figure}

Figure \ref{subfig:search_total_surplus} shows the total surplus as function of $\theta$ when commission rate is 15\%, i.e., $\tau =0.15$.\footnote{This is the rate for most product categories on Amazon.com}  When consumer search costs are high, compared with the full competition case, using learning algorithms always increase the total surplus. Because compared to the outside option, products on the platform become cheaper and hence more attractive. Figure \ref{subfig:commission_total_surplus} plots the total surplus when the platform adjusts its commission rate, given a high value of $\theta$. Algorithmic pricing consistently yields a higher total surplus than the full competition case, even when the platform is adjusting the commission rate.

% \textbf{Proof?} 
% $$\frac{\partial TS}{\partial p}<0$$

Thus, even if the platform's own single-period profit is lower, considering the sellers' profit and consumer surplus, the platform will not make any adjustments, and the beneficial outcomes for sellers and consumers will remain.

\webappsection{Additional Empirical Results}

\webappsubsection{Distribution of Sponsored Product Ads}

\begin{figure}[!ht]
\begin{center}
    \caption{Distribution of Sponsored Product Ads by Page Layout}
\label{fig:by_position_sponsored_heatmap_combined}

\begin{subfigure}[t]{0.476\textwidth}
    \centering
        \caption{Page layout with 60 products.}

    \includegraphics[width=\textwidth]{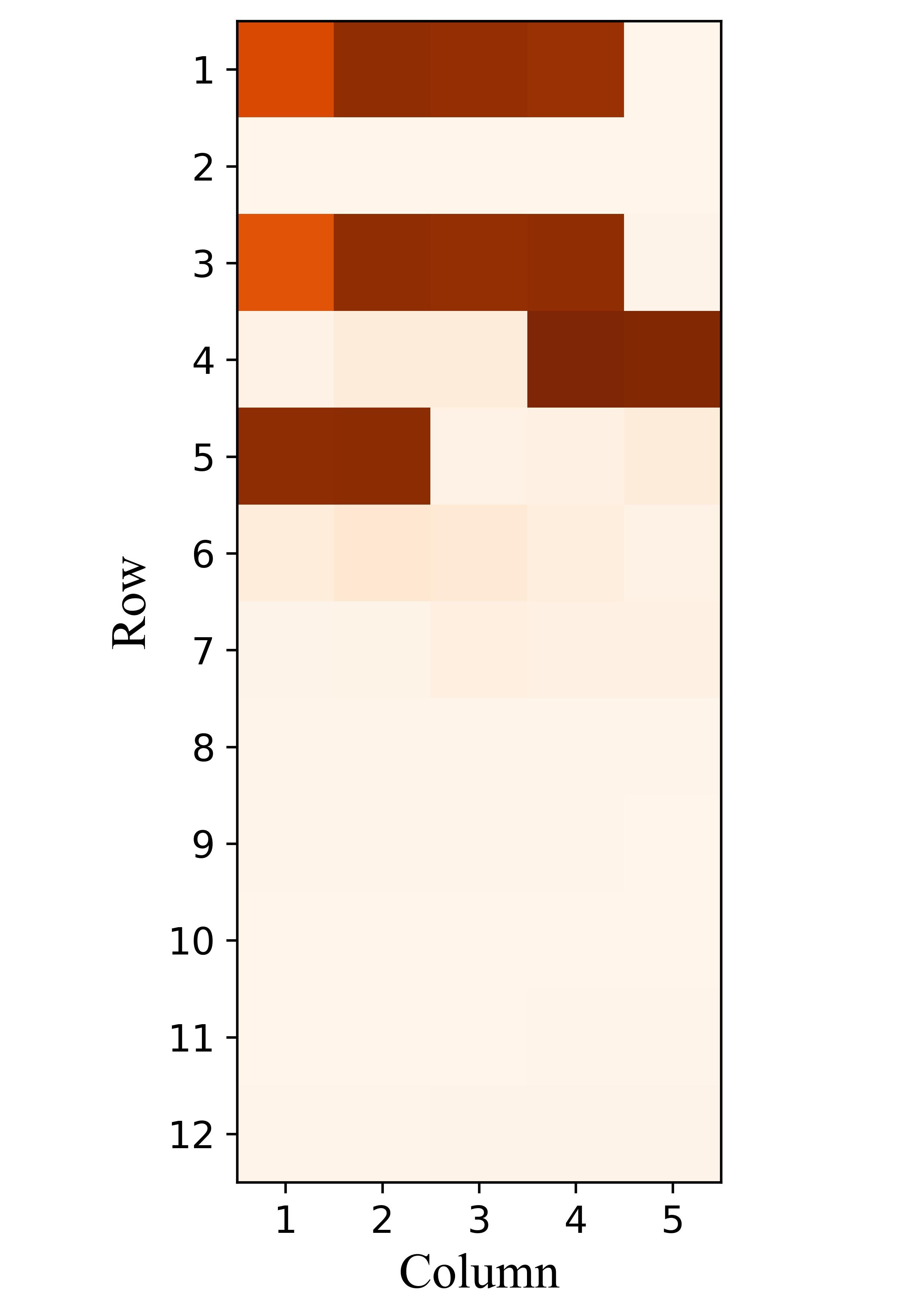}
    \label{fig:heatmap_60}
\end{subfigure}
\hfill
\begin{subfigure}[t]{0.34\textwidth}
    \centering
        \caption{Page layout with 22 products.}

    \includegraphics[width=\textwidth]{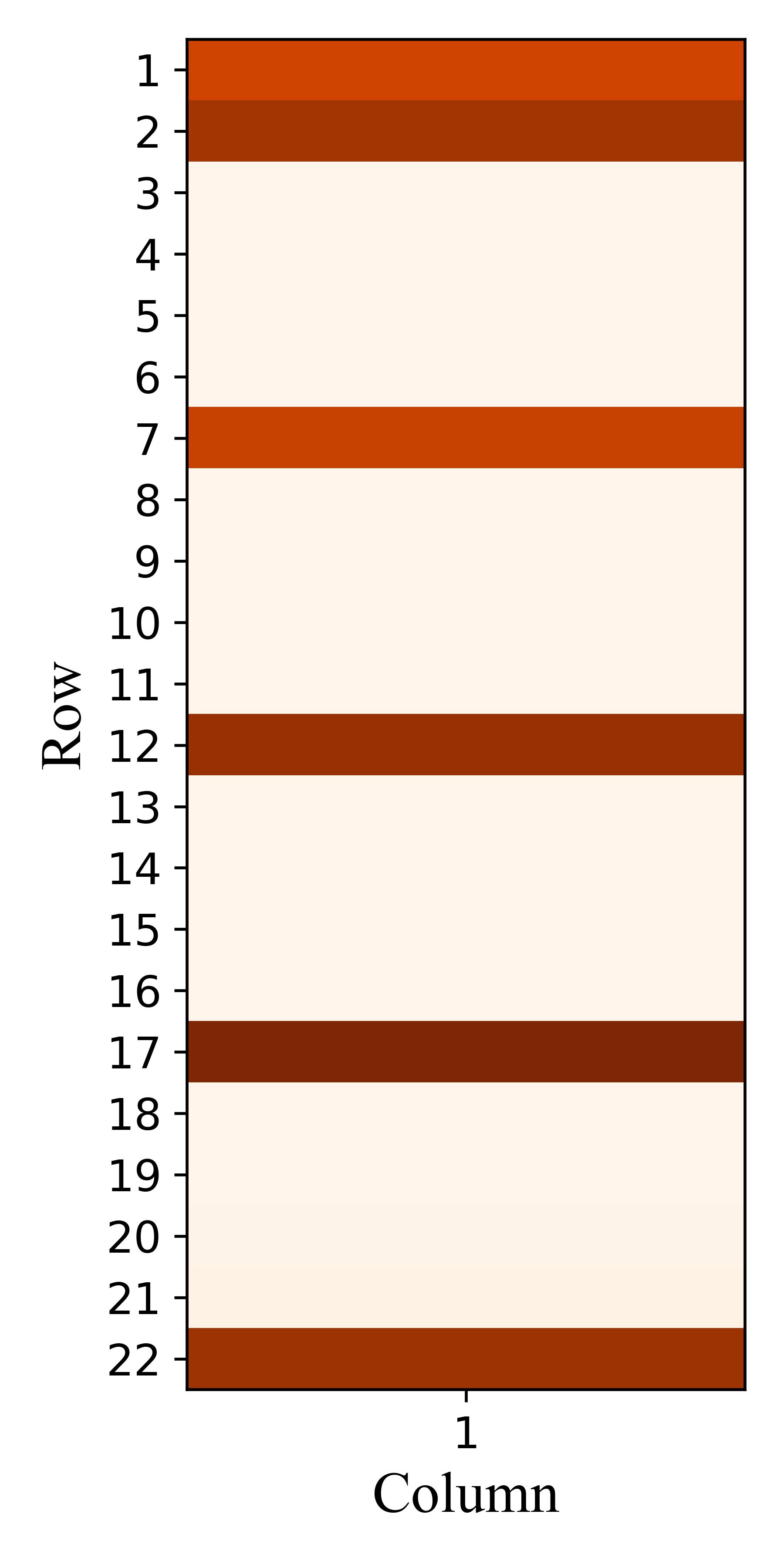}
    \label{fig:heatmap_22}
\end{subfigure}
\end{center}

This figure shows the heatmap plots of the ratios of sponsored products by position for product result pages containing 60 and 22 products.

\end{figure}

The positions that are most likely to be sponsored positions in our data are 1-4, 11-14, and 19-22 for a product page with 60 products, and 1, 2, 7, 12, 17, and 22 for a product page with 22 products.
Figure \ref{fig:by_position_sponsored_heatmap_combined} present the corresponding heatmaps of the density of sponsored products within a product result page.

\begin{figure}[!ht]
\begin{center}
\caption{Distribution of Sponsored Product Ads by Position for Different Page Layouts}
\label{fig:by_position_sponsored_combined}
\begin{subfigure}[b]{0.48\textwidth}
    \centering
        \caption{Page Layout with 60 products}
    \includegraphics[width=\textwidth]{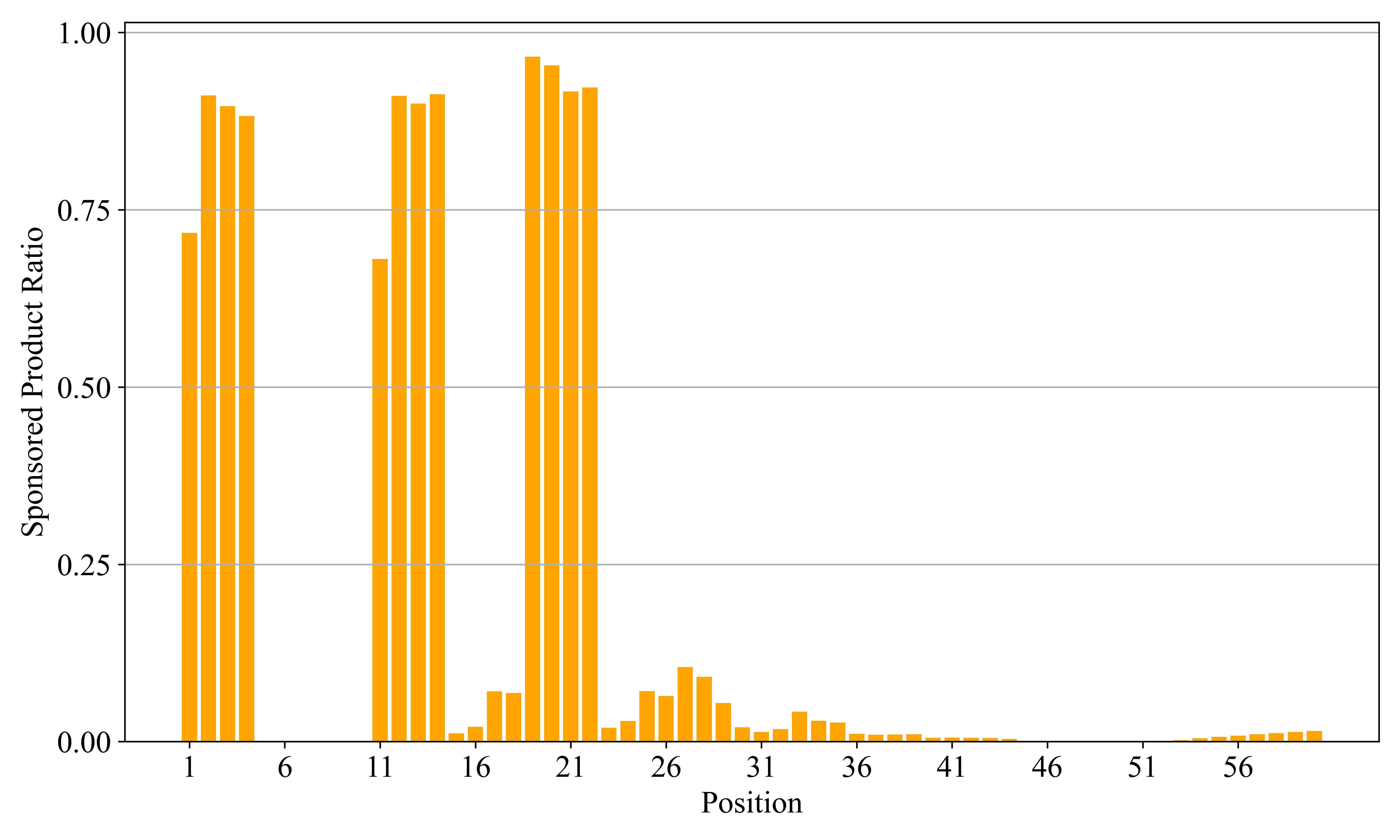}
    \label{fig:by_position_sponsored_60}
\end{subfigure}
\hfill
\begin{subfigure}[b]{0.48\textwidth}
    \centering
        \caption{Page Layout with 22 products}
    \includegraphics[width=\textwidth]{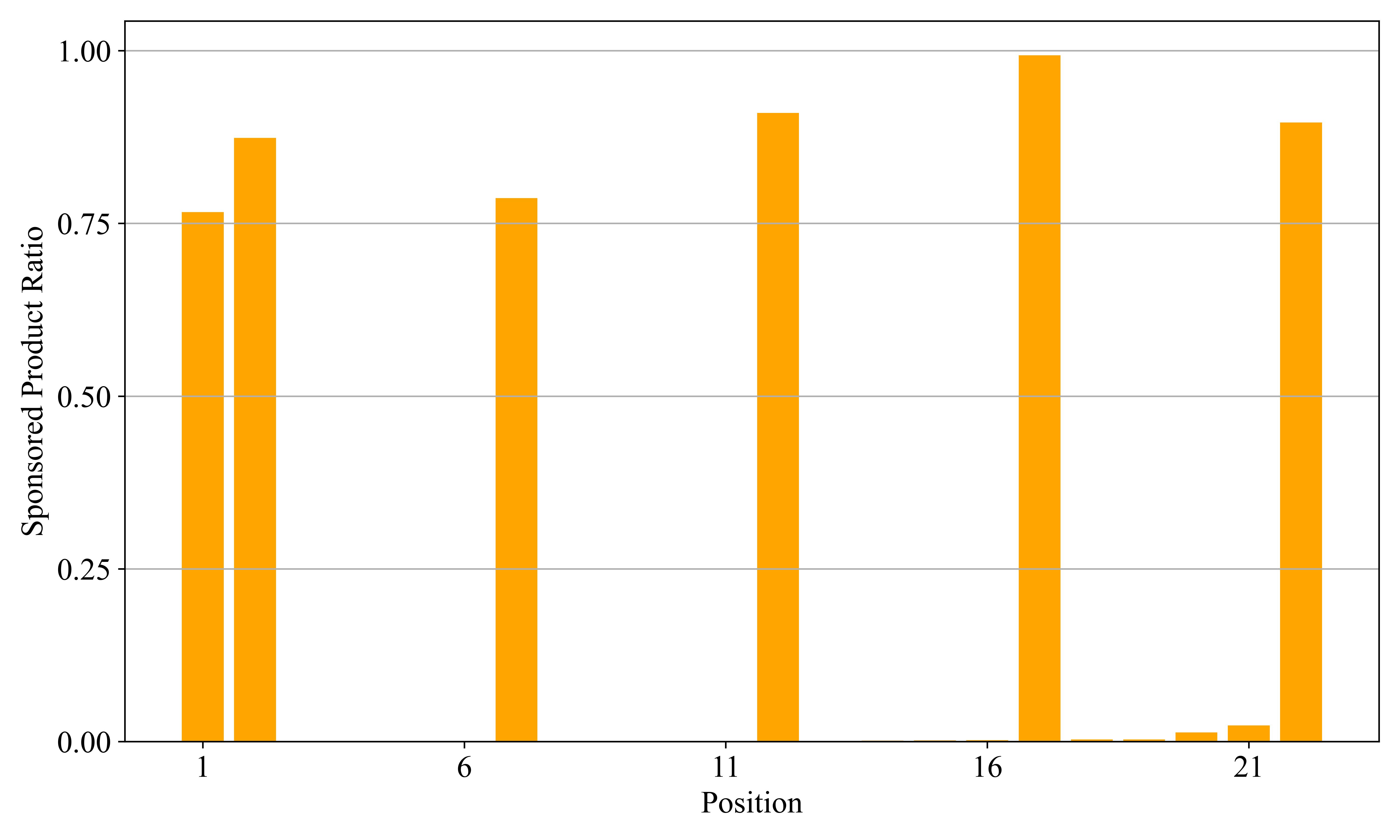}
    \label{fig:by_position_sponsored_22}
\end{subfigure}
\end{center}
This figure shows the ratios of sponsored products by position for product result pages with 60 and 22 products.
\end{figure}

\webappsubsection{Summary Statistics at Keyword Level}

Table     \ref{tab:summary_statistics_keyword} presents the summary statistics at keyword level.

\begin{table}[!ht]
    \centering
        \caption{Summary Statistics at Keyword Level}
    \begin{tabular}{lcccc} \hline
  & Mean & SD & Min & Max \\ \hline
Price & 56.83 & 86.24 & 6.576 & 1,073 \\
CPC & 1.393 & 0.896 & 0.420 & 7.390 \\
Monthly Sales & 5,041	& 3,903	& 207.2	& 28,097\\
Monthly search volume  & 284,196 & 396,570 & 50,008 & 4,167,001 \\

\hline
N = 2,382
\end{tabular}
    \label{tab:summary_statistics_keyword}
\end{table}

\webappsubsection{Comparing Sponsored and Organic Products}
\label{appsubsec:sponsored_organic_comparison}

\begin{figure}[ht]
\begin{center}
\caption{Position vs Observables}
\label{fig:by_position_price} 
\begin{subfigure}{0.4\textwidth}
    \centering
\caption{Price}
    \includegraphics[width=\textwidth]{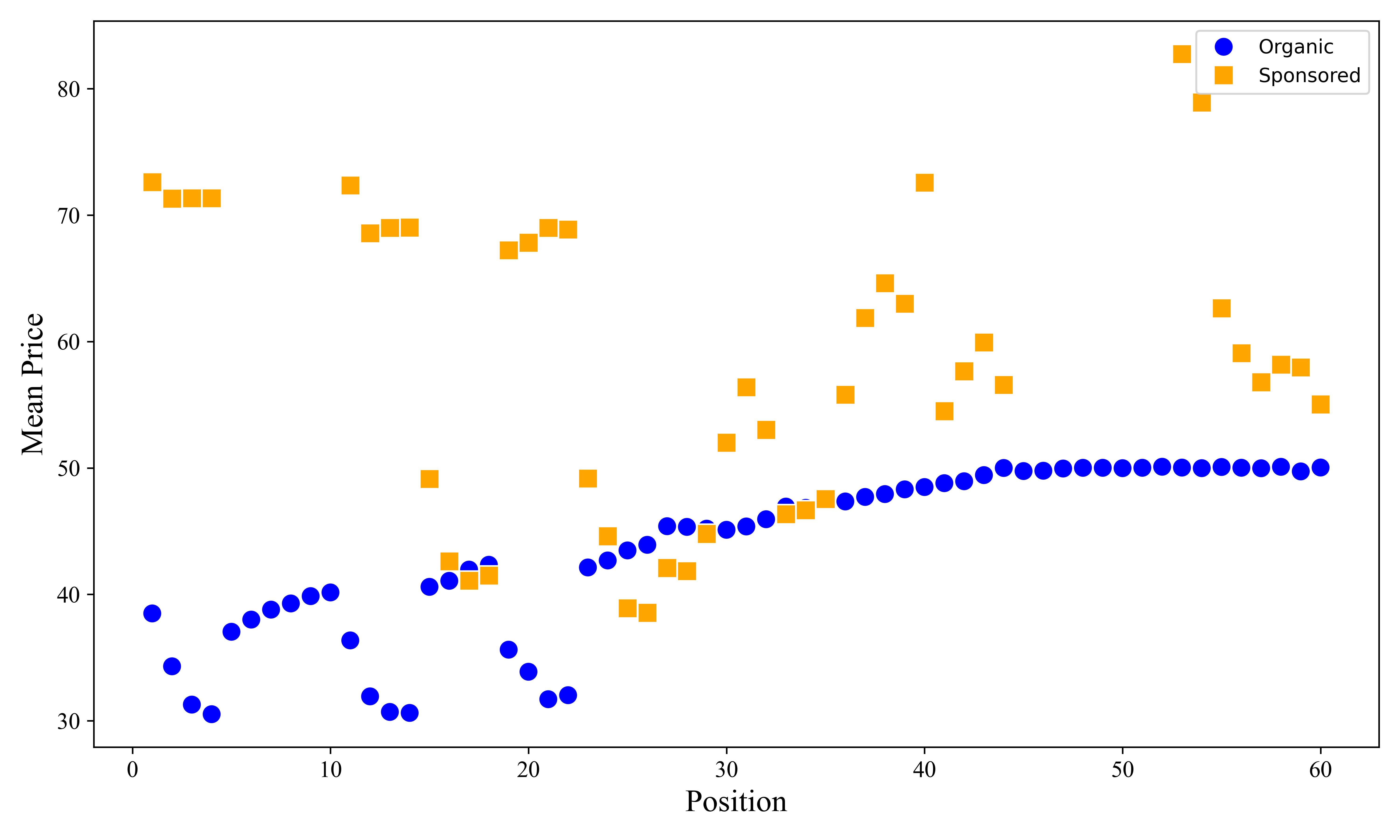}
\end{subfigure}
\hfill
\begin{subfigure}{0.4\textwidth}
\centering
\caption{Number Bought Last Month}
\includegraphics[width=\textwidth]{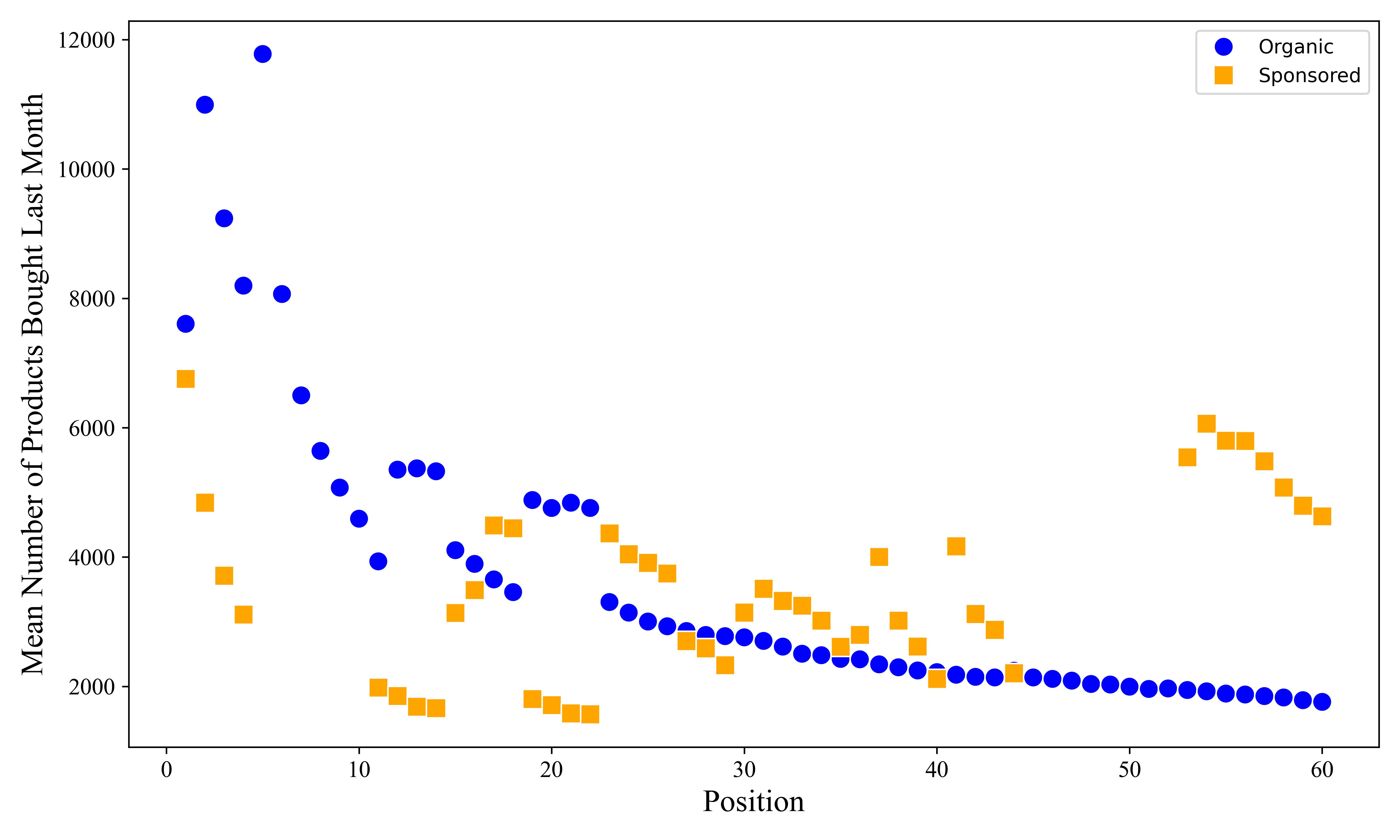}
\end{subfigure}
\begin{subfigure}{0.4\textwidth}
\centering
\caption{Number of Reviews}
\includegraphics[width=\textwidth]{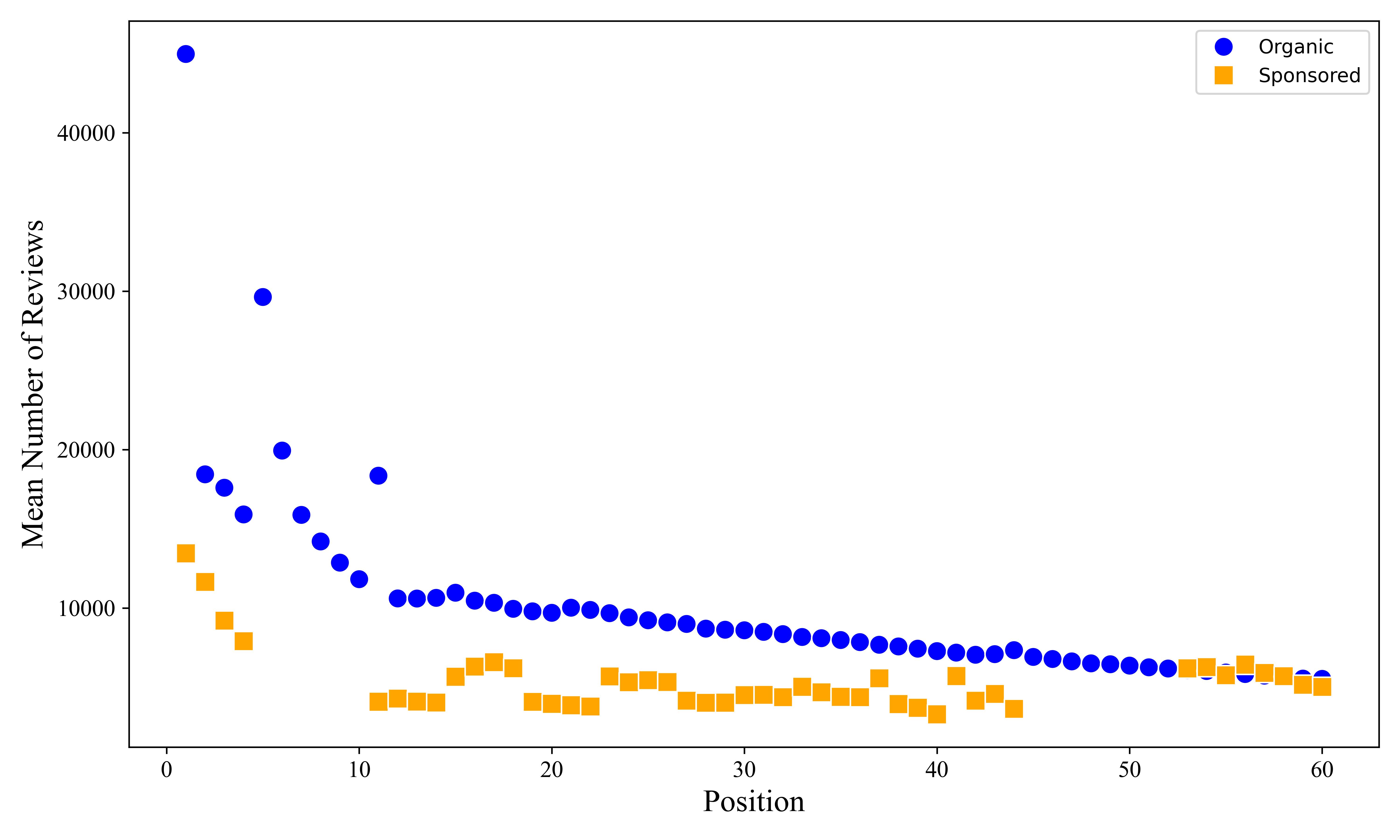}
\end{subfigure}
\hfill
\begin{subfigure}{0.4\textwidth}
\centering
\caption{Ratings}
\includegraphics[width=\textwidth]{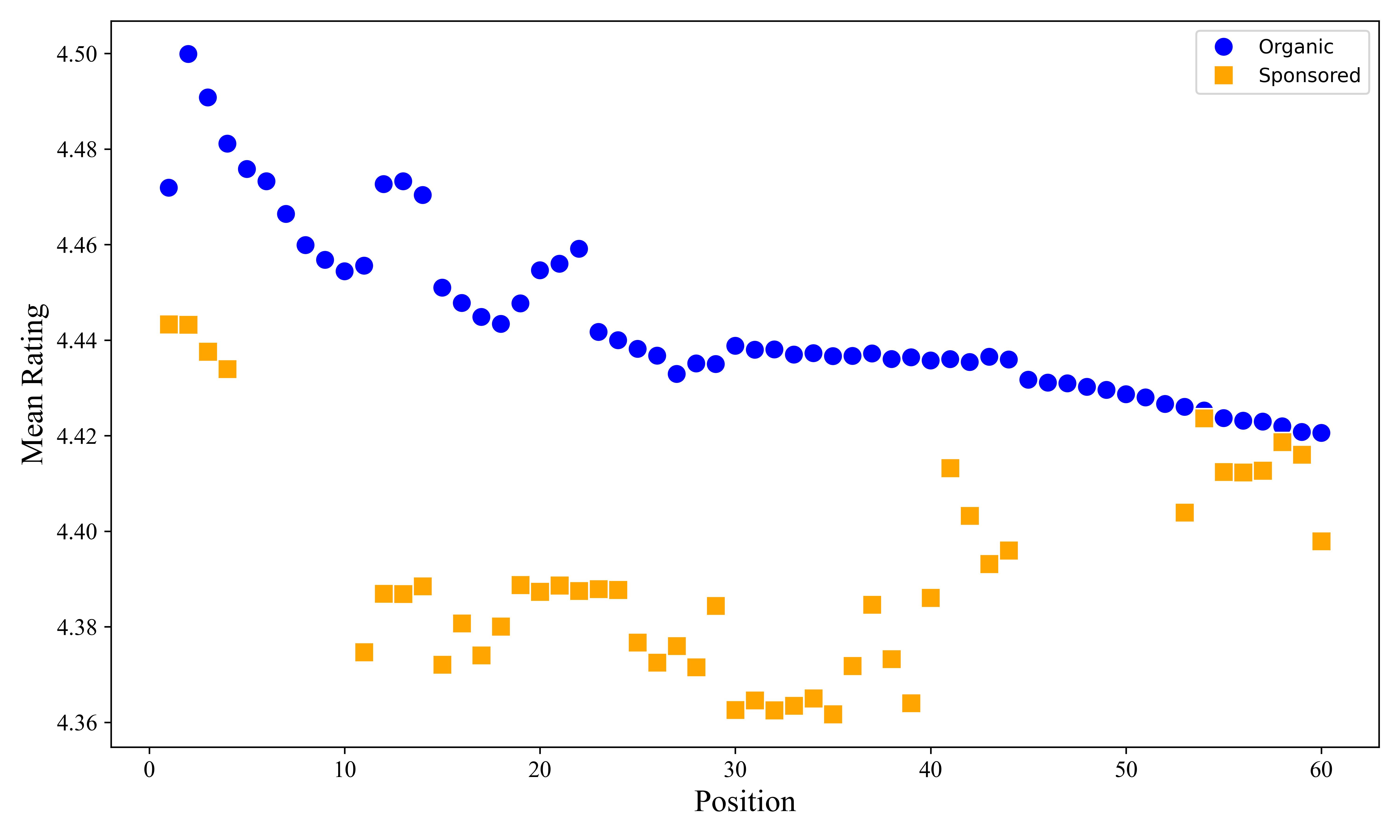}
\end{subfigure}

\end{center}
\small The panels (a)–(d) show the mean product prices, number of products bought last month, number of reviews, and ratings by each position for sponsored and organic products. The blue circle represents the organic products, while yellow squares denote the sponsored products.
\end{figure}

We plot the average prices, number of ratings, ratings, and number bought last month for a given position in Figure \ref{fig:by_position_price}. %Compared with organic products in the same positions, the sponsored products have higher prices, fewer reviews, and lower ratings, and fewer number  bought last month.  This intuitively makes sense, as sponsored product ads are additional costs for sellers, which would increase their marginal costs and, consequently, the prices in the market. This is also consistent with our theoretical prediction that sponsored product ads will increase equilibrium prices compared with the scenario without ads.
Compared with their organic counterparts in the same positions, sponsored products have lower number of reviews and number of products bought last month, indicating that they are more likely to be less established products and might be buying ads to gain awareness from consumers. The average ratings decrease with position, with sponsored products having lower average ratings than organic products in the same position. However, ratings might be a noisy indicator of quality, and sellers might be buying fake reviews \citep{he2022market}. We are cautious about drawing the conclusion that sponsored products have lower mean quality.

\webappsubsection{Regression: Negative Interaction}

We run the following equivalent regressions, controlling for category fixed effects.
\begin{equation}
price_{k} =\beta_0 + \beta_1 \cdot search \text{ } cost \text{ }  high_k +  \beta_2 \cdot algo_\text{ }high_k + \beta_3 \cdot search \text{ } cost \text{ }  high_k \cdot  algo_\text{ }high_k  +\epsilon_{k}
\end{equation}

Table \ref{tab:algo_search_inter} presents the results. The interaction of algorithm usage and consumer search costs on prices is significantly negative when controlling for category fixed effects.

\begin{table}[!ht]
    \centering
        \caption{Search Cost and Algorithm Usage Interaction}
    \begin{tabular}{lcc} \hline
  \multicolumn{3}{c}{Price}\\ \hline
  \\
Search cost high & 63.36*** & 42.55*** \\
 & (5.942) & (5.809) \\
Algo high & 2.502 & 1.797 \\
 & (5.942) & (5.727) \\
Algo high $\cdot$ search cost high & -22.41*** & -26.83*** \\
 & (8.404) & (7.819) \\
 constant & 30.45*** & 42.31*** \\
 & (4.077) & (3.900) \\
 &  &  \\
% Observations & 1,565 & 1,565 \\
% R-squared & 0.043 & 0.233 \\
 Category FE & No & Yes \\ \hline
\multicolumn{3}{c}{Standard errors in parentheses *** p$<$0.01, ** p$<$0.05, * p$<$0.1} \\
\end{tabular}
    \label{tab:algo_search_inter}
\end{table}

\end{document}